\documentclass[10pt]{article}
\usepackage{wrapfig}
\usepackage{cancel,amsthm,lscape, amssymb, amsmath,mathrsfs}

\usepackage{mathdots}
\usepackage{color,tikz}
\usetikzlibrary{patterns}
\usepackage{pgfplots}
\usetikzlibrary{decorations.markings,arrows, calc}
\usepackage{color,todonotes}
\usepackage{graphicx,framed,verbatim,caption}
\usepackage[colorlinks=true, pdfstartview=FitV, linkcolor=darkblue, citecolor=darkblue, urlcolor=darkblue]{hyperref}
\definecolor{shadecolor}{rgb}{0.9, 0.9, 0.86}
\definecolor{darkgreen}{rgb}{0.2, 0.5,  0}
\definecolor{darkblue}{rgb}{0.1,0.1,0.45}

\def\Re{\mathrm {Re}\,}

\def \Ai {\mathrm {Ai}}

\def \ref#1{\ref{#1})}
\def \& {&\hspace{-10pt}}
\def\Ai{ {\mathrm {Ai}}}

\def \wt{\widetilde}
\def\t{ {\mathbf t}}
\newcommand{\G}{\Gamma}
\newcommand{\g}{\gamma} 
\renewcommand{\d}{\mathrm d}
\newcommand{\pa}{\partial}       
\newtheorem{theorem}{Theorem}[section]
\newtheorem{example}[theorem]{Example}
\newtheorem{exercise}[theorem]{Exercise}

\newtheorem{conjecture}[theorem]{Conjecture}
\newtheorem{lemma}[theorem]{Lemma}
\newtheorem{remark}[theorem]{Remark}
\newtheorem{problem}[theorem]{Riemann--Hilbert Problem}
\newtheorem{fproblem}[theorem]{Formal Riemann-Hilbert Problem}

\newtheorem{proposition}[theorem]{Proposition} 
\newtheorem{corollary}[theorem]{Corollary} 
 
\newtheorem{definition}[theorem]{Definition}

\def\le{\left}
\def\ri{\right}

\def\res{\mathop{\mathrm {res}}\limits_}

\def\bt{\begin{theorem}}
\def\et{\end{theorem}}
\def\bc{\begin{corollary}}
\def\ec{\end{corollary}}
\def\bx{\begin{example}}
\def\ex{\end{example}}
\def\bxr{\begin{exercise}\small}
\def\exr{\end{exercise}}
\def\bl{\begin{lemma}}
\def\el{\end{lemma}}
\def\bd{\begin{definition}}
\def\ed{\end{definition}}
\def\bp{\begin{proposition}}
\def\ep{\end{proposition}}

\def\br{\begin{remark}}
\def\er{\end{remark}}

\def\be{\begin{equation}}
\def\ee{\end{equation}}
\def \Tr {\mathrm{Tr}\,}
\def\&{\hspace{-15pt}&}
\def\bea{\begin{eqnarray}}
\def\eea{\end{eqnarray}}
\def\beas{\begin{eqnarray*}}
\def\eeas{\end{eqnarray*}}

\def \pa{\partial}

\def\E{\mathrm E}

\def\N{{\mathbb N}}
\def\T{{\mathbf T}}
\def\t{{\mathbf t}}
\def\e{{\mathrm e}}
\def\wh{\widehat}

\def\Z{{\mathbb Z}}

\def\l{ \lambda }
\def\1{{\bf 1}}
\def\i{{\mathrm{i}}}
 
\def\z{\zeta}
\def\diag{{\mathrm{diag}}}
\def\Id{{\mathrm{Id}}}
\def\SL{{\mathrm{SL}}}
\def\Mat{{\mathrm{Mat}}}

\def\QED {\hfill $\blacksquare$\par\vskip 3pt}

\renewcommand{\theequation}{\arabic{section}.\arabic{equation}}

\makeatletter
\@addtoreset{equation}{section}
\makeatother

\begin{document}

\baselineskip 14pt plus 1pt minus 1pt

\begin{flushright}
\end{flushright}
\vspace{0.2cm}
\begin{center}
\begin{Large}
\textbf{ 
 The Kontsevich--Penner matrix integral, isomonodromic tau functions and open intersection numbers
} 
\end{Large}
\end{center}
\bigskip
\begin{center}
M. Bertola$^{\dagger\ddagger \clubsuit}$\footnote{Marco.Bertola@\{concordia.ca, sissa.it\}},  
G. Ruzza $^{\ddagger}$ \footnote{giulio.ruzza@sissa.it}.
\\
\bigskip
\begin{minipage}{0.7\textwidth}
\begin{small}
\begin{enumerate}
\item [${\dagger}$] {\it  Department of Mathematics and
Statistics, Concordia University\\ 1455 de Maisonneuve W., Montr\'eal, Qu\'ebec,
Canada H3G 1M8} 
\item[${\ddagger}$] {\it SISSA, International School for Advanced Studies, via Bonomea 265, Trieste, Italy }
\item[${\clubsuit}$] {\it Centre de recherches math\'ematiques,
Universit\'e de Montr\'eal\\ C.~P.~6128, succ. centre ville, Montr\'eal,
Qu\'ebec, Canada H3C 3J7} 
\end{enumerate}
\end{small}
\end{minipage}
\vspace{0.5cm}
\end{center}
\bigskip
%%%%%%%%%%%%%%%%  Abstract  %%%%%%%%%%%%%%%%
\begin{center}
\begin{abstract}
We identify the Kontsevich--Penner matrix integral, for finite size $n$, with the isomonodromic tau function of a $3\times 3$ rational connection on the Riemann sphere with $n$ Fuchsian singularities placed  in correspondence with the eigenvalues of the external field of the matrix integral. 
By formulating the isomonodromic system in terms of an appropriate Riemann--Hilbert boundary value problem, we can pass to the limit $n\to\infty$ (at a formal level) and identify  an isomonodromic system  
in terms of the Miwa variables, which play the role of times of a KP hierarchy. 
This allows to derive  the String and Dilaton equations via a purely Riemann--Hilbert approach.

The expression of the formal limit of the partition function as an isomonodromic tau function  allows us to derive explicit closed  formul\ae\ for the correlators of this matrix model in terms of the solution of the Riemann Hilbert problem with all times set to zero.  These correlators  have been conjectured to describe the intersection numbers for Riemann surfaces with boundaries, or {\it open intersection numbers}.
 \end{abstract}
\end{center}
\tableofcontents
%\end{titlepage}
\section{Introduction and results}
\subsection{The Kontsevich--Penner matrix integral and open intersection numbers}
The {\it Kontsevich--Penner matrix integral} is expressed as
\be
\label{kontsevichpennermatrixintegral}
\mathcal{Z}_n(Y;N) := \frac{\det(\i Y)^{N}}{\int_{H_n}\mathrm{d} M \exp\Tr(-YM^2)}\,\int_{H_n}\mathrm{d}M\,\frac{\exp\Tr\left(\frac{\i}{3}M^3-YM^2
\right)}{\det(M+\i Y)^N}.
\ee
Here $H_n$ is the space of $n\times n$ hermitian matrices, $N$ is a nonnegative integer in our applications but could be taken as arbitrary integer or even real parameter. The matrix  $Y=\diag(y_1,...,y_n)$ is a diagonal matrix satisfying $\Re y_j>0$ so that the integrals in  \eqref{kontsevichpennermatrixintegral} converge absolutely.

This matrix integral belongs to the family of the {\it generalized Kontsevich models} \cite{KhMaMiMoZa1992}; the choice of the potential as in \eqref{kontsevichpennermatrixintegral} \cite{Ko1992,Pe1988} has recently attracted some interest \cite{Al2015a,Al2015b,BrHi2015} as it is conjectured \cite{AlBuTe2017} that the correlators of the model described by \eqref{kontsevichpennermatrixintegral} provide {\it open intersection numbers}.

The matrix integral $\mathcal{Z}_n(Y;N)$ admits a regular asymptotic expansion as $y_j\to\infty$\footnote{By a regular asymptotic expansion as $y_j\to\infty$ we mean an asymptotic expansion involving negative integer powers of $y_j$ only and no exponential term.} within the sector $\Re y_j>0$, (see Paragraph \ref{paragraphwronskian}). There exists a formal power series $\mathcal{Z}(\T;N)$ in an infinite set of times $\T=(T_1,T_2,...)$ uniquely determined by
\begin{equation}
\label{formal}
\log\mathcal{Z}_n(Y;N)\sim \log\mathcal{Z}(\T;N),\;\;\;\;T_k=\begin{cases}\frac{1}{k}\Tr\le(Y^{-k}\ri) & k=1,...,n \\
0 & k\geq n+1\end{cases}
\end{equation}
as $n\to\infty$. The asymptotic equality \eqref{formal} should be understood as follows; the left hand side admits a (formal) Taylor expansion in the {\it Miwa variables} $T_{1},\dots, T_{n}$ (i.e. regular asymptotic expansion in the symmetric polynomials of $y_j^{-1}$); the coefficients of this expansion depend on $n$ but they {\it stabilize} as $n\to \infty$. Namely, each Taylor coefficient depends only on a finite number of $T_j$'s and as soon as $n$ is sufficiently large, the coefficient becomes independent on $n$; this is a general feature of generalized Kontsevich models \cite{KhMaMiMoZa1992}.

Consider now the rescaled times $\t=(t_1,t_2,...)$ defined by\footnote{We use a different normalization of the Kontsevich--Penner matrix integral with respect to \cite{AlBuTe2017}; this explains why our normalization of times is different.}
\begin{equation}
\label{scaledtimes}
t_k:=(-1)^{k}\,k!!\,2^{-\frac{k}{3}}\,T_k
\end{equation}
where we have used the double factorial $k!!:=\prod\limits_{j=0}^{\lfloor\frac{k-1}{2}\rfloor}(k-2j)$ for any nonnegative integer $k$.

\begin{conjecture}
\label{conjecture}\cite{AlBuTe2017}
The coefficients of the formal power series $\log\mathcal{Z}(\t;N)$ are the {\it open intersection numbers}.
\end{conjecture}
The open intersection numbers \cite{PaSoTe2014,Te2015,Bu2016,BuTe2017} are a generalization of the closed intersection numbers. The latter were introduced by Witten \cite{Wi1991};
\begin{equation}
\label{closedintersectionnumbers}
\langle\tau_{r_1}\cdots\tau_{r_n}\rangle_{c}:=\int_{\overline{\mathcal{M}_{h,n}}}\psi_1^{r_1}\wedge\cdots\wedge\psi_n^{r_n}
\end{equation}
where $c$ stands for {\it closed}, $r_i\in\mathbb{Z}_{\geq 0}$, $\overline{\mathcal{M}_{h,n}}$ is the moduli space of stable Riemann surfaces of genus $h$ with $n$ marked points and $\psi_i\in H^2\le(\overline{\mathcal{M}_{h,n}},\mathbb{Q}\ri)$ are tautological classes. The genus $h$ is read off the dimensional constraint $\sum_{i=1}^nr_i=\dim_\mathbb{C}\overline{\mathcal{M}_{h,n}}=3h-3+n$. In order to prove the Witten Conjecture \cite{Wi1991}, Kontsevich showed \cite{Ko1992} that
\begin{equation}
\label{kontsevich}
\langle\tau_{r_1}\cdots\tau_{r_n}\rangle_c=\le.\frac{\pa^n}{\pa t_{2r_1+1}\cdots\pa t_{2r_n+1}}\log\mathcal{Z}(\t;N=0)\ri|_{\t=0}.
\end{equation}
The generalization consists in considering moduli spaces of {\it open} Riemann surfaces, i.e. Riemann surfaces with boundary, with $n$ marked points which may belong to the boundary. The open intersection numbers $\le\langle\tau_{\frac{d_1}{2}}\cdots\tau_{\frac{d_n}{2}}\ri\rangle_o$ are polynomials in $N$, defined in terms of a combinatorial formula \cite{Te2015}. Here $o$ stands for {\it open}, $d_i\in\mathbb{Z}_{\geq 0}$, the even $d_i$'s label the interior marked points and the odd $d_i$'s label the boundary marked points. The dimensional constraint now reads as $\sum_{i=1}^nd_i=3g-3+2n$ where $g:=2h+b-1$ is the {\it doubled genus}\label{dg}, $h$ being the number of handles and $b$ the number of boundary components of the Riemann surfaces with boundary; the coefficient in front of $N^b$ in $\le\langle\tau_{\frac{d_1}{2}}\cdots\tau_{\frac{d_n}{2}}\ri\rangle$ is the actual intersection number over the moduli space of open Riemann surfaces with $b$ boundary components.

The Conjecture \ref{conjecture} can be stated as
\begin{equation}
\label{131}
\left\langle\tau_{\frac{d_1}{2}}\cdots\tau_{\frac{d_n}{2}}\right\rangle_o=\le.\frac{\pa^n}{\pa t_{d_1+1}\cdots\pa t_{d_n+1}}\log\mathcal{Z}(\t;N)\ri|_{\t=0}
\end{equation}
which would be a generalization of the Kontsevich's identity \eqref{kontsevich}.

\begin{remark}
\label{rmksafnuk}
In \cite{Sa2016} the author provides an alternative construction of the open intersection numbers and proves that their generating function is precisely $\log\mathcal{Z}(\t;N)$. The relation between the two definitions is still not clear.
\end{remark}

The ultimate goal of the paper is to provide explicit (closed) expressions for the generating functions of  the numbers \eqref{131} in the same spirit as in \cite{BeDuYa2016}. To do so we first need to identify $\mathcal Z_n$ with an isomonodromic tau function, as in \cite{BeCa2017} for the case $N=0$, and then identify a suitable (formal) limit of the resulting isomonodromic system. In the next subsection we provide more details on what we mean.

The final explicit formul\ae\ stemming from our approach can be concisely reported; 
consider the sequence of polynomials $P_{a,b}^k=P_{a,b}^k(N)$ in $N$ ($k=0,1,2,...$, $a,b=0,\pm 1$) defined by the generating functions
\begin{equation}
\label{polynomialsK}
\begin{gathered}
\renewcommand*{\arraystretch}{1.25}
\sum_{m\geq 0}\frac{\G\le(\frac{a-b+1}{2}\ri)}{\G\le(\frac{a-b+1+6m}{2}\ri)}P_{a,b}^{2m}(N)Z^m={\e^{\frac{Z}{3}}}\,_2F_2\le(\le.\begin{matrix}
\frac{1-a-b-2N}{2} & \frac{1+a+b+2N}{2} \\
\frac{1}{2} & \frac{1+a-b}{2}\end{matrix}\ri|-\frac{Z}{4}\ri)
\\
\renewcommand*{\arraystretch}{1.25}
\sum_{m\geq 0}\frac{ \G\le(\frac{a-b+2}{2}\ri)}{\G\le(\frac{a-b+4+6m}{2}\ri)}P_{a,b}^{2m+1}(N)Z^m=-\frac{2N+a+b}{2}\e^{\frac{Z}{3}}\,_2F_2\le(\le.\begin{matrix}
\frac{2-a-b-2N}{2} &\frac{2+a+b+2N}{2} \\
\frac{3}{2} & \frac{2+a-b}{2}\end{matrix}\ri|-\frac{Z}{4}\ri)
\end{gathered}
\end{equation}
where $_2F_2\le(\le.\begin{smallmatrix}\alpha&\beta \\ \gamma&\delta\end{smallmatrix}\ri|\z\ri):=\sum_{n\geq 0}\frac{(\alpha)_n(\beta)_n}{(\g)_n(\delta)_n}\frac{\z^n}{n!}$ is a generalized hypergeometric series\footnote{
Note that  for $a-b+1=0$ or $a-b+2=0$ both sides of \eqref{polynomialsK} have simple poles, and then the meaning of the identity is that of the residue. }, and form the matrix
\begin{equation}
\label{matrixA}
A(\l):=\begin{bmatrix}
N\sum\limits_{k\geq 0}P_{1,-1}^k(N)\l^{-\frac{3k+2}{2}} & \sum\limits_{k\geq 0}P_{-1,-1}^k(N)\l^{-\frac{3k}{2}} & \sum\limits_{k\geq 0}P_{0,-1}^k(N)\l^{-\frac{3k+1}{2}}\\
N\sum\limits_{k\geq 0}P_{1,0}^k(N)\l^{-\frac{3k+1}{2}} & \sum\limits_{k\geq 0}P_{-1,0}^k(N)\l^{-\frac{3k-1}{2}} & \sum\limits_{k\geq 0}P_{0,0}^k(N)\l^{-\frac{3k}{2}}\\
N\sum\limits_{k\geq 0}P_{1,1}^k(N)\l^{-\frac{3k}{2}} & \sum\limits_{k\geq 0}P_{-1,1}^k(N)\l^{-\frac{3k-2}{2}} & \sum\limits_{k\geq 0}P_{0,1}^k(N)\l^{-\frac{3k-1}{2}}
\end{bmatrix}.
\end{equation}

The formul\ae\ we obtain, taking \eqref{131} as definition for the open intersection numbers, are (Thm. \ref{thm2})

\begin{shaded}
\be
\label{1point0}
\sum_{d_1,...,d_n\geq 0}\le\langle\prod_{i=1}^n\frac{(-1)^{d_i+1}(d_i+1)!!}{2^{\frac{d_i+1}{3}}\l_i^{\frac{d_i}{2}+1}}\tau_{\frac{d_i}{2}}\ri\rangle_{o} =
\begin{cases}
-\sum\limits_{g\geq 1}\frac{2}{3g+2}P_{0,0}^{g+1}(N)\l_1^{-\frac{3g+1}{2}} & n=1 \\
\\
-\frac{1}{n}\sum\limits_{i\in S_n}\Tr\frac{A(\l_{i_1})\cdots A(\l_{i_n})}{(\l_{i_1}-\l_{i_2})\cdots(\l_{i_n}-\l_{i_1})}- \frac{\delta_{n,2}}{\le(\l_1^{\frac{1}{2}}-\l_2^{\frac{1}{2}}\ri)^{2}}& n\geq 2.
\end{cases}
\ee
\end{shaded}

\begin{example}\label{ex1point}
We remind the celebrated result of Itzykson and Zuber \cite{ItZu1992} expressing in a concise form the generating function for the  closed one--point intersection numbers;
\be
\label{closed1pf}
\sum_{r\geq 0}\langle\tau_{r-2}\rangle_c\, X^{r}=\e^{\frac{X^3}{24}}\ \ \ \Rightarrow \ \  \langle\tau_{3h-2}\rangle_c = \frac {1}{24^h h!}
\ee
where we have added the unstable closed intersection number $\le\langle\tau_{-2}\ri\rangle_c=1$. The formula \eqref{1point0} with the polynomial $P_{0,0}^{m}$ defined in \eqref{polynomialsK} gives then an appealing and simple form for the  analogous generating function of the {\it open} intersection numbers;
\begin{shaded}
\bea
\label{bellaformula}
\sum_{d\geq 0}\langle\tau_{\frac{d}{2}-2}\rangle_{_o}\, X^{\frac d2}=\e^{\frac{X^3}{6}}\le(_2F_2\le(\le.\begin{matrix}
\frac{1}{2}-N &\frac{1}{2}+N \\
\frac{1}{2} & \frac{1}{2}\end{matrix}\ri|-\frac{X^3}{8}\ri)+N \,X^\frac 32 \,_2F_2\le(\le.\begin{matrix}
1-N & 1+N \\
1 & \frac{3}{2}\end{matrix}\ri|-\frac{X^3}{8}\ri)\ri)
\eea 
\end{shaded}
\noindent We have added the unstable open intersection numbers $\le\langle\tau_{-2}\ri\rangle_{_o}:=1$, $\le\langle\tau_{-\frac{1}{2}}\ri\rangle_{_o}:=N$ for convenience in writing the generating function. 
The first values are (see also Appendix \ref{appendixtable})
\begin{equation}
\begin{gathered}
\left\langle\tau_1\right\rangle_{_o}=\frac{1+12N^2}{24},\;\;\left\langle\tau_{\frac{5}{2}}\right\rangle_{_o}=\frac{N+N^3}{12},\;\;\left\langle\tau_{4}\right\rangle_{_o}=\frac{1+56N^2+16N^4}{1152},\;\;\le\langle\tau_{\frac{11}{2}}\ri\rangle_{_o}=\frac{12N+25N^3+3N^5}{2880}.
\end{gathered}
\end{equation}
Note that the series only contains powers of $X$ that are  multiple of $\frac 32 $,  and  the doubled genus $g$ (defined on pag. \pageref{dg}), is related to $d$ by $g =\frac d 3 -1$.
If we set $N=0$ the first hypergeometric function in \eqref{bellaformula} becomes ${\rm e}^{-\frac {X^3}8}$ and we fall back to the standard closed intersection number case recovering the known formula \eqref{closed1pf}.
\end{example}
\subsection{Dressing, Schlesinger transformations and the isomonodromic tau function}\label{introductionschlesinger}
The first goal of the paper is to identify the matrix integral \eqref{kontsevichpennermatrixintegral} with an isomonodromic tau function, in the sense of Jimbo, Miwa and Ueno \cite{JiMiUe1981}, following a similar logic to the one used in \cite{BeCa2017}; historically the first time that matrix models of 2D gravity (which are, however, one-matrix models without external source) were related on a formal level to isomonodromic tau functions can be traced to the works \cite{Moore1,Moore2}.
 Consider the third order linear ODE ({\it bare equation})
\be
\label{scalarODE}
(\pa_\l^3-\l\pa_\l-N)\psi(\l)=0
\ee
which we can express equivalently as a first order matrix linear ODE
\begin{equation}
\label{bareODE}
\pa_\l\Psi(\l)=A(\l)\Psi(\l),\;\;\;\;A(\l):=\begin{bmatrix}
0&1&0 \\ 0&0&1 \\ N&\l&0
\end{bmatrix}.
 \end{equation}

In the following we will leave the dependence on $N$ implicit, for the sake of brevity.

The general solution to  \eqref{scalarODE} can be written explicitly in terms of Fourier--Laplace contour integrals (see \eqref{definitionair} and \eqref{defg}); the solution has an irregular singularity at $\l=\infty$ \cite{Wa2002} and exhibits a Stokes phenomenon (Proposition \ref{propbare}) not dissimilar from the prototypical example of the Airy equation (in fact \eqref{bareODE} is closely related to it).

In \ref{paragraphstokes} the $3\times 3$ fundamental matrix solution to  \eqref{bareODE} is expressed as a solution to  a Riemann--Hilbert matrix problem for a sectionally analytic matrix--valued function $\Psi$ in four sectors $I,...,IV$ of the $\l$-plane, see \eqref{Psi}. The restrictions $\Psi_I,...,\Psi_{IV}$ of $\Psi$ to the sectors admit the {\it same} asymptotic expansion, in the corresponding sector:
\be
\label{asymptoticpsi}
\begin{gathered}
\Psi(\l)=\G(\l)\e^{\vartheta(\l)},
\\
\G(\l)\sim\l^S \, G \, Y(\l) \, \l^{L},\;\;\;\l\to\infty\ \ \text{ in }\ \ I\cup\cdots\cup IV
\end{gathered}
\ee
where
\begin{equation}
\label{SGT}
\begin{gathered}
S:=\diag\le(-\frac{1}{2},-\frac{1}{2},0\ri),\;\;\;
G:=\begin{bmatrix}
1&0&0 \\ 0&\frac{1}{\sqrt{2}}&\frac{1}{\sqrt{2}} \\ 0&-\frac{1}{\sqrt{2}}&\frac{1}{\sqrt{2}}
\end{bmatrix}
\\
L:=\diag\le(-N+\frac{1}{2},\frac{N}{2}+\frac{1}{4},\frac{N}{2}+\frac{1}{4}\ri),\;\;\;
\vartheta(\l):=\diag\le(0,-\frac{2}{3}\l^{\frac{3}{2}},\frac{2}{3}\l^{\frac{3}{2}}\ri)
\end{gathered}
\end{equation}
and $Y(\l)=\1+\mathcal{O}\le(\l^{-\frac{1}{2}}\ri)$ is a formal power series in $\l^{-\frac{1}{2}}$ whose coefficients do not depend on the sector.

The matrix-valued function $\Psi$ is analytic in $I\cup\cdots\cup IV$ and its non-tangential boundary values on the rays separating adjacent sectors are related by right multiplication with constant invertible matrices; this is due to the fact that  $\Psi_I,...,\Psi_{IV}$ solve the same ODE \eqref{bareODE}. 
 
These constant multipliers are usually called {\it generalized monodromy matrices}; in this case they are {\it Stokes matrices} and {\it formal monodromy matrices}, see \eqref{stokes}.

\begin{remark}
All roots of $\lambda$ are principal roots, with $\arg(\lambda)\in (-\pi,\pi)$.
\end{remark}

The notion of {\it Schlesinger transformations} was formalized in \cite{JiMi1980} and is known also under the name of {\it dressing method}. In the present case the gist of the idea is as follows. Consider the bare ODE \eqref{bareODE} and fix $n$ points $\vec{\l}=(\l_1,...,\l_n)$ in the $\l$-plane; one is tasked to find a rational gauge transformation $R_n=R_n(\lambda;\vec{\l})$, with poles only at $\l\in\vec{\l}$, such that the generalized monodromy data of the gauge transformed system $\Psi_n := R_n\Psi$ is the same as  that of the bare ODE. In particular, the gauged connection matrix $A_n := R_n A R_n ^{-1}- \pa_\l R_n R^{-1}_n$ may have additional poles at $\l\in \vec \l$ but the induced monodromy representation around these poles must be trivial, on account that $A$ did not have already a pole there.

In \cite{JiMiUe1981} it is shown that this {\it isomonodromy} constraint implies that $\Psi_n$ satisfies a compatible system of linear differential equations
\begin{equation}
\label{isomonodromy0}
\begin{cases}
\pa_\l\Psi_n(\l;\vec{\l})=A_n(\l;\vec{\l})\Psi_n(\l;\vec{\l}) \\
\pa_{\l_j}\Psi_n(\l;\vec{\l})=\Omega_{n,j}(\l;\vec{\l})\Psi_n(\l;\vec{\l})&(j=1,...,n)
\end{cases}
\end{equation}
with {\it zero curvature equations}
\begin{equation}
\label{compatibility0}
\pa_{\l_j}A_n-\pa_\l\Omega_{n,j}=[\Omega_{n,j},A_n],\;\;\,\;
\pa_{\l_j}\Omega_{n,k}-\pa_{\l_k}\Omega_{n,j}=[\Omega_{n,j},\Omega_{n,k}]\;\;\,\;\;(j,k=1,...,n).
\end{equation}
To such an {\it isomonodromic system} one can associate a (locally defined) function $\tau_n(\vec{\l};N)$, called {\it isomonodromic tau function} as in \cite{JiMiUe1981} (see below for the definition in the present case).

It was already shown in \cite{JiMi1980} that a tau function arising in such a way from Schlesinger transformations is the determinant of some explicit $n\times n$ {\it characteristic matrix}.

We take here a different point of view (the same as in \cite{BeCa2017}). The notion of Schlesinger transformation can be reformulated entirely in terms of Riemann--Hilbert data, i.e. in terms of monodromy data; this allows to control certain asymptotic limits where the degree of $R$ tends to infinity, namely the limit $n\to\infty$ (we are not considering this detailed analysis in the present paper but it was done in loc. cit.).

More precisely, there exists a diagonal  matrix $D_n=D_n(\l;\vec{\l})$ (rational in $\sqrt{\l}$, see \eqref{defD} below for an explicit expression), with zeros only at $\l\in\vec{\l}$ and such that $R_n\Psi D_n$ is analytic at $\l\in\vec{\l}$. Under such a correspondence, the Schlesinger transformation $\Psi\mapsto R_n\Psi$ corresponds to the {\it dressing} of the monodromy data $\wt{M}\mapsto D^{-1}_n\wt{M}D_n$ (more details in Sec. \ref{sectionisomonodromic}).

In \cite{Be2010} a general formula for the isomonodromic tau function was found in terms of the monodromy data; adapted to the present case it implies that the tau function $\tau_n(\vec{\l};N)$ of the system \eqref{isomonodromy0}--\eqref{compatibility0} is defined by
\be
\label{malgrange0}
\pa_{\l_j}\log\tau_n(\vec{\l};N)=\int_\Sigma\frac{\d\l}{2\pi\i}\ \Tr\le(\G_{n-}^{-1}\ \pa_\l\G_{n-}\ \pa_{\l_j}M_n\ M_n^{-1}\ri)
\ee
where $\Psi_n=\G_n D^{-1}_n\e^{\vartheta}$ ($\vartheta$ as in \eqref{SGT}), $M_n$ are the dressed jump matrices for $\G_n$, defined along $\Sigma$ which is the union of the oriented rays separating the sectors $I,...,IV$ described above (see figure \ref{figurejumps}); the subscript $-$ indicates the choice of boundary value at the oriented contour $\Sigma$.

In general, the expression in the right-hand side of \eqref{malgrange0} is termed {\it Malgrange differential}.

The effect of dressing the jump matrices by a rational diagonal matrix on the Malgrange differential, and consequently on the tau function, was studied in \cite{BeCa2015}; the result is a direct argument for the identification, up to some explicit rational expression, of the tau function for the system \eqref{isomonodromy0}-\eqref{compatibility0} with the determinant of the {\it characteristic matrix} of \cite{JiMi1980}, along with an interpretation of the linear map represented by this matrix.

By a careful analysis of this determinant (carried over in Appendix \ref{appendixpropb}) one obtains the following Theorem, which will be proved as a special case of a slightly more general theorem (Thm. \ref{Thmmaximus}) in Section \ref{proofthm1}.

\begin{shaded}
\begin{theorem}
\label{thm1}
The Kontsevich--Penner matrix integral \eqref{kontsevichpennermatrixintegral} coincides with the tau function \eqref{malgrange0};
\be
\mathcal{Z}_n(Y;N)=\tau_n (\vec{\l};N)
\ee
with the identification $\l_j=y_j^2$.
\end{theorem}
\end{shaded}
\subsection{The limiting isomonodromic system}\label{introductionlimiting}
The matrix $D_n(\l;\vec{\l})$ is such that up to a constant scalar matrix the following {\it formal} identity holds
\be
D_n^{-1}\propto\exp\sum_{k\geq 1}T_k\lambda^{\frac{k}{2}}\theta_k,\;\;\;\;\;\theta_k=\diag(0,(-1)^k,1)
\ee
where $\T=(T_1,T_2,...)$ are the aforementioned Miwa variables $T_k=\frac{1}{k}\sum_{j=1}^n\lambda_j^{-\frac{k}{2}}$. This suggests to introduce directly the matrix  $\Psi(\l;\T)$, as a  function of the Miwa times $\T$, as we now explain. For the sake of definiteness, one has to assume that we have truncated the Miwa times, i.e. $T_k=0$ for $k\gg 1$. Then  $\Psi(\l;\T)$ is a matrix-valued function, sectionally analytic in the sectors $I,...,IV$, satisfying the same jump condition as $\Psi(\l)$ on $\Sigma$ and having an asymptotic expansion 
\be
\begin{gathered}
\label{BAfunction}
\Psi(\l;\T)=\G(\l;\T)\e^{\Theta(\l;\T)},\;\;\;\;\Theta(\l;\T)=\sum\limits_{k\geq 1}\le(T_k+\frac{2}{3}\delta_{k,3}\ri)\lambda^{\frac{k}{2}}\theta_k,
\\
\G(\l;\T)\sim\l^S\,G\,\le(\1+\mathcal{O}\le(\l^{-\frac{1}{2}}\ri)\ri)\,\l^L,\;\;\;\;\l\to\infty\text{ in }I,\cdots,IV
\end{gathered}
\ee
with $S,G,L$ as in \eqref{SGT}. More details are provided in Sec \ref{section3}.

To $\Psi(\l;\T)$ one can associate a tau function $\tau(\T;N)$ as in \cite{JiMiUe1981} by
\begin{equation}
\label{taujapanese}
\pa_{T_k}\log\tau(\T;N)=-\res{\l=\infty}\Tr\le(\l^{\frac{k}{2}}\ \G^{-1}\ \pa_\l\G\ \theta_k\ri)
\end{equation}
(compare with Proposition \eqref{propjapanese}).

By the  {\it isomonodromic method} we refer to the identification of  the (formal)  limit  of the Kontsevich--Penner matrix integral  $\mathcal Z_n(Y;N)$   with the above tau function $\tau(\T;N)$ (see Prop. \ref{proplimiting}). 
This identification  allows us to exploit a rich formalism to compute its logarithmic derivatives of arbitrary order; this is best achieved by arranging them into suitable {\it generating functions} (usually called {\it correlation functions}). To this end we define the {\it correlators} as the expressions below:

\begin{equation}
\begin{gathered}
\left\langle\left\langle\tau_{\frac{d_1}{2}}\cdots\tau_{\frac{d_n}{2}}\right\rangle\right\rangle:=\frac{\pa^n\log\tau(\t;N)}{\pa t_{d_1+1}\cdots\pa t_{d_n+1}}\;\;\;\;(n\geq 1, \,\,d_j\geq 0)
\\
\left\langle\tau_{\frac{d_1}{2}}\cdots\tau_{\frac{d_n}{2}}\right\rangle:=\le.\left\langle\left\langle\tau_{\frac{d_1}{2}}\cdots\tau_{\frac{d_n}{2}}\right\rangle\right\rangle\ri|_{\t=0}\;\;\;\;(n\geq 1,\,\,d_j\geq 0).
\end{gathered}
\end{equation}

The aforementioned correlation functions are then given by

\be
\begin{gathered}
S_n(\l_1,...,\l_n;\t):=\sum_{d_1,...,d_n\geq 0}\left\langle\left\langle\tau_{\frac{d_1}{2}}\cdots\tau_{\frac{d_n}{2}}\right\rangle\right\rangle \frac{(-1)^{d_1+1}(d_1+1)!!}{2^\frac{d_1+1}{3}\l^{\frac{d_1}{2}+1}_1}\cdots\frac{(-1)^{d_n+1}(d_n+1)!!}{2^\frac{d_n+1}{3}\l_n^{\frac{d_n}{2}+1}}
\;\;\;\;(n\geq 1)
\\
S_n(\l_1,...,\l_n):=S_n(\l_1,...,\l_n;\t=0)\;\;\;\;(n\geq 1).
\end{gathered}
\end{equation}

\begin{shaded}
\begin{theorem}
\label{thm2}
The correlation functions are
\begin{equation}
\label{1pointresult}
S_1(\l)=-\sum_{g\geq 1}\frac{2}{3g+2}P_{0,0}^{g+1}(N)\l^{-\frac{3g+1}{2}}
\end{equation}
\begin{equation}
\label{npointresult}
S_n(\l_1,...,\l_n)=-\frac{1}{n}\sum_{i\in S_n}\Tr\frac{A(\l_{i_1})\cdots A(\l_{i_n})}{(\l_{i_1}-\l_{i_2})\cdots(\l_{i_n}-\l_{i_1})}- \frac{\delta_{n,2}}{\le(\l_1^{\frac{1}{2}}-\l_2^{\frac{1}{2}}\ri)^{2}} \ \ \ \ \ (n\geq 2)
\end{equation}
where the polynomials $P_{0,0}^{k}(N)$ are as in \eqref{polynomialsK} and the matrix $A(\l)$ is as in \eqref{matrixA}.
\end{theorem}
\end{shaded}

The proof is given in Section \ref{secproofthm2}.

\begin{example}
For $n=1$ the formula \eqref{1pointresult} can be expressed as in Example \ref{ex1point}. For $n=2,3$ the formula \eqref{npointresult} can be written in simple form  exploiting the cyclic property of the trace, as
\begin{equation}
S_2(\l_1,\l_2)=\Tr\frac{A(\l_1)A(\l_2)}{(\l_1-\l_2)^2}- \frac 1{\le(\l_1^{\frac{1}{2}}-\l_2^{\frac{1}{2}}\ri)^{2}}
\end{equation}
\begin{equation}
S_3(\l_1,\l_2,\l_3)=-\Tr\frac{A(\l_1)A(\l_2)A(\l_3)-A(\l_2)A(\l_1)A(\l_3)}{(\l_1-\l_2)(\l_2-\l_3)(\l_3-\l_1)}
\end{equation}
\end{example}

\vskip 0.3cm
A second application of the isomonodromic method is the derivation of the {\it String} and {\it Dilaton} equations for the Kontsevich--Penner model (already established in the literature in \cite{BrHi2012}, \cite{Al2015b}) as a consequence of  translation and dilation co-variance of the system \eqref{TkODE}--\eqref{compatibility3}:

\begin{proposition}
\label{propositionstringdilaton}
The isomonodromic tau function $\tau(\T;N)$, and consequently the partition function $\mathcal Z(\T;N)$, satisfies
\begin{equation}
\label{stringeq}
\left(\sum_{k\geq 3}\frac{k}{2}T_k\frac{\pa}{\pa T_{k-2}}+\frac{\pa}{\pa T_1}+\frac{T_1^2}{4}+NT_2\ri)\tau(\T;N)=0
\end{equation}
\begin{equation}
\label{dilatoneq}
\left(\sum_{k\geq 1}\frac{k}{2}T_k\frac{\pa}{\pa T_k}+\frac{\pa}{\pa T_3}+\frac{1}{16}+\frac{3N^2}{4}\right)\tau(\T;N)=0.
\end{equation}
\end{proposition}
The proof is found in Section \ref{proofprop18}.

\begin{remark}
Our normalization for the main variables differs slightly from the literature, e.g. the set of times used in \cite{Al2015b} is related to our time variables $T_k$ as $(-1)^k\,2^{-\frac{k}{3}} T_k$.
\end{remark}

\paragraph{Organization of the paper.}
In Section \ref{sectionisomonodromic} we study the ODE \eqref{bareODE} and define the suitable Schlesinger transformations in terms of a Riemann--Hilbert problem along the lines of \cite{Be2010}. Exploiting the theory developed in \cite{BeCa2015}, already applied to the case $N=0$ in \cite{BeCa2017}, we prove Theorem \ref{thm1} by explicit computation of the tau function. In section \ref{section3} we study the formal limiting Riemann--Hilbert problem and prove Theorem \ref{thm2} and Proposition \ref{propositionstringdilaton}. Technical parts of the proofs are postponed to Appendices \ref{proofA} and \ref{appendixpropb}.
In Appendix \ref{appendixtable} we collect a table of some open intersection numbers, up to six--point, obtained with the formul\ae\ of Theorem \ref{thm2}; we observe that these ``intersection numbers'' are polynomials in $N$ whose constant term was verified to coincide with the standard (closed) intersection numbers.
\paragraph {\bf Acknowledgements.} The research of M. B.   was supported in part by the Natural Sciences and Engineering Research Council of Canada grant
RGPIN-2016-06660. G. R. wishes to thank the Department of Mathematics and Statistics at Concordia University for hospitality during which the work was completed. 

\section{The Kontsevich--Penner integral as isomonodromic tau function}
\label{sectionisomonodromic}
\paragraph{Wronskian representation of the Kontsevich--Penner matrix integral.}
\label{paragraphwronskian}
The matrix integral \eqref{kontsevichpennermatrixintegral} can be rewritten using the following chain of equalities:
\be \label{zn}
\begin{gathered}
\int_{H_n}\mathrm{d} M\,\exp\Tr\left(\frac{\i}{3}M^3-YM^2-N\log
(M+\i Y)\right)=\\
\stackrel{(1)}{=}\exp\le(\frac{2}{3}\Tr Y^3\ri)\int_{H_n}\mathrm{d}M'\,\exp\Tr\left(\i \frac{{M'}^3}{3}+\i M'Y^2-N\log M'\right)=\\
\stackrel{(2)}{=}C_n\,\exp\le(\frac{2}{3}\Tr Y^3\ri)\int_{\mathbb{R}^n}\Delta^2(X)\prod_{j=1}^n\frac{\exp\frac{\i x_j^3}{3}}{x_j^N}\mathrm{d}x_j\int_{U(n)}\mathrm{d}U\exp\Tr\le(\i Y^2UXU^\dag\ri)=\\
\stackrel{(3)}{=}\widetilde{C_n}\,\exp\le(\frac{2}{3}\Tr Y^3\ri)\int_{\mathbb{R}^n}\frac{\Delta(X)\det\le[\exp\le(\i x_jy_k^2\ri)\ri]_{j,k=1}^n}{\Delta(Y^2)}\prod_{j=1}^n\frac{\exp\frac{\i x_j^3}{3}}{x_j^N}\mathrm{d}x_j=\\
\stackrel{(4)}{=}n!\,\widetilde{C_n}\,\frac{\exp\le(\frac{2}{3}\Tr Y^3\ri)}{\Delta(Y^2)}\det\left[\int_\mathbb{R}x^{j-1-N}\exp\le(\frac{\i x^3}{3}+\i xy_k^2\ri)\mathrm{d}x\right]_{j,k=1}^n
\end{gathered}
\ee
In (1) we perform a shift $M':=M+\i Y$ and an analytic continuation: the integral is now only conditionally convergent, it is absolutely convergent only when understood as integration over $H_n+\i\epsilon\1$ for any $\epsilon>0$. In (2) we apply Weyl integration formula and we use the notation $X=\diag(x_1,...,x_n)$,  with $\Delta(X)=\det[x_j^{k-1}]_{1 \leq j,k\leq n}=\prod_{1\leq j<k\leq n}(x_k-x_j)$ being the Vandermonde determinant and $\mathrm{d}U$ the Haar measure on $U(n)$. In (3) we apply Harish-Chandra formula and in (4) Andreief identity. $C_n,\widetilde{C_n}$ denote constants depending only on $n$ that we need not make explicit (compare with \eqref{eq}).

\begin{definition}
For $N\in \Z$ we define the functions 
\be \label{definitionair}
f(\l;N):=\frac{\i^N}{\sqrt{2\pi}}\int_{\mathbb{R}+\i\epsilon}\frac{\exp\le(\frac{\i x^3}{3}+\i x\l\ri)}{x^N}\mathrm{d}x
\ee
The integral is absolutely convergent for any $\epsilon>0$ and it defines an entire function of $\l$ (independent of $\epsilon$).
\end{definition}
\begin{remark}
$f(\l;0)=\sqrt{2\pi}\,\Ai(\l)$, where $\Ai(\l)$ is the Airy function. The sequence of functions $\big (f(\l;N)\big)_{N\in \Z}$ satisfies the simple recurrence relation
\be
\label{sequence}
\pa_\l f(\l;N)=-f(\l;N-1).
\ee
Moreover each function in \eqref{definitionair} satisfies the differential equation
\begin{equation}
\label{diffeqf}
\le(\pa_\l^3-\l\pa_\l+N-1\ri)\,f(\l;N)=0
\end{equation}
as it is easily verified by using integration by parts.
Combining  \eqref{sequence}  and \eqref{diffeqf}, we also obtain the following recurrence relation 
\begin{equation}
\label{basicidentity}
f(\l;N-3)-\l f(\l;N-1)-(N-1) f(\l;N)=0.
\end{equation}
\end{remark}

\begin{proposition}\label{propode0}
When $\l\to\infty$ within the sector $-\pi<\arg\l<\pi$ we have
\begin{equation}
\label{asymptoticsrealwithallcoefficients}
f(\l;N)\sim\frac{\exp\le(-\frac{2}{3}\l^{\frac{3}{2}}\ri)}{\sqrt{2} \l^{\frac{N}{2}+\frac{1}{4}}}F_-(\l;N)
\end{equation}
where
\begin{equation}
\label{defCjN}
F_-(\l;N)=1+ \sum_{j\geq 1}(-1)^jC_j(N)\l^{-\frac{3j}{2}},\;\;\;C_j(N):=\sum_{b=0}^{2j}\frac{(-1)^b}{3^bb!}{-N\choose 2j-b}\frac{\G\le(\frac{1}{2}+j+b\ri)}{\sqrt{\pi}}.
\end{equation}
\end{proposition}

The proof is based on a formal steepest descent argument and is contained in Appendix \ref{proofA}.
\begin{remark}
\label{remarkexpansion}

In different sectors (e.g. in $\pi<\arg\l<3\pi$) a formal analytic continuation of the expression in the right-hand side of \eqref{asymptoticsrealwithallcoefficients} is needed, so we shall consider also the power series
\begin{equation}
\label{Fpiu}
F_+(\l;N)=1 + \sum_{j\geq 1}C_j(N)\l^{-\frac{3j}{2}}.  
\end{equation}
\end{remark}
\begin{remark}
As a  corollary of the recurrence relation \eqref{basicidentity} we obtain the following recurrence relation for the formal series $F_\pm(\;N)$:
\be
\label{basicidentity2}
F_\pm(\l;N-2)-F_\pm(\l;N)\pm N\l^{-\frac{3}{2}}F_\pm(\l;N+1)=0.
\ee
\end{remark}

\begin{proposition}\label{computationzn}
The Kontsevich--Penner matrix integral can be expressed as follows:
\be
\label{eq}
\mathcal{Z}_n(Y;N)=2^{\frac{n}{2}}\,\frac{\le(\det\,Y^{N+\frac{1}{2}}\ri)\,\exp\le(\frac{2}{3}\Tr Y^3\ri)}{\Delta(Y)}\det\left[\pa_\l^{j-1}\,f(\l;N)\big|_{\l=y_k^2}\right]_{j,k=1}^n \ .
\ee
\end{proposition}
\noindent {\bf Proof.}  
From \eqref{zn} and the Gaussian integral formula
\be
\int_{H_n}\exp\Tr\le(-YM^2\ri)=\frac{{\sqrt{\pi}}^{n^2}}{\prod\limits_{j=1}^n \sqrt{y_j}\prod\limits_{1\leq j<k\leq n}(y_j+y_k)}
\ee
we have only to recover the proportionality constant in \eqref{eq}. In the limit $Y\to\diag(+\infty,...,+\infty)$ we have $\mathcal{Z}_n(Y;N)\to 1$ and, by \eqref{sequence} and \eqref{asymptoticsrealwithallcoefficients},
\begin{equation}
\det\left[\pa_\l^{j-1}\,f(\l;N)\big|_{\l=y_k^2}\right]_{j,k=1}^n\sim
\Delta(Y)\prod_{j=1}^n\frac{\exp\le(-\frac{2}{3}y_j^3\ri)}{\sqrt{2}\,y_j^{N+\frac{1}{2}}}
\end{equation}
so that this constant is eventually found to be $2^{\frac{n}{2}}$.
\QED 
\subsection{Stokes phenomenon for the bare system}\label{paragraphstokes}
Consider the ODE \eqref{bareODE} and introduce
\begin{equation}
\label{defg}
g(\l;0):=1,\ \ 
g(\l;N):=\frac{(-\i)^N}{\G(N)}\int_0^{+\infty\e^{\i\epsilon}}x^{N-1}\exp\le(\frac{\i x^3}{3}+\i x\l\ri)\,\mathrm{d}x
\ ,\qquad \ \ N = 1,2,\dots.
\end{equation}
(Note that $g(\l,0)$ is also the limit as $\nu\to 0$ of $ g(\l;\nu)$).
The integral is absolutely convergent for any $0<\epsilon<\frac{\pi}{3}$ so it defines an entire function of $\l$ (independent of $\epsilon$).
Using integration by parts it is easy to check that $g(\l;N)$ satisfies the differential equation
\begin{equation}
\label{diffeqg}
(\pa_\l^3-\l\pa_\l-N)\,g(\l;N)=0.
\end{equation}

\begin{proposition} \label{propode}
When $\l\to\infty$ within the sector $-\frac{\pi}{3}<\arg\l<\pi$
\begin{equation}
\label{asymptoticsg}
g(\l;N)\sim \l^{-N}\le(1+\mathcal{O}\le(\l^{-3}\ri)\ri).
\end{equation}
\end{proposition}
The proof is contained in Appendix \ref{proofA}.

\begin{definition}
\label{defsectors}
Fix three angles $\beta_\pm,\beta_0$ such that
\begin{equation}
-\pi<\beta_-<-\frac{\pi}{3},\;\;\;\;-\frac{\pi}{3}<\beta_0<\frac{\pi}{3},\;\;\;\;\frac{\pi}{3}<\beta_+<\pi
\end{equation}
and define four sectors $I,II,III,IV$ in the complex $\l$-plane, with $-\pi<\arg\l<\pi$, as follows
\begin{equation}
\label{beta}
\begin{gathered}
\l\in I\iff-\pi<\arg\l<\beta_-,\;\;\;\;
\l\in II\iff\beta_-<\arg\l<\beta_0,\\
\l\in III\iff\beta_0<\arg\l<\beta_+,\;\;\;\;
\l\in II\iff\beta_+<\arg\l<\pi\ .\\
\end{gathered}
\end{equation}
Let $\Sigma:=\mathbb{R}_-\sqcup\le(\bigsqcup_{j\in\{0,\pm\}}\e^{\i\beta_j}\mathbb{R}_+\ri)$ be the oriented contour delimiting the sectors $I,...,IV$, as in figure \ref{figurejumps}.

Let $\omega:=\e^{\frac{2\pi\i}{3}}$, $\nabla:=\begin{bmatrix} 1 & \pa_\l & \pa_\l^2 \end{bmatrix}^\top$, and define
\begin{equation}
\label{Psi}
\Psi(\l)=\begin{cases}
\le[\begin{array}{c|c|c}
\omega^{-N}\,\nabla g(\omega^{-1}\l;N) & \omega^{-N}\,\nabla f(\omega^{-1}\l;1-N) & \i\omega^{-\frac{N}{2}}\,\nabla f(\omega\l;1-N)
\end{array}\ri]
& \l\in I \\ & \\
\le[\begin{array}{c|c|c}
\omega^N\,\nabla g(\omega\l;N) & -\nabla f(\l;1-N) & \i\omega^{-\frac{N}{2}}\,\nabla f(\omega\l;1-N)
\end{array}\ri]
& \l\in II \\ & \\
\le[\begin{array}{c|c|c}
\nabla g(\l;N) & -\nabla f(\l;1-N) & -\i\omega^{\frac{N}{2}}\,\nabla f(\omega^{-1}\l;1-N) 
\end{array}\ri]
& \l\in III \\ & \\
\le[\begin{array}{c|c|c} 
\omega^{-N}\,\nabla g(\omega^{-1}\l;N) & \omega^{N}\,\nabla f(\omega\l;1-N) & -\i\omega^{\frac{N}{2}}\,\nabla f(\omega^{-1}\l;1-N)
\end{array}\ri]
&\l\in IV.
\end{cases}
\end{equation}
\end{definition}

\begin{figure}[htbp]
\centering
\hspace{1cm}
\begin{tikzpicture}

\draw[->] (-3,0) -- (3,0);
\draw[->] (0,-3) -- (0,3);

\node at (-2,0.9) {$IV$};
\node at (1.8,1.8) {$III$};
\node at (1.8,-1.8) {$II$};
\node at (-2,-0.9) {$I$};

\node at (2,0.55) {$-$};
\node at (2,0.15) {$+$};
\node at (1.2,0.4) {$\mathcal{S}_0$};
\node at (3.4,0.4) {$\e^{\i\beta_0}\mathbb{R}_+$};
\draw[->,very thick,rotate=10] (0,0) -- (2,0);
\draw[very thick,rotate=10] (2,0) -- (2.5,0);

\begin{scope}[rotate=130]
\node at (2,0.25) {$-$};
\node at (2,-0.25) {$+$};
\node at (1.2,0.2) {$\mathcal{S}_+$};
\end{scope}

\node at (-1.3,2.4) {$\e^{\i\beta_+}\mathbb{R}_+$};
\draw[->,very thick,rotate=130] (0,0) -- (2,0);
\draw[very thick,rotate=130] (2,0) -- (2.5,0);

\begin{scope}[rotate=-110]
\node at (2,0.25) {$-$};
\node at (2,-0.25) {$+$};
\node at (1.2,0.25) {$\mathcal{S}_-$};
\end{scope}

\node at (-1.3,-2.7) {$\e^{\i\beta_-}\mathbb{R}_+$};
\draw[->,very thick,rotate=-110] (0,0) -- (2,0);
\draw[very thick,rotate=-110] (2,0) -- (2.5,0);

\node at (-2,0.25) {$+$};
\node at (-2,-0.25) {$-$};
\node at (-1.2,-0.2) {$\mathcal{M}$};

\node at (-2.6,0.3) {$\mathbb{R}_{-}$};
\draw[->,very thick,rotate=180] (0,0) -- (2,0);
\draw[very thick,rotate=180] (2,0) -- (2.5,0);
\end{tikzpicture}
\caption{Jump $\wt{M}$ of $\Psi$ along $\Sigma$: $\Psi_+=\Psi_-\wt{M}$.}
\label{figurejumps}
\end{figure}

\begin{proposition}\label{propbare}
$\Psi(\l)$ solves \eqref{bareODE} in all sectors $I,...,IV$; furthermore it has the same asymptotic expansion in all sectors $I,...,IV$
\begin{equation}
\label{eqq}
\Psi(\l)\sim \l^S\, G\, \le(\mathbf{1}+\mathcal{O}\le(\l^{-\frac{1}{2}}\ri)\ri) \, \l^L\,\e^{\vartheta(\l)}
\end{equation}
where $S,G,L,\vartheta$ have been defined in \eqref{SGT}, and satisfies a jump condition along $\Sigma$
\begin{equation}
\Psi_+(\l)=\Psi_-(\l)\wt{M},\;\;\;\l\in\Sigma
\end{equation}
where boundary values are taken with respect to the orientation of $\Sigma$ shown in figure \ref{figurejumps} and $\wt{M}:\Sigma\to\SL(3,\mathbb{C})$ is defined piecewise as
\begin{equation}
\label{stokes}
\begin{gathered}
\wt{M}:=\begin{cases}
\mathcal{S}_{0,\pm},&\l\in\e^{\i\beta_{0,\pm}}\mathbb{R}_+
\\
\mathcal{M},&\l\in\mathbb{R}_-
\end{cases}
\\
\mathcal{S}_-:=\begin{bmatrix}
1&0&0 \\ 0&1&0 \\ \frac{(-1)^N\sqrt{2\pi}}{\G(N)} &\i(-1)^N &1
\end{bmatrix},\;\;\;\;
\mathcal{S}_0:=\begin{bmatrix}
1&0&0 \\ -\frac{\i\sqrt{2\pi}}{\G(N)} &1&\i(-1)^N \\ 0&0& 1
\end{bmatrix},
\\
\mathcal{S}_+:=\begin{bmatrix}
1&0&0 \\ 0&1&0 \\ -\frac{(-1)^{N}\sqrt{2\pi}}{\G(N)} &\i(-1)^N  & 1
\end{bmatrix},\;\;\;\;
\mathcal{M}:=\begin{bmatrix}
1&0&0 \\ 0&0&-\i(-1)^N  \\ 0&-\i(-1)^N &0
\end{bmatrix}.
\end{gathered}
\end{equation}
\end{proposition}
\noindent {\bf Proof.}  
The differential equation follows from \eqref{diffeqf} and \eqref{diffeqg}. The asymptotic expansion \eqref{eqq} follows by analytic continuation of the expansions \eqref{asymptoticsrealwithallcoefficients} and \eqref{asymptoticsg}. For the jump use the following identities, consequence of the Cauchy Theorem
\begin{equation}
\begin{gathered}
f(\l;1-N)+\omega^Nf(\omega\l;1-N)+\omega^{-N}f(\omega^{-1}\l;1-N)=0,
\\
g(\l;N)-\omega^Ng(\omega\l;N)=-\frac{\i\sqrt{2\pi}}{\G(N)}f(\l;1-N).
\end{gathered}
\end{equation}
\QED 
\begin{remark}
The identity $\det\Psi(\l)=1$ holds identically in all sectors.
\end{remark}

\begin{remark}
In the terminology of linear complex ordinary differential equations $\mathcal{S}_{\pm,0}$ are the Stokes matrices and $\mathcal{M}$ the formal monodromy of the singularity $\l=\infty$ of \eqref{bareODE}. Notice the {\it no-monodromy condition} $\mathcal{M}\mathcal{S}_+\mathcal{S}_0\mathcal{S}_-=\1$.
\end{remark}

\subsection{Extension of the Kontsevich--Penner matrix integral}\label{paragraphextension}
By \eqref{eq} and Proposition \ref{propode0} we see that $\mathcal{Z}_n(Y;N)$ admits a regular asymptotic expansion for large $Y$ when $\Re y_j>0$. As $f(\l;N)$ are entire functions we could try to analytically continue $\mathcal{Z}_n(Y;N)$ to the region $\Re y_j<0$ via the right-hand side of \eqref{eq}. However, this would result in the fact that  $\mathcal{Z}_n(Y;N)$ does not admit a regular asymptotic expansion in the region where some $\Re y_j<0$. 

The main motivation for the definition of the Kontsevich--Penner model is the construction of a {\it generating function} (in the sense of formal series). 
In order to have a regular expansion near infinity also in  the sector $\Re y_j<0$ (and, in fact, the {\it same} expansion) we need to modify the definition of $\mathcal{Z}_n(Y;N)$. To this end we start from the Wronskian representation \eqref{zn} in terms of the function $f(\l,N)$ \eqref{definitionair};  in the left plane we replace them by other solution to  the ODE \eqref{diffeqf} in appropriate way so as to preserve the regularity of the asymptotic expansion. The logic is completely parallel to the one used in \cite{BeCa2015} and is forced on us by the Stokes phenomenon of the solutions to the ODE \eqref{diffeqf}, which is closely related to the Airy differential equation. 

\begin{definition}\label{defextend}
We order the variables $y_j$ so that $\Re y_j> 0$ for $j=1,...,n_1$ and $\Re y_j<0$ for $j=n_1+1,...,n_1+n_2=n$. We denote $\vec{\l}=(\l_1,...,\l_{n_1})$ and $\vec{\mu}=(\mu_1,...,\mu_{n_2})$ with $y_j=\sqrt{\l_j}$ for $j=1,...,n_1$ and $y_{n_1+j}=-\sqrt{\mu_j}$ for $j=1,....,n_2$, all roots being principal. We define the {\it extended Kontsevich--Penner partition function} by the expression (for the meaning of the sectors, see Def. \ref{defsectors} and Fig. \ref{figurejumps})
\begin{equation}
\label{defextendeq}
\renewcommand*{\arraystretch}{1.55}
\mathcal{Z}_n(\vec{\l},\vec{\mu};N):=2^{\frac{n}{2}}\e^{Q(\vec{\l},\vec{\mu})}\,\Delta(\vec{\l},\vec{\mu};N)\,\det\left[\begin{array}{c}
\left[\omega^{N+\frac{1}{2}}f^{(j-1)}(\omega^{-1}\l_k;N)\right]_{1\leq k\leq n_1,\;\l_k\in I} 
\\ 
\hline
\left[f^{(j-1)}(\l_k;N)\right]_{1\leq k\leq n_1,\;\l_k\in II\cup III} 
\\ 
\hline
\left[\omega^{-N-\frac{1}{2}}f^{(j-1)}(\omega\l_k;N)\right]_{1\leq k\leq n_1,\;\l_k\in IV} 
\\ 
\hline
\left[\omega^{\frac{N}{2}+\frac{1}{4}}f^{(j-1)}(\omega\mu_k;N)\right]_{n_1+1\leq k\leq n,\;\mu_k\in I\cup II} 
\\ 
\hline
\left[\omega^{-\frac{N}{2}-\frac{1}{4}}f^{(j-1)}(\omega^{-1}\mu_k;N)\right]_{n_1+1\leq k\leq n,\;\mu_k\in III\cup IV} 
\end{array}\right]_{1\leq j\leq n}
\end{equation}
where
\begin{equation}
\label{Q}
Q(\vec{\l};\vec{\mu}):=\frac{2}{3}\sum_{j=1}^{n_1}\l_j^{\frac{3}{2}}-\frac{2}{3}\sum_{j=1}^{n_2}\mu_j^{\frac{3}{2}}
\end{equation}
and
\begin{equation}
\label{DELTA}
\Delta(\vec{\l},\vec{\mu};N):=\frac{\prod\limits_{j=1}^{n_1}\l_j^{\frac{1}{4}+\frac{N}{2}}\prod\limits_{j=1}^{n_2}(-\mu_j)^{\frac{1}{4}+\frac{N}{2}}}{\prod\limits_{1\leq j<k\leq n_1}\le(\sqrt{\l_k}-\sqrt{\l_j}\ri)\prod\limits_{1\leq j<k\leq n_2}\le(\sqrt{\mu_{j}}-\sqrt{\mu_{k}}\ri)
\prod\limits_{j=1}^{n_1}\prod\limits_{k=1}^{n_2}\le(\sqrt{\l_j}+\sqrt{\mu_k}\ri)}.
\end{equation}
\end{definition}
We deduce that $\mathcal{Z}_n(\vec{\l},\vec{\mu};N)$ as defined in \eqref{defextendeq} has a regular asymptotic expansion when $\l_j,\mu_j\to\infty$  in the indicated sectors. This regular asymptotic expansion coincides with the already discussed regular asymptotic expansion of $\mathcal{Z}_n(Y;N)$ for $\Re y_k=\Re \sqrt{\l_k}\ge 0$. As analytic functions, $\mathcal{Z}_n(\vec{\l},\vec{\mu};N)=\mathcal{Z}_n(Y;N)$ provided that $n_2=0$,  $\l_k\in II\cup III$ and $y_k=\sqrt{\l_k}$ for all $k=1,..,n$.

We point out that the definition \eqref{defextendeq} depends not only on the belonging of $y_j$ to the left/right half-planes but also on the placement of the boundaries between the sectors $I$--$IV$ in Def. \ref{defsectors}. If we move the boundaries within the bounds of Def. \ref{defsectors} then this yields different functions $\mathcal Z_n(\vec \lambda, \vec \mu;N)$ but all admitting the same asymptotic expansion as $\vec \l, \vec \mu$ tend to infinity within the respective sectors. We opted to leave this dependence on the sectors understood, without explicit reference in the symbol of the function, in the interest of lighter notations.  
\subsection{Dressed Riemann-Hilbert problem}\label{paragraphdressing}
\begin{definition}
Recall $\vec{\l}=(\l_1,...,\l_{n_1})$ and $\vec{\mu}=(\mu_1,...,\mu_{n_2})$ from Definition \ref{defextend}. Introduce
\begin{equation}
\label{defD}
\begin{gathered}
D_n(\l;\vec{\l},\vec{\mu}):=\diag\le(\alpha,\pi_+,\pi_-\ri)
\\
\alpha:=\prod_{j=1}^{n_1}\sqrt{\l_j}\prod_{j=1}^{n_2}\sqrt{\mu_j},\;\;\;\pi_\pm:=\prod_{j=1}^{n_1}\le(\sqrt{\l_j}\pm\sqrt{\l}\ri)\prod_{j=1}^{n_2}\le(\sqrt{\mu_j}\mp\sqrt{\l}\ri)
\end{gathered}
\end{equation}
and $M_n:\Sigma\to\SL(3,\mathbb{C})$
\begin{equation}\label{defm}
M_n:=\le(D_{n}^{-1}\,\e^\vartheta\ri)_-\,\wt{M}\,\le(\e^{-\vartheta}\,D_n\ri)_+
\end{equation}
$\wt{M}$ and the notation $\pm$ for boundary values being as in \eqref{stokes}.
\end{definition}
The boundary value specifications $\pm$ in \eqref{defm} give different values along the cut $\mathbb{R}_-$ only. In particular it is easy to check that  $M_n|_{\mathbb{R}_-}$ does not depend on $\vec{\l},\vec{\mu}$. The angles $\beta_{0,\pm}$ are chosen so that none of zeros of $D_n$ occur along the three rays $\e^{\i \beta_{0,\pm}}\mathbb{R}_+$. 

The construction is such that along the three rays $\e^{\i \beta_{0,\pm}}\mathbb{R}_+$ the jump matrix $M_n$ is exponentially close to the identity matrix; $M_n(\l)=\1+\mathcal{O}\le(\l^{-\infty}\ri)$ as $|\l|\to\infty$.

We now formulate the dressed Riemann--Hilbert problem.
\begin{problem}\label{RHP}
Find a $\Mat(3,\mathbb{C})$-valued function $\G_n=\G_n(\l;\vec{\l},\vec{\mu})$ analytic in $\l\in\mathbb{C}\setminus\Sigma$, admitting non-tangential boundary values ${\G_n}_\pm$ at $\Sigma$ (as in figure \ref{figurejumps}) such that
\begin{equation}
\begin{cases}
{\G_n}_+(\l)={\G_n}_-(\l) M_n(\l)&\l\in\Sigma \\
\G_n(\l)\sim \l^S\, G \, Y_n(\l) \, \l^L &\l\to\infty
\end{cases}
\end{equation}
where $S,G,L$ are as in \eqref{SGT}, $M_n$ as in \eqref{defm} and $Y_n(\l)$ a formal power series in $\l^{-\frac{1}{2}}$ satisfying the normalization
\begin{equation}
\label{normalizationinfty}
Y_n(\l)=\1+\begin{bmatrix}
0&a_n&-a_n \\ 0&c_n&0 \\ 0&0&-c_n
\end{bmatrix}\lambda^{-\frac{1}{2}}+\mathcal{O}(\l^{-1}).
\end{equation}
\end{problem}
We will see (Rem. \ref{solutionto}) that the existence of the solution to the Riemann--Hilbert problem \ref{RHP} depends on the non-vanishing of a function of $\vec \l, \vec \mu$ which is (restriction of an) entire function. Hence the singular locus  in the parameter space where the problem is unsolvable, is a divisor ({\it Malgrange divisor}) and the problem is generically solvable.
\begin{remark}
We observe that we can analytically continue $\Gamma_n|_{IV}$ beyond $\arg\l=\pi$ so that the asymptotic expansion $\G_n\sim \l^S G Y_n \l^L$ remains valid in a sector up to $\arg\l=\pi+\epsilon$. Similarly said for $\G_n|_{I}$, in a sector from $\arg \l = -\pi-\epsilon$. By matching the expansions in the overlap sector, we obtain
\begin{equation}
\label{formalmonodromy}
\l^S \, \e^{2\pi\i S} \, G \, Y_n(\l\e^{2\pi\i }) \, \l^L \, \e^{2\pi\i L}=\l^S \, G \, Y_n(\l) \, \l^L \, \mathcal{M}.
\end{equation}
 By trivial algebra \eqref{formalmonodromy} implies the following symmetry relation for the formal power series $Y_n(\l)$
\begin{equation}
\label{eigenspaces}
Y_n(\l\e^{2\pi\i })=\begin{bmatrix}
1 & 0& 0 \\ 0& 0&1 \\ 0& 1&0
\end{bmatrix} Y_n(\l) \begin{bmatrix}
1 & 0& 0 \\ 0& 0&1 \\ 0& 1&0
\end{bmatrix}.
\end{equation}
In terms of the  coefficients of the expansion of $Y_n$,   we find that the coefficients of the fractional powers must be {\it odd} under the conjugation \eqref{eigenspaces}, while those of the integer powers must be {\it even}. 
In particular this implies the following form for $Y_n$
\begin{equation}
Y_n(\l)=\1+ \begin{bmatrix}
0&a_n& - a_n \\ b_n&c_n&d_n \\ -b_n&-d_n&-c_n
\end{bmatrix} \l^{-\frac{1}{2}}+\mathcal{O}(\l^{-1})\ .
\end{equation}
\end{remark}
\begin{remark}\label{rmksymmetrynormalization}
The normalization condition \eqref{normalizationinfty} is necessary to ensure the uniqueness of the solution to  the Riemann-Hilbert problem \ref{RHP}. To explain this, consider  the identity
\begin{equation}
\begin{bmatrix}
1 &0&0 \\ 0&1 &0 \\ \alpha &\beta &1
\end{bmatrix} \l^S G=\l^S G\le(\1+\begin{bmatrix}
0&0&0 \\- \frac{\alpha}{\sqrt{2}} &-\frac{\beta}{2}&-\frac{\beta}{2} \\ \frac{\alpha}{\sqrt{2}} &\frac{\beta}{2}&\frac{\beta}{2} 
\end{bmatrix}\l^{-\frac{1}{2}}\ri).
\label{318}
\end{equation}
This identity shows that the simple requirement $Y_n(\l)=\1+\mathcal{O}\le(\l^{-\frac{1}{2}}\ri)$ leaves the freedom of multiplying on the left by the two-parameter family of matrices indicated in \eqref{318}. The normalization \eqref{normalizationinfty} $b_n=0,d_n=0$ fixes uniquely the gauge arbitrariness implied by \eqref{318}.
\end{remark}

\subsection{Proof of Theorem \ref{thm1}}
\label{proofthm1}
\subsubsection{The Malgrange differential and the tau function}

If $\Psi_0(\lambda) = \Psi(\lambda)$ is the sectionally analytic solution to  the ODE \eqref{bareODE} defined in \eqref{Psi} and $\vartheta$ is as in \eqref{SGT}, then by Proposition \ref{propbare} $\Gamma_0:=\Psi_0 \e^{-\vartheta}$ solves the Riemann-Hilbert Problem \ref{RHP} for $n=0$ (the gauge factor explained in \ref{rmksymmetrynormalization} is $\1$).
Then, the matrix $\Psi_n=\Psi_n(\l;\vec{\l},\vec{\mu})$ defined as
\begin{equation}
\Psi_n:=\Gamma_n D^{-1}_n \e^\vartheta
\end{equation}
$\vartheta$ as in \eqref{SGT}, satisfies a Riemann-Hilbert problem of the form
\begin{equation}
\label{RHPPSI}
\begin{cases}
{\Psi_n}_+(\l)={\Psi_n}_-(\l) \wt{M} &\l\in\Sigma \\
\Psi_n(\l)\sim \l^S \, G \, Y_n(\l) \, \l^L \, D^{-1}_n \, \e^{\vartheta(\l)} &\l\to\infty.
\end{cases}
\end{equation}

It is now a simple exercise (see Prop. \ref{propexistR} below) to show that $\Psi_n(\lambda) = R_n(\lambda) \Psi_0(\lambda)$ with $R_n(\lambda)$ rational in $\l$.
This is (with the generalization of the parameters $\vec \mu$) the matrix $\Psi_n$ mentioned in the Introduction (Sec. \ref{introductionschlesinger}).

Since the jumps $\wt{M}$ in \eqref{RHPPSI} do not depend on $\l,\vec{\l},\vec{\mu}$ it follows also that $\Psi_n$ satisfies a compatible system of linear differential equations together with its zero-curvature equations
\begin{equation}
\label{isomonodromy}
\begin{cases}
\pa_\l\Psi_n=A_n\,\Psi_n \\
\delta\Psi_n=\Omega_n\,\Psi_n
\end{cases}
\;\;\;\;\;\;\;\;\,
\begin{cases}
\delta A_n=\pa_\l\Omega_n+[\Omega_n,A_n]\\
\delta\Omega_n=\Omega_n\wedge\Omega_n
\end{cases}
\end{equation}
where  we have  introduced the differential in the parameters $\vec{\l},\vec{\mu}$
\begin{equation}
\label{diffisomontimes}
\delta:=\sum\limits_{j=1}^{n_1}\mathrm{d}\l_j\,\pa_{\l_j}+\sum\limits_{j=1}^{n_2}\mathrm{d}\mu_j\,\pa_{\mu_j}.
\end{equation}
The system \eqref{isomonodromy} is an isomonodromic system in the sense of \cite{JiMiUe1981}.
To this system we associate the {\it Malgrange differential} in the $\{\vec{\l},\vec{\mu}\}$-space
\begin{equation}
\label{malgrange}
\omega_n:=\int_\Sigma\frac{\mathrm{d}\l}{2\pi\i }\,\Tr\le({\Gamma_n^{-1}}_- {\Gamma_n'}_- \delta M_n  M^{-1}_n\ri)\ ,\qquad ':=\pa_\l.
\end{equation}
The Malgrange differential $\omega_n$ is $\delta-$closed \cite{Be2010} and we introduce the isomonodromic tau function $\tau_n(\vec{\l},\vec{\mu};N)$ by
\begin{equation}
\label{tau}
\delta\log\tau_n=\omega_n.
\end{equation}
This definition generalizes the original definition of \cite{JiMiUe1981}; we refer to \cite{Be2010} for the relation between the two definitions and from now on we work with the definition \eqref{tau} (however, compare with Prop. \ref{propjapanese}).
\begin{remark}
\label{rmksigma}
Since the jump $M_n$ along $\mathbb{R}_-$ does not depend on any of the parameters, the integration in \eqref{malgrange} extends to the three rays $\e^{\i \beta_{0,\pm}}\mathbb{R}_+$ only.
\end{remark}
\subsubsection{The Schlesinger transform and the characteristic matrix}
Introduce the so--called {\it Schlesinger transform matrix} $R_n:=\G_n  D^{-1}_n \G^{-1}_0$, so that $\G_n=R_n \G_0  D_n$.

\begin{proposition}\label{propexistR}
The matrix $R_n$ is a rational function of $\l$, with simple poles at $\l\in\vec{\l},\vec{\mu}$ only.
\end{proposition}
\noindent {\bf Proof.} 
First of all the matrix $R_n$ has no jump discontinuities for $\l\in\Sigma$, as
\begin{equation}
R_{n+}=\G_{n+}\,D^{-1}_{n+}\,\G^{-1}_{0+}=\G_{n-}\,M_n\, D_{n+}^{-1}\,M^{-1}_0\,\G_{0-}^{-1}=\G_{n-}\,D^{-1}_{n-}\,\G_{0-}^{-1}=R_{n-}\ .
\end{equation}
Secondly, from its definition it immediately follows that the points  $\l_j, \mu_j$ are simple poles since both $\G_n$ and $\G^{-1}_0$ are locally analytic at these points. It follows that $R_n$ is a meromorphic function with simple poles only as indicated. 
Note also that, using the symmetry \eqref{eigenspaces} for the asymptotic expansions of $\G_n, \G_0$ and the explicit expression of $D_n$ \eqref{defD}, one can directly  check that $R_n(\l)$ has an expansion in integer powers with  at most linear  growth in $\l$ at $\l=\infty$. The conclusion follows from Liouville's theorem.
\QED 

\begin{figure}[htbp]
\centering
\hspace{1cm}
\begin{tikzpicture}[scale=0.7]

\draw[very thick,rotate=10,->] (3,0) -- (3.5,0);
\draw[very thick,rotate=130,->] (3,0) -- (3.5,0);
\draw[very thick,rotate=-110,->] (3,0) -- (3.5,0);
\draw[very thick,rotate=180,->] (3,0) -- (3.5,0);
\draw[very thick,rotate=10] (3.5,0) -- (4,0);
\draw[very thick,rotate=130] (3.5,0) -- (4,0);
\draw[very thick,rotate=-110] (3.5,0) -- (4,0);
\draw[very thick,rotate=180] (3.5,0) -- (4,0);

\draw[very thick, fill=white!90!black,decoration={markings, mark=at position 0.2 with {\arrow{>}}, mark=at position 0.42 with {\arrow{>}}, mark=at position 0.58 with {\arrow{>}}, mark=at position 0.88 with {\arrow{>}}}, postaction={decorate}] (0,0) circle (3);
\filldraw[fill=white, very thick,decoration={markings, mark=at position 0.3 with {\arrow{<}}}, postaction={decorate}] (1,0.8) circle (0.3);
\filldraw[fill=white, very thick,decoration={markings, mark=at position 0.3 with {\arrow{<}}}, postaction={decorate}] (1.7,0.2) circle (0.3);
\filldraw[fill=white, very thick,decoration={markings, mark=at position 0.3 with {\arrow{<}}}, postaction={decorate}] (1.3,-2) circle (0.3);

\draw[->] (-4.2,0) -- (4.2,0);
\draw[->] (0,-4.2) -- (0,4.2);

\end{tikzpicture}
\caption{Contour $\sigma$ for Riemann--Hilbert problem \ref{RHPR}. The shaded region is $\mathbf{D}_-$.}
\label{figureR}
\end{figure}
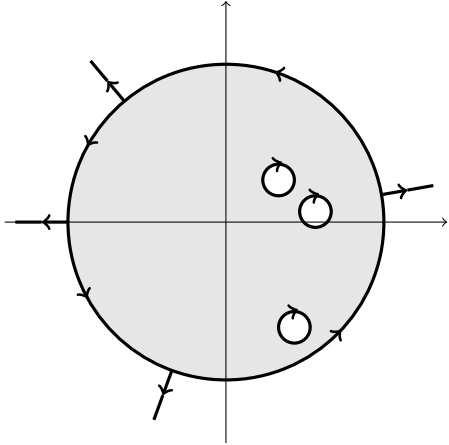

Next, we relate the Schlesinger transform matrix $R:=R_n$ (we drop the subscript $_n$ for brevity) to the solution to  a suitable Riemann-Hilbert problem. Define disks on the Riemann sphere
\begin{equation}
\mathbf{D}_\zeta:=\{|\l-\zeta|<\rho\},\;\;\;\zeta\in\vec{\l},\vec{\mu},\;\;\;\;\mathbf{D}_\infty:=\left\lbrace\le|\l\ri|>\rho^{-1}\right\rbrace
\end{equation}
where we take $\rho>0$ small enough so that these disks are pairwise disjoint and disjoint from $\Sigma$. Let us define also
\begin{equation}
\mathbf{D}_+:=\mathbf{D}_\infty\sqcup\le(\bigsqcup_{\zeta\in\vec{\l},\vec{\mu}}\mathbf{D}_\zeta\ri),\;\;\;\;\mathbf{D}_-:=\mathbb{C}\setminus\overline{\mathbf{D}_+},\;\;\;\;
\sigma:=\pa\mathbf{D}_-\sqcup(\Sigma\cap\mathbf{D}_\infty)
\end{equation}
with the natural orientation, as shown in figure \ref{figureR}.

It is convenient to relate the matrix $R$  to the following Riemann--Hilbert problem.
\begin{problem}\phantomsection \label{RHPR}
Find a $\Mat(3,\mathbb{C})$-valued function $\mathbf{R}(\l)$ analytic in $\l\in\mathbb{C}\setminus\sigma$ admitting non-tangential boundary values $\mathbf{R}_{\pm}(\l)$ at $\sigma$ satisfying
\be
\begin{cases}
\mathbf{R}_+(\l)=\mathbf{R}_-(\l) \mathbf{J}(\l)&\l\in\sigma
\\ \mathbf{R}(\l)=\1+\mathcal{O}(\l^{-1})&\l\to\infty
\end{cases}
\ee
where the jump matrix $\mathbf{J}(\l)=\mathbf{J}_n(\l;\vec{\l},\vec{\mu})$ is piecewise defined on $\sigma$ as
\be \label{J}
\mathbf{J}(\l)=\begin{cases}
\mathbf{J}_\zeta(\l):=\Gamma_0(\l) (\zeta-\l)^{\E_{33}},&\l\in \pa\mathbf{D}_\zeta,\;\zeta\in\vec{\l}\\
\mathbf{J}_\zeta(\l):=\G_0(\l) (\zeta-\l)^{\E_{22}},&\l\in \pa\mathbf{D}_\zeta,\;\zeta\in\vec{\mu} \\
\mathbf{J}_\infty(\l):=\G_0(\l)  D_n(\l) \l^{-L}  G^{-1}  \l^{-S},&\l\in\pa\mathbf{D}_\infty \\
\mathbf{J}_\Sigma(\l):=\l_-^S  G  \l_-^L  M_n(\l)  \l_+^{-L}  G^{-1}  \l_+^{-S}, & \l\in\Sigma\cap\mathbf{D}_\infty.
\end{cases}
\ee
Hereafter, $\E_{ab}$ denotes the elementary unit matrix $(\E_{ab})_{ij}:=\delta_{ia}\delta_{jb}$.
\end{problem}

\begin{remark}
Notice that there is no jump due to formal monodromy, i.e. $\mathbf{J}|_{\mathbb{R}_-\cap\mathbf{D}_\infty}=\1$. Furthermore, we have $\mathbf{J}|_{\Sigma\cap\mathbf{D}_\infty}=\1+\mathcal{O}\le(\l^{-\infty}\ri)$.
\end{remark}
The explicit relation of $R$ and $\mathbf R$ is provided in the next proposition.
\begin{proposition}
\label{proprhpr}
The solution $\mathbf{R}(\l)$ to Riemann-Hilbert problem \ref{RHPR}, if it exists,  is unique. The Schlesinger transform $R$ satisfies $R|_{\mathbf{D}_-}=(\1+c_n\E_{32}) \mathbf{R}|_{\mathbf{D}_-}$, with $c_n$ as in \eqref{normalizationinfty} and $\E_{32}$ the elementary unit matrix.
\end{proposition}
\noindent {\bf Proof.}  
Given two different solutions, their matrix ratio has no jump on $\sigma$ so it is entire in the $\l$-plane and asymptotic to $\1$ at $\l=\infty$; it follows that it equals identically $\1$ by Liouville Theorem. Next we define
\be \label{Rtilde}
\widetilde{\mathbf{R}}(\l):=\begin{cases}
\Gamma_n(\l)  D^{-1}_n(\l) (\zeta-\l)^{\E_{33}},&\l\in\mathbf{D}_\zeta,\;\zeta\in\vec{\l} \\
\Gamma_n(\l)  D^{-1}_n(\l) (\zeta-\l)^{\E_{22}},&\l\in\mathbf{D}_\zeta,\;\zeta\in\vec{\mu} \\
\Gamma_n(\l)  D^{-1}_n(\l) \Gamma_0^{-1}(\l), &\l\in\mathbf{D}_- \\
\Gamma_n(\l)  \l^{-L}  G^{-1}  \l^{-S},&\l\in\mathbf{D}_\infty\setminus\Sigma.
\end{cases}
\ee
$\wt{\mathbf{R}}$ satisfies the jump condition $\wt{\mathbf{R}}_+=\wt{\mathbf{R}}_- \mathbf{J}$ along $\sigma$. From the last line of \eqref{Rtilde} we infer that as $\l\to\infty$ in any of the sectors $I,...,IV$
\begin{equation}
\wt{\mathbf{R}}(\l)\sim \l^S \, G \, Y_n(\l) \, G^{-1} \, \l^{-S}=(\1+c_n\E_{32}) (\1+\mathcal{O}(\l^{-1}))
\end{equation}
with $c_n$ is the coefficient appearing in the expansion \eqref{normalizationinfty}.
Hence $\mathbf{R}:=(\1-c_n\E_{32}) \, \wt{\mathbf{R}}$ solves Riemann-Hilbert problem \ref{RHPR}. The proof is completed by observing directly from \eqref{Rtilde} that $\wt{\mathbf{R}}|_{\mathbf{D}_-}=R|_{\mathbf{D}_-}$.
\QED 

\begin{lemma}\label{lemma}
The jump matrix $\mathbf{J}(\l)$ \eqref{J} admits formal meromorphic extension in $\mathbf{D}_+$. More precisely
\be
\begin{cases}
\mathbf{J}_\z(\l)=\Gamma_0(\z) (\l-\z)^{\E_{33}} (\1+\mathcal{O}(\l-\z)), & \z\in\vec{\l}\\
\mathbf{J}_\z(\l)=\Gamma_0(\z) (\l-\z)^{\E_{22}} (\1+\mathcal{O}(\l-\z)), & \z\in\vec{\mu}\\
\mathbf{J}_\infty(\l)=G_\infty(\l)  H_\infty(\l)&
\\
\mathbf{J}_\Sigma(\l)=\1+\mathcal{O}\le(\l^{-\infty}\ri)&
\end{cases}
\ee
where
\be
\label{Hinfty}
H_\infty:=\l^S\, G\, 
\begin{bmatrix}1 & 0 & 0 \\ 0&(-1)^n \l^{\frac{n}{2}} & 0 \\ 0 & 0 &  \l^{\frac{n}{2}} \end{bmatrix}
%\diag(1,(-1)^n\l^{\frac{n}{2}},\l^{\frac{n}{2}}) 
\, G^{-1} \, \l^{-S}
=\begin{cases}
\begin{bmatrix}1 & 0 & 0 \\ 0&\l^{\frac{n}{2}} & 0 \\ 0 & 0 & \l^{\frac{n}{2}} \end{bmatrix}
& n\text{ even} \\
\begin{bmatrix} 1&0&0\\ 0&0&\l^{\frac{n-1}{2}} \\ 0&\l^{\frac{n+1}{2}}&0 
\end{bmatrix} & n\text{ odd}
\end{cases}
\ee
and $G_\infty:=\mathbf{J}_\infty  H_\infty^{-1}$ is formally analytic at $\l=\infty$.
\end{lemma}
\noindent {\bf Proof.}  
The only nontrivial expansion is that of $\mathbf{J}_\infty$. However, using the symmetry \eqref{eigenspaces}, it is easy to check that the matrix $G_\infty$
\bea
\label{Ginfty}
G_\infty=\mathbf{J}_\infty H_\infty^{-1}=\G_0 \, \l^{-L}\,\widetilde{D}_n \, G^{-1}\, \l^{-S}\sim \l^S \, G \, Y_0 \, \wt{D}_n \, G^{-1} \, \l^{-S}
\\
\wt{D}_n:=D_n \, \diag(1,(-1)^n\l^{-\frac{n}{2}},\l^{-\frac{n}{2}})
\eea
has an expansion in integer powers of $\l$ only.
\QED

We need to apply the results of \cite{BeCa2015}, which we recall now. Let us introduce $\mathcal{H}:=L^2(\pa\mathbf{D}_+,|\mathrm{d}\l|)\otimes\mathbb{C}^3$, where $\mathbb{C}^3$ are row-vectors. 
The space $\mathcal{H}$ is isomorphic to the direct sum of $n+1$ copies of $L^2(S^1)\otimes\mathbb{C}^3$, i.e. $\mathcal{H}$ has a basis  given by
\be \label{cons}
(\l-\z)^r\,\chi_{\pa\mathbf{D}_\z}(\l)\,\mathbf{e}_j^\top,\;\;\;\;\l^{-r-1}\,\chi_{\pa\mathbf{D}_\infty}(\l)\,\mathbf{e}_j^\top,\;\;\,\;\;r\in\mathbb{Z},\;j\in\{1,2,3\},\;\zeta\in\vec{\l},\vec{\mu}
\ee
where $\mathbf{e}_j$ is the standard basis of column-vectors in $\mathbb{C}^3$ and $\chi_X(\l)$ the indicator function of the set $X$. Consider the subspace $\mathcal{H}_+$ consisting of row-vectors which are analytic in $\mathbf{D}_+$ and vanish at $\l=\infty$; equivalently, $\mathcal{H}_+$ has a basis given by \eqref{cons} restricted to $r\geq 0$. Let  $ \mathcal{\mathcal{C}_\pm}:\mathcal{H}\to\mathcal{H}$ the projectors defined by the Cauchy integrals
\begin{equation}
\mathcal{C}_\pm[f](\l):=\oint_{\pa\mathbf{D}_+}\frac{\mathrm{d}\l'}{2\pi\i }\frac{f(\l')}{\l'-\l_\pm}.
\end{equation}
The range of $\mathcal C_+$ is $\mathcal H_+$ and we denote by $\mathcal H_-$ the range of $\mathcal C_-$, namely, functions that admit analytic extension to $\mathbf D_-$; from the Sokhotski-Plemelji formula $\mathcal C_+ + \mathcal C_-=\Id$,  it follows that $\pm \mathcal C_\pm :\mathcal{H}\to\mathcal{H}_\pm$ are complementary projectors.   Introduce the following subspaces of $\mathcal{H}_-$
\begin{equation}
V:=\mathcal{C}_-[\mathcal{H}_+\mathbf{J}^{-1}],\;\;\;\;W:=\mathcal{C}_-[\mathcal{H}_+\mathbf{J}].
\end{equation}
Then $\{v_\z\}_{\z\in\vec{\l},\vec{\mu}}$ and $\{w_\ell\}_{\ell=1}^{n}$ defined as
\begin{equation}
\label{basis}
v_\z:=\begin{cases}
\mathbf{e}_{3}^\top\frac{\Gamma_0^{-1}(\z)}{\l-\z}, & \z\in\vec{\l} \\
\mathbf{e}_{2}^\top\frac{\Gamma_0^{-1}(\z)}{\l-\z}, & \z\in\vec{\mu}
\end{cases}\;\;\;\;\;\,\,\,\,\,\begin{cases}w_{2m+1}:=\l^{m}\mathbf{e}_{2}^\top, &m=0,...,\lfloor \frac{n-1}{2}\rfloor \\
w_{2m+2}:=\l^{m}\mathbf{e}_{3}^\top, &m=0,...,\lfloor\frac{n-2}{2}\rfloor \end{cases}
\end{equation}
are bases of $V$ and $W$ respectively. To prove that $\{w_\ell\}_{\ell=1}^n$ is a basis of $W$ we use that $G_\infty(\l)$ is formally analytic with formally analytic inverse at $\l=\infty$ so that $W=\mathcal{C}_-[\mathcal{H}_+H_\infty]$, where $H_\infty(\l)$ is as in \eqref{Hinfty}. The linear operator
\begin{equation}
\label{mathbbG}
\mathbb{G}:V\to W:v\mapsto\mathcal{C}_-[v\mathbf{J}]
\end{equation}
is well defined. 
The invertibility of $\mathbb G$ is equivalent to the existence of the solution $\mathbf R$ of the Riemann--Hilbert problem \ref{RHPR} (equivalently, the existence of the solution to  the Riemann--Hilbert problem \ref{RHP}) \cite{BeCa2015}; in fact the inverse is  given by
\begin{equation}
\mathbb{G}^{-1}:V\to W:w\mapsto\mathcal{C}_-[w\mathbf{J}^{-1}\mathbf{R}^{-1}]\mathbf{R}.
\end{equation}
By expressing the operator $\mathbb G$ in the bases \eqref{basis} we  obtain the  {\it characteristic matrix} $\mathbf{G}=\le(\mathbf{G}_{k,\ell}\ri)_{k,\ell=1}^n$
\begin{equation}
\label{characteristicmatrix}
\mathbf{G}_{k,\ell}:=
\left[\begin{array}{c}
\renewcommand*{\arraystretch}{1.9} 
\left[\res{\l=\infty}\frac{\l^{\left\lfloor\frac{\ell-1}{2}\right\rfloor}}{\l-\l_k}
\,\mathbf{e}_{3}^\top\,\Gamma_0^{-1}(\l_k)\,G_\infty(\l)\,\mathbf{e}_{2+(\ell \!\!\!\!\mod 2)}\right]_{k=1,...,n_1}
\\
\hline
\\
\left[\res{\l=\infty}\frac{\l^{\left\lfloor\frac{\ell-1}{2}\right\rfloor}}{\l-\mu_k}\,\mathbf{e}_{2}^\top\,\Gamma_0^{-1}(\mu_k)\,
G_\infty(\l)\,\mathbf{e}_{2+(\ell\!\!\!\!\mod 2)}\right]_{k=1,...,n_2}
\end{array}\right]_{\ell=1,...,n}.
\end{equation}
Therefore the solvability of the Riemann--Hilbert problems \ref{RHPR}, \ref{RHP} is equivalent to the non-vanishing of  the determinant of the matrix $\mathbf G$.  

The result which we need to recall, adapted to our case, is the following
\begin{theorem}
The following variational formula holds
\begin{equation}
\label{bertolacafasso}
\delta\,\log\det\mathbf{G}=\int_{\pa\mathbf{D}_-}\frac{\mathrm{d}\l}{2\pi\i }\,\Tr\le(\mathbf{R}^{-1}\,\mathbf{R}'\,\delta\mathbf{J}\,\mathbf{J}^{-1}\ri)+\sum_{\z\in\vec{\l},\vec{\mu}}\res{\l=\z}\Tr\le(\G_0^{-1}\G_0'\delta U_\z \, U_\z^{-1}\,\d\l\ri)
\end{equation}
where $\delta$ is the variation \eqref{diffisomontimes} and 
\begin{equation}
\label{defU}
U_\z:=\begin{cases}(\l-\z)^{\E_{33}},&\z\in\vec{\l}
\\
(\l-\z)^{\E_{22}}, & \z\in\vec{\mu}
\end{cases}
\end{equation}
and $\int_{\pa\mathbf{D}_-}\frac{\mathrm{d}\l}{2\pi\i }$ is understood as the sum over the (formal) residues at $\l\in\{\vec{\l},\vec{\mu},\infty\}$.
\end{theorem}
For the proof see \cite[Theorem B.1]{BeCa2015}.

Central to the proof of Theorem \ref{thm1}, and more generally of Theorem \ref{Thmmaximus}, is the following Proposition.
\begin{proposition}
\label{propositionb}
The following formula holds
\begin{equation}
\label{thm1b}
\renewcommand*{\arraystretch}{1.55}
\det\mathbf{G}=\pm\e^{Q(\vec{\l},\vec{\mu})}\det
\left[\begin{array}{c}
\left[\omega^{N+\frac{1}{2}}f^{(j-1)}(\omega^{-1}\l_k;N)\right]_{1\leq k\leq n_1,\;\l_k\in I} 
\\ 
\hline
\left[f^{(j-1)}(\l_k;N)\right]_{1\leq k\leq n_1,\;\l_k\in II\cup III} 
\\ 
\hline
\left[\omega^{-N-\frac{1}{2}}f^{(j-1)}(\omega\l_k;N)\right]_{1\leq k\leq n_1,\;\l_k\in IV} 
\\ 
\hline
\left[\omega^{\frac{N}{2}+\frac{1}{4}}f^{(j-1)}(\omega\mu_k;N)\right]_{n_1+1\leq k\leq n,\;\mu_k\in I\cup II} 
\\ 
\hline
\left[\omega^{-\frac{N}{2}-\frac{1}{4}}f^{(j-1)}(\omega^{-1}\mu_k;N)\right]_{n_1+1\leq k\leq n,\;\mu_k\in III\cup IV} 
\end{array}\right]_{1\leq j\leq n}
\end{equation}
where $Q(\vec{\l},\vec{\mu})$ has been defined in \eqref{Q} and the explicit sign is irrelevant to our purposes.
\end{proposition}
The proof is found in Appendix \ref{appendixpropb}.

\begin{remark}
\label{solutionto}
The determinant on the right side of \eqref{thm1b} is the restriction of an entire function of $\vec \l, \vec \mu$ to specific sectors; therefore its zero-locus is a divisor that includes the ``diagonals'' $\l_j=\l_k$, $\mu_j=\mu_k$ ($j\neq k$). In general it contains other components.  Recall that $\mathbf G$ is a matrix representing the operator
$\mathbb G$ \eqref{mathbbG},  and that its invertibility is equivalent to the solvability of Riemann--Hilbert problems \ref{RHP}, \ref{RHPR}. Therefore the Riemann--Hilbert problem \ref{RHP} is generically solvable.
\end{remark} 

\subsubsection{Conclusion of the proof}
\begin{figure}[htbp]
\centering
\hspace{1cm}
\begin{tikzpicture}
\begin{scope}[shift={(-4.3,0)}]
\draw[->] (-2.6,0) -- (2.6,0);
\draw[->] (0,-2.6) -- (0,2.6);

\node at (1.5,0.4) {$-$};
\node at (1.5,0.1) {$+$};
\draw[->,dashed, very thick,rotate=10] (0,0) -- (2,0);

\begin{scope}[rotate=130]
\node at (1.5,0.2) {$-$};
\node at (1.5,-0.2) {$+$};
\end{scope}

\draw[->,very thick,dashed, rotate=130] (0,0) -- (2,0);

\begin{scope}[rotate=-110]
\node at (1.5,0.25) {$-$};
\node at (1.5,-0.25) {$+$};
\end{scope}

\draw[->,very thick,dashed, rotate=-110] (0,0) -- (2,0);

\draw[->,very thick] (2,0.6) to[out=190, in=-50] (-1,1.9);
\draw[->,very thick,rotate=130] (2,0.6) to[out=190, in=-60] (-0.6,1.9);
\draw[->,very thick,rotate=-110] (2,0.6) to[out=190, in=-60] (-0.6,1.9);

\node at (1,1) {\textbullet};
\node at (0.4,1.5) {\textbullet};
\node at (0.8,-1.4) {\textbullet};
\end{scope}

\begin{scope}[shift={(4.3,0)}]
\draw[->] (-2.6,0) -- (2.6,0);
\draw[->] (0,-2.6) -- (0,2.6);

\begin{scope}[rotate=10]
\draw[very thick,->] (2.5,0) -- (2,0) arc (0:115:2cm) coordinate (x) to
 ($(x)+ (120:0.5)$);

%\draw[very thick] (2.5,0) -- (2,0) arc (0:115:2cm);
%\draw[very thick,rotate=115,->] (2,0)--(2.5,0); 
\end{scope}

\begin{scope}[rotate=130]
\draw[very thick,->] (2.5,0) -- (2,0) arc (0:115:2cm) coordinate (x) to
 ($(x)+ (120:0.5)$);
%\draw[very thick,rotate=115,->] (2,0)--(2.5,0); 
\end{scope}

\begin{scope}[rotate=250]
\draw[very thick,->] (2.5,0) -- (2,0) arc (0:115:2cm) coordinate (x) to
 ($(x)+ (120:0.5)$);
%\draw[very thick] (2.5,0) -- (2,0) arc (0:115:2cm);
%\draw[very thick,rotate=115,->] (2,0)--(2.5,0); 
\end{scope}

\draw[very thick,decoration={markings, mark=at position 0.3 with {\arrow{<}}}, postaction={decorate}] (1,1) circle (0.25cm);

\draw[very thick,decoration={markings, mark=at position 0.3 with {\arrow{<}}}, postaction={decorate}] (0.4,1.5) circle (0.25cm);

\draw[very thick,decoration={markings, mark=at position 0.3 with {\arrow{<}}}, postaction={decorate}] (0.8,-1.4) circle (0.25cm);

\node at (1,1) {\textbullet};
\node at (0.4,1.5) {\textbullet};
\node at (0.8,-1.4) {\textbullet};

\end{scope}
\end{tikzpicture}
\caption{On the left, contour $\Sigma'$ (dashed) and $\wt{\Sigma}$ such that $\int_{\Sigma'}\mathrm{d}\l\,(f_+-f_-)=\int_{\wt{\Sigma}}\mathrm{d}\l\,f$; $\wt{\Sigma}$ must leave all poles of $f$ (the dots in the picture) on the right. On the right, the deformation of $\wt{\Sigma}$ using the Cauchy Theorem.}
\label{figurecauchy}
\end{figure}
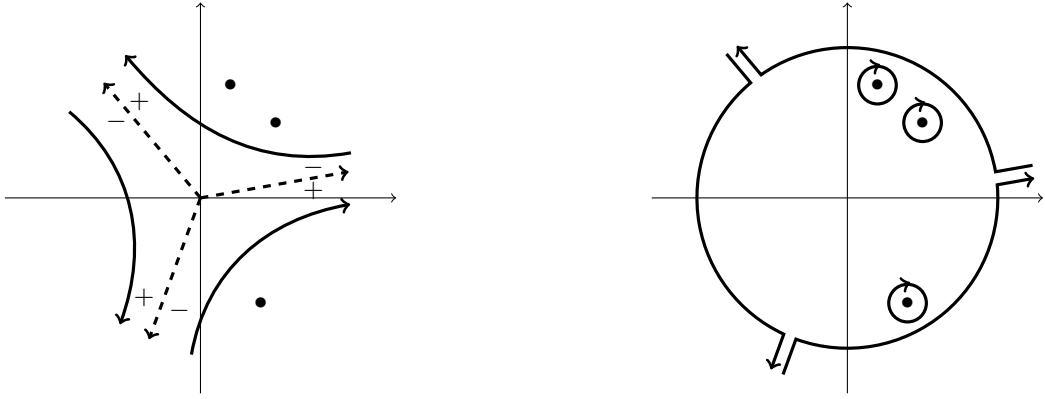

\begin{proposition}
The Malgrange differential $\omega_n$ introduced in \eqref{malgrange} is
\begin{equation}
\label{thm1a}
\omega_n=\delta\log\le(\Delta(\vec{\l},\vec{\mu};N)\,\det\mathbf{G}\ri)
\end{equation}
where $\Delta(\vec{\l},\vec{\mu};N)$ has been defined in \eqref{DELTA}.
\end{proposition}
\noindent {\bf Proof.}  
Denote $R:=R_n,\,D:=D_n$. From $\G_n=R \, \G_0 \, D$ and $M_n=D^{-1} M_0 D$ one obtains
\be \label{formula}
\begin{gathered}
\G_{n-}^{-1}  \G_{n-}'=D^{-1}\G_{0-}^{-1}R^{-1}R'\G_{0-}D+D^{-1}\G_{0-}^{-1}\G_{0-}'D+D^{-1}D'
\\
\delta M_n  M^{-1}_n=D^{-1}M_0\delta D D^{-1}M^{-1}_0D - D^{-1}\delta D
\end{gathered}
\ee
so that using \eqref{formula} and the cyclicity of the trace,
\begin{equation}
\begin{gathered}
\Tr({\G_n}_-{\G_n}_-'\delta M_nM_n^{-1})=\Tr(\G_{0-}^{-1}R^{-1}R'\G_{0-}M_0\delta D D^{-1}M^{-1}_0-D^{-1}\G_{0-}^{-1}R^{-1}R'\G_{0-}\delta D+
\\
+\G_{0-}^{-1}\G_{0-}'M_0\delta D D^{-1}M^{-1}_0-D^{-1}\G_{0-}^{-1}\G_{0-}'\delta D+D'D^{-1}M_0\delta D D^{-1}M^{-1}_0-D^{-1}D'D^{-1}\delta D)
\end{gathered}
\end{equation}
It is easy to check, thanks to the block--triangular structure of $M_0$ in \eqref{defm}, that the last two terms above are traceless and thus drop out. The remaining terms can be rewritten, using $\G_{0+}=\G_{0-}M_0$, $\G_{0+}'=\G_{0-}'M_0+\G_{0-}M'_0$, as
\begin{equation}
\begin{gathered}
\Tr(R^{-1}R'\G_{0+}\delta D D^{-1}\G_{0+}^{-1}-R^{-1}R'\G_{0-}\delta DD^{-1}\G_{0-}^{-1}+\G_{0+}^{-1}\G_{0+}'\delta D D^{-1}-M^{-1}_0M'_0\delta DD^{-1}+\\
-\G_{0-}^{-1}\G_{0-}'\delta DD^{-1})=\Delta\le[\Tr(R^{-1}R'\G_0\delta DD^{-1}\G_0^{-1}+\G^{-1}_0\G_0'\delta DD^{-1})\ri]
\end{gathered}
\end{equation}
where $\Delta$ is the jump operator $\Delta[f]=f_+-f_-$ and we have used $\Tr(M^{-1}_0M'_0\delta DD^{-1})=0$. Let us call $\Sigma':=\Sigma\setminus\mathbb{R}_-$ (see Remark \ref{rmksigma}) and let $\wt{\Sigma}$ be the contour depicted in figure \ref{figurecauchy}, which has the property that $\int_{\Sigma'}\mathrm{d}\l\,\Delta[f]=\int_{\wt{\Sigma}}\mathrm{d}\l\,f$, so that
\begin{equation}
\begin{gathered}
\omega_n=\int_{\Sigma'}\frac{\mathrm{d}\l}{2\pi\i }\,\Delta\le[\Tr(R^{-1}R'\G_0\delta DD^{-1}\G^{-1}_0+\G^{-1}_0\G'_0\delta DD^{-1})\ri]= \\ =\int_{\wt{\Sigma}}\frac{\mathrm{d}\l}{2\pi\i }\,\Tr(R^{-1}R'\G_0\delta DD^{-1}\G^{-1}_0+\G^{-1}_0\G'_0\delta DD^{-1}).
\end{gathered}
\end{equation}

Applying Cauchy's Theorem we can deform $\wt{\Sigma}$ as in figure \ref{figurecauchy} so that finally
\begin{equation}
\label{lastexpression}
\omega_n=\int_{\pa\mathbf{D}_-}\frac{\mathrm{d}\l}{2\pi\i }\,\Tr(R^{-1}R'\G_0\delta DD^{-1}\G^{-1}_0+\G^{-1}_0\G'_0\delta DD^{-1})
\end{equation}
with the understanding that $\int_{\pa\mathbf{D}_-}\frac{\mathrm{d}\l}{2\pi\i }$ is the sum over the (formal) residues at $\l\in\{\vec{\l},\vec{\mu},\infty\}$. We want  to compare now the last expression \eqref{lastexpression} for $\omega_n$ with \eqref{bertolacafasso}. To this end we recall the identities
\begin{equation}
\delta \mathbf{J}_\infty  \mathbf{J}_\infty^{-1}=\G_0 \delta DD^{-1}\G^{-1}_0\;\;\;\;\;\;
\delta \mathbf{J}_\z \mathbf{J}_\z^{-1}=\G_0 \delta U_\z U^{-1}_\z \G^{-1}_0
\end{equation}
where $U_\zeta$ has been defined in \eqref{defU}, and the identities
\begin{equation}
\res{\l=\z}\G^{-1}_0\G'_0\delta D D^{-1}=\res{\l=\z}\G^{-1}_0\G'_0\delta U_\z U_\z^{-1},\;\;\;\;\z\in\vec{\l},\vec{\mu}
\end{equation}
which follow from $\frac{\d\sqrt{\z}}{\sqrt{\z}-\sqrt{\l}}=\frac{\d\z}{\z-\l}+\mathcal{O}(1)$ as $\l\to\z$. Replacing  $R=(\1+c\E_{32})\mathbf{R}$ (Proposition \ref{proprhpr}) we obtain
\begin{equation}
\label{miracolo}
\begin{gathered}
\omega_n-\delta\log\det\mathbf{G}=-\sum_{\l\in\vec{\l},\vec{\mu}}\res{\l=\z}\Tr\le(R^{-1}R'\G_0\le(\delta DD^{-1}-\delta U_\z U_\z^{-1}\ri)\G^{-1}_0\ri)+\res{\l=\infty}\Tr\le(\G^{-1}_0\G'_0\delta D D^{-1}\ri).
\end{gathered}
\end{equation}
A simple computation for the last term in \eqref{miracolo} shows that
\begin{equation}
\res{\l=\infty}\Tr\le(\G^{-1}_0\G'_0\delta DD^{-1}\ri)=-\Tr((S+L)\E_{11})\delta\alpha\,\alpha^{-1}=N\sum_{\z\in\vec{\l},\vec{\mu}}\frac{\d\sqrt{\z}}{\sqrt{\z}}.
\end{equation}
Define $T_\z:=U^{-1}_\z D$ and $R_+:=R\G_0 U_\z$, for $\z\in\vec{\l},\vec{\mu}$. Notice that $T_\z,R_+$ are analytic at $\l\in\vec{\l},\vec{\mu}$ and that $\delta DD^{-1}-\delta U_\z U_\z^{-1}=\delta T_\z T_\z^{-1}$ and so for all $\z\in\vec{\l},\vec{\mu}$
\begin{equation}
\begin{gathered}
\res{\l=\z}\Tr\le(R^{-1}R'\G_0\delta T_\z T^{-1}_\z\G^{-1}_0\ri)=\res{\l=\z}\Tr\le((U_\z^{-1}\G^{-1}_0R^{-1})(R'\G_0 U_\z)\delta T_\z T^{-1}_\z\ri)=
\\
=\res{\l=\z}\Tr\le((U_\z^{-1}\G^{-1}_0R^{-1})((R\G_0 U_\z)'-R\G'_0 U_\z-R\G_0 U_\z')\delta T_\z T^{-1}_\z\ri)=
\\
=\underbrace{\res{\l=\z}\Tr\le(R_+^{-1}R'_+\delta T_\z T^{-1}_\z\ri)}_{=0}-
\underbrace{\res{\l=\z}\Tr\le(\G^{-1}_0\G'_0\delta T_\z T^{-1}_\z\ri)}_{=0}-
\res{\l=\z}\Tr\le(U_\z^{-1}U'_\z\delta T_\z T^{-1}_\z\ri)=
\\
=\res{\l=\z}\frac{1}{\l-\z}\sum_{\z'\in\vec{\l},\vec{\mu}}\frac{\mathrm{d}\sqrt{\z'}}{\sqrt{\z'}-\sqrt{\l}}.
\end{gathered}
\end{equation}
To summarize,   we have proved
\begin{equation}
\begin{gathered}
\omega_n=\delta\log\det\mathbf{G}+\sum_{\z\in\vec{\l},\vec{\mu}}\res{\l=\z}\frac{1}{\l-\z}\sum_{\z'\in\vec{\l},\vec{\mu}}\frac{\mathrm{d}\sqrt{\z'}}{\sqrt{\z'}-\sqrt{\l}}+N\sum_{\z\in\vec{\l},\vec{\mu}}\frac{\mathrm{d}\sqrt{\z}}{\sqrt{\z}}=\delta\log\det\mathbf{G}+\delta\log\Delta
\end{gathered}
\end{equation}
which completes the proof.
\QED 

\begin{theorem}
\label{Thmmaximus}
The extended Kontsevich--Penner partition function coincides with the isomonodromic tau function:
\be
\mathcal{Z}_n(\vec{\l},\vec{\mu};N)=\tau_n(\vec{\l},\vec{\mu};N).
\ee
\end{theorem}
\noindent {\bf Proof.}  Comparing \eqref{thm1b} and \eqref{thm1a} with \eqref{defextendeq} we get $\delta\log\tau_n(\vec{\l},\vec{\mu};N)=\omega_n=\delta\log \mathcal{Z}_n(\vec{\l},\vec{\mu};N)$ so that $\tau_n(\vec{\l},\vec{\mu};N)$ coincides with $\mathcal{Z}_n(\vec{\l},\vec{\mu};N)$ up to multiplicative constants. The proof is complete, as the tau function is defined only up to arbitrary multiplicative constants.
\QED 
This in particular proves Theorem \ref{thm1}, which is the case $n_2=0, n=n_1$ of Thm. \ref{Thmmaximus}.
\section{The limiting isomonodromic system}\label{section3}
\subsection{The limiting isomonodromic system, and the limiting tau function}
\label{sectionprooflimiting}
As outlined in Sec. \ref{introductionlimiting} of the Introduction, we formulate an  isomonodromic system such that its  isomonodromic tau function $\tau(\T;N)$ coincides with the asymptotic expansion $\mathcal{Z}(\T;N)$
(see Prop.  \ref{proplimiting}). 

The arguments which follow are of a formal nature; however they are intended only to provide motivation for the formulation of the Riemann--Hilbert problem. After the formal discussion we explain some minimal requirements such that the Riemann--Hilbert problem \ref{rhp3times3} has a consistent analytic meaning and show how to identify it with the formal limit and hence also identify the corresponding tau function with the limiting partition  function $\mathcal Z(\T;N)$.

The products $\pi_\pm$ in \eqref{defD} can be rewritten formally as
\be
\label{31}
\frac{\alpha}{\pi_+} = \exp\sum_{k\geq 1} (-1)^k T_k \lambda^{\frac{k}{2}},\;\;\;\;\;\;
\frac{\alpha}{\pi_-} = \exp\sum_{k\geq 1} T_k \lambda^{\frac{k}{2}}
\ee
where $\alpha$ is as in \eqref{defD} and we have introduced the {\it Miwa variables} $\T=(T_1,T_2,...)$
\begin{equation}
T_k(\vec{\l},\vec{\mu}):=\frac{1}{k}\sum_{j=1}^{n_1}\le(\frac{1}{\sqrt{\l_j}}\ri)^k+\frac{1}{k}\sum_{j=1}^{n_2}\le(\frac{1}{-\sqrt{\mu_j}}\ri)^k =\frac 1 k  \sum_{j=1}^n \frac 1{y_j^k} = \frac 1 k \Tr Y^{-k}.
\end{equation}
Consequently, the matrix $D_n$ can be rewritten formally as
\begin{equation}
\label{dlim}
D^{-1}_n=\alpha^{-1}\,\exp\sum_{k\geq 1}T_k\l^{\frac{k}{2}}\theta_k\;\;\,\;\;\theta_k:=\diag(0,(-1)^k,1).
\end{equation}
More precisely, the expression above is actually convergent for $|\l|<\min\{|\l_j|, |\mu_j|\}$.

Note that $D_n$ acts by conjugation on the jumps of $\G_n$ and hence the scalar constant $\alpha$ in \eqref{dlim} is irrelevant.
In the limit $n\to\infty$ we can formally consider the variables $T_1,T_2,...$ as independent.
We then arrive at a (formal) limit of the Riemann--Hilbert problem \eqref{RHPPSI} (dropping $\alpha$ as explained above) for the matrix 
\be
\Psi (\l;\T)=\Gamma (\l;\T) \e^{\Theta(\l;\T)},\;\;\;\,\;
\Theta(\l;\T):=\sum_{k\geq 1}\left(T_k+\frac{2}{3}\delta_{k,3}\right)\l^{\frac{k}{2}}\theta_k.
\ee
Consequently, the matrix $\Gamma(\l;\T)$ solves  a new Riemann--Hilbert problem as follows:
\begin{fproblem}
\label{rhp3times3}
Let $\T$ denote the infinite set of variables $\T=(T_1,T_2,...)$. The {\it formal} Riemann-Hilbert problem amounts to finding a $3\times 3$ analytic matrix-valued function $\G=\G(\l;\T)$ in $\l\in\mathbb{C}\setminus\Sigma$ admitting non-tangential boundary values $\G_\pm$ at $\Sigma$ such that
\begin{equation}
\begin{cases}
{\G}_+(\l;\T)={\G}_-(\l;\T) M(\l;\T)&\l\in\Sigma \\
\G(\l;\T)\sim \l^S \, G \, Y(\l;\T) \, \l^L &\l\to\infty
\end{cases}
\end{equation}
where $M(\l;\T):=\e^{\Theta(\l;\T)_-} \wt{M} \e^{-\Theta(\l;\T)_+}$, $\wt M$ as in \eqref{stokes}, and $Y(\l;\T)$ is a formal power series in $\l^{-\frac{1}{2}}$ satisfying the normalization
\begin{equation}
\label{normalizationinftyinfty}
Y(\l;\T)=\1+\begin{bmatrix}
0&a&-a \\ 0&c&0 \\ 0&0&-c
\end{bmatrix}\l^{-\frac{1}{2}}+\mathcal{O}\le(\l^{-1}\ri)
\end{equation}
for some functions $a=a(\T),c=c(\T)$.
\end{fproblem}
\begin{remark}
Remark \ref{rmksymmetrynormalization} applies  here as well for the uniqueness of the solution to  the Riemann-Hilbert problem \ref{rhp3times3}. Moreover, the symmetry relation \eqref{eigenspaces} holds true similarly here, namely
\be
\label{symmetry}
Y(\l\e^{2\pi\i};\T)=
\begin{bmatrix}
1&0&0 \\ 0&0&1 \\ 0&1&0
\end{bmatrix}
Y(\l;\T)
\begin{bmatrix}
1&0&0 \\ 0&0&1 \\ 0&1&0
\end{bmatrix}
\ee
\end{remark}
We now explain a meaningful setup where the Riemann--Hilbert problem \ref{rhp3times3} can be given a completely rigorous analytic meaning. 
The driving idea is that of truncating the time variables to some finite (odd) number. 

Fix now $K\in \N$ and assume that $T_{\ell}=0$ for all $\ell \geq 2K+2$.
Set $\T= (T_1,\dots, T_{2K+1},0,\dots)$ with $T_{2K+1}\not=0$. In addition, the angles $\beta_{0,\pm}$ (satisfying \eqref{beta}) and the argument of $T_{2K+1}$ must satisfy the following condition:
\begin{equation}
\label{consist}
\begin{cases}
\Re\le(\l^{\frac{2K+1}{2}}T_{2K+1}\ri)<0,&\l\in\e^{\i\beta_\pm}\mathbb{R}_+
\\
\Re\le(\l^{\frac{2K+1}{2}}T_{2K+1}\ri)>0,&\l\in\e^{\i\beta_0}\mathbb{R}_+.
\end{cases}
\end{equation}

Under this assumption, given the particular triangular structure of the Stokes matrices $\mathcal{S}_{0,\pm}$, the jumps $M=\e^{\Theta_-} \wt{M} \e^{-\Theta_+}$ are exponentially close to the identity matrix  along the rays $\e^{\i\beta_{0,\pm}}\mathbb{R}_+$.

\begin{proposition}\phantomsection
\label{proplimiting}
\begin{enumerate}
\item
Under the assumptions above, there is a unique analytic solution to  Riemann--Hilbert problem \ref{rhp3times3}, 
which is holomorphic in $T_1,\dots, T_{2K+1}$ in a neighbourhood of the origin $T_\ell=0,  \ \ \forall\ell$ and with the argument of $T_{2K+1}$ restricted to  a suitable sector of the complex plane as implied by \eqref{consist}.
\item 
Moreover, setting $T_1=0,...,T_{2K}=0$ and then letting $T_{2K+1}\to 0$ in the appropriate sector then $\G(\l;\T)\to\G_0(\l)$. 
\item
In the same limit, all the logarithmic derivatives of $\tau(\T;N)$ converge to the logarithmic derivatives of $\mathcal{Z}(\T;N)$ with respect to the corresponding Miwa variables evaluated at $\T=0$. In particular  $\log\,\tau(\T;N)$ admits a formal Taylor series expansion at $\T=0$ which is the asymptotic expansion of the truncation $(T_1,\dots, T_{2K+1}, 0 ,\dots)$ for arbitrary integer $K$ in an appropriate sector of the time  $T_{2K+1}$ near $0$. 
\end{enumerate}
\end{proposition}

{
The proof of points 1,2 of Prop. \ref{proplimiting} follows the same lines as the proof of Prop. 3.5 in \cite{BeCa2017}. 
The independence on $K$ of the formal Taylor expansion  mentioned in point 3 is proven as in Prop. 3.6 in loc. cit.; 

The convergence of the function $\tau_n(\vec \l, \vec \mu;N)$ in terms of Miwa variables to $\tau(\T;N)$ could be proved with a rigorous  analytic approach (at least for subsequences of $n$) following the lines of Thm. 1.9 in \cite{BeCa2017}. 
We recall the main steps that could be followed for a  rigorous analytic proof.
For a given $2K+1$--tuple of times $(T_1,\dots, T_{2K+1})$, the equation \eqref{31} suggests that we should consider the Taylor approximations 
\be
P_n(\z) + \mathcal O(\z^{n+1}) =\exp\le[{\sum_{j=1}^{2K+1} T_j \z^j}\ri],\ \ \z \to 0.
\ee
Denote the set of roots of the polynomials $P_n(y)$ by $\mathcal Y_n(\T)$ and partition them into the ones in the left/right half planes:$\mathcal Y_n(\T) =
\mathcal Y_n^{L}(\T)\sqcup \mathcal Y_n^{R}(\T)$ ( the purely imaginary ones conventionally assigned to the right plane). 

The squares of the zeroes in $\mathcal Y_n^{L}(\T)$ define the $\vec \mu$ and similarly, the squares of the zeroes in $\mathcal Y_n^{R}(\T)$ define the $\vec \l$. This, in turn, defines the products $\pi_\pm^{(n)}$ in \eqref{31} (we emphasize the dependence on $n$ by the superscript). 

A rigorous analytical proof then should analyze the uniform convergence on  compact sets of the $\l$--plane of  the expressions 
\be
\frac {\pi_{\pm}^{(n)} (\l)}
{\pi_{\pm}^{(n)} (0)} \to \exp\le[{\sum_{j=1}^{2K+1}(\mp)^j T_j \l^\frac j2}\ri]
\ee
so that the jump matrices of the Riemann--Hilbert problem \ref{RHP} converge in the appropriate norms to the jump matrices of the Riemann--Hilbert problem \ref{rhp3times3}. 

This was accomplished in \cite{BeCa2017} in complete rigour for the special set of times $(0,\dots, 0, T_{2K+1}, 0,\dots)$ in the case $N=0$; a general proof requires detailed analysis of the convergence and location of zeroes as $n\to \infty$ for arbitrary $\T$. We consider this a technical issue beyond the scope of the present work, since we are mostly concerned with the formal structure of the expansion. 

 }

\paragraph{Relation with a hierarchy of isomonodromic systems}
We now briefly review the  usual arguments  showing that $\Psi(\l;\T)=\G(\l;\T)\e^{\Theta(\l;\T)}$ satisfies a Riemann--Hilbert problem with jump matrices $\wt M$ independent of $\lambda,\T$ so that we fall within the purview of the theory of isomonodromic deformations. Namely, $\Psi(\l;\T)$ is a solution to  a {\it polynomial} ODE in the variable $\l$ (see Remark \ref{rmkA})
\be
\label{lambdaODE}
\pa_\l \Psi(\l;\T)=A(\l;\T) \Psi(\l;\T)
\ee
satisfying also a compatible system of deformation equations with respect to the times $\T$ (see Lemma \ref{lemmaomega})
\be
\label{TkODE}
\pa_{T_k}\Psi(\l;\T)=\Omega_k(\l;\T) \Psi(\l;\T), \ \ \ \ k=1,...,2K+1
\ee
with zero--curvature equations
\begin{equation}
\label{compatibility3}
\pa_{T_k} A-\pa_\l \Omega_k=[\Omega_k,A],\;\;\;\;\;\pa_{T_k}\Omega_j-\pa_{T_j} \Omega_k=[\Omega_k,\Omega_j], \ \ \ \ j,k=1,...,2K+1.
\end{equation}

The equations \eqref{lambdaODE} and \eqref{TkODE} thus provide an isomonodromic system, in the sense of \cite{JiMiUe1981}. More precisely we have the following

\begin{lemma}
\label{lemmaomega}
The matrices $\Omega_k$ in \eqref{TkODE} are polynomials in $\l$ of degree $\lceil\frac{k}{2}\rceil$. They can be written as:
\begin{equation}
\Omega_k(\l;\T)=\frac{\pa c(\T)}{\pa T_k}\E_{32}+\res{\nu=\infty}\frac{\nu^{\frac{k}{2}}\Psi(\nu;\T) \theta_k \Psi^{-1}(\nu;\T)\d \nu}{\l-\nu}
\end{equation}
where $\theta_k$ is in \eqref{dlim}, $c=c(\T)$ is as in \eqref{normalizationinftyinfty}, and $\E_{32}$ denotes the elementary unit matrix.
\end{lemma}
\noindent {\bf Proof.}  
According to \eqref{lambdaODE} we have $\Omega_k(\l;\T) = \pa_{T_k} \Psi(\l;\T) \Psi(\l;\T)^{-1}$.  Since the jump matrices of $\Psi$ are independent of $\T$, and $\det \Psi\equiv 1$, it follows that $\Omega_k$ is an entire function of $\l$. Using the asymptotic expansion for $\Psi$ that follows from the Riemann--Hilbert problem \ref{rhp3times3}, as well as the symmetry \eqref{symmetry} one can verify that it has polynomial growth at $\l=\infty$ of degree $\lceil\frac{k}{2}\rceil$.
By Liouville's theorem $\Omega_k$ coincides with the polynomial part of $\pa_{T_k} \Psi(\l;\T) \Psi(\l;\T)^{-1}$. 
We can compute it as
\begin{equation}
\begin{gathered}
\Omega_k= \le(\pa_{T_k} \Psi\, \Psi^{-1}\ri)_+= \le(\l^S\,G \,\frac{\pa Y}{\pa T_k} \, Y^{-1} \, G^{-1}\,\l^{-S}+\l^{\frac{k}{2}}\,\l^S \, G \, Y\, \theta_k  Y^{-1} \, G^{-1} \, \l^{-S}\ri)_+=
\\
=\frac{\pa c}{\pa T_k}\E_{32}+\le(\l^{\frac{k}{2}}\Psi(\l) \theta_k \Psi^{-1}(\l)\ri)_+=\frac{\pa c}{\pa T_k}\E_{32}+\res{\nu=\infty}\frac{\nu^{\frac{k}{2}}\Psi(\nu) \theta_k \Psi^{-1}(\nu)\d\nu}{\l-\nu}
\end{gathered}
\end{equation}
where we have denoted the polynomial part of the expansion at $\l=\infty$ by the subscript $+$. We remark that the numerator inside the residue admits a formal series expansion in $\nu$ with only integer powers, thanks to the symmetry \eqref{symmetry}, with only finitely many positive powers in $\nu$; therefore the (formal) residue is well defined.
\QED 

\begin{remark}
\label{rmkA}
Arguments entirely analogous to those used in Lemma \ref{lemmaomega} show that $A(\l;\T)$ is also a polynomial in $\l$, of degree $K$.
\end{remark}

Associated with this isomonodromic system we have the isomonodromic tau function $\tau(\T;N)$:

\begin{proposition}
\label{propjapanese}
The Malgrange differential
\begin{equation}
\omega_M:=\int_\Sigma\frac{\d\l}{2\pi\i}\,\Tr\le(\G^{-1}_-(\l;\T)\,\G'_-(\l;\T)\,\delta M(\l;\T)\,M^{-1}(\l;\T)\ri)
\end{equation}
coincides with the Jimbo-Miwa-Ueno differential
\begin{equation}
\omega_{JMU}:=-\res{\l=\infty}\mathrm{d}\l\,\Tr\le(\G^{-1}(\l;\T)\,\G'(\l;\T)\,\delta\Theta(\l;\T)\ri)
\end{equation}
where the differential $\delta:=\sum\limits_{k\geq 1}\mathrm{d}T_k\,\pa_{T_k}$.
\end{proposition}
\noindent {\bf Proof.} 
The formal residue in the Jimbo-Miwa-Ueno differential can be computed in the present case as
\begin{equation}
\omega_{JMU}=-\res{\l=\infty}\d\l\,\Tr\le(\G^{-1} \G' \delta\Theta\ri)=\int_{\wt\Sigma}\frac{\d\l}{2\pi\i}\,\Tr\le(\G^{-1} \G' \delta\Theta\ri)
\end{equation}
where $\wt\Sigma$ is the contour depicted in figure \ref{figurecauchy}, without taking into account the small circles around the poles. Exactly as already explained in figure \ref{figurecauchy} the last integral can be rewritten thanks to the Cauchy Theorem as:
\be
\omega_{JMU}=\int_\Sigma\frac{\d\l}{2\pi\i}\,\Delta\le[\Tr\le(\G^{-1} \G' \delta\Theta\ri)\ri]
\ee
Finally, using $\G_+=\G_-M$ and $\Tr(M^{-1}M'\delta\Theta)=0$,
\be
\begin{gathered}
\Delta\le[\Tr\le(\G^{-1} \, \G' \,\delta\Theta\ri)\ri]=\Tr\le(\le(M^{-1}\G_-^{-1}\G_-'M-\G_-^{-1}\G_-'\ri)\delta\Theta\ri)=
\\
=\Tr\le(\G_-^{-1}\G_-'\le(M\,\delta\Theta\, M^{-1}-\delta\Theta\ri)\ri)=\Tr\le(\G_-^{-1}\,\G_-'\,\delta M\, M^{-1}\ri)
\end{gathered}
\ee
thus $\omega_{JMU}=\omega_M$.
\QED

As $\delta\omega_{JMU}=0=\delta\omega_M$ \cite{JiMiUe1981,Be2010} the isomonodromic tau function $\tau(\T;N)$ can be introduced according to
\be
\delta\log\tau=\omega_M=\omega_{JMU}.
\ee
This definition is spelled out as \eqref{taujapanese}.

Finally we remark that in the formal limit in which $\T$ is no longer assumed to have only a finite number of nonzero entries, we lose the $\l$-ODE \eqref{lambdaODE} but we can consider \eqref{TkODE} as an infinite hierarchy of PDEs in the times $\T$, with coefficients which are polynomials in $\l$; the zero curvature equations \eqref{compatibility3} still make sense by themselves by  matching the corresponding coefficients of the powers of $\l$. 

The formula \eqref{taujapanese} still makes sense, so we may regard $\tau(\T;N)$ as a formal function of the infinitely many variables $\T=(T_1,T_2,...)$.
\subsection{Correlations functions: proof of Theorem \ref{thm2}}\label{secproofthm2}
In this paragraph we prove Theorem \ref{thm2}. In particular we need a simplification of certain products of asymptotic expansions of the functions $f(\l;N)$, as in the following Lemma.

\begin{lemma}\label{lemmalaplace}
For $a,b\in\{0,\pm 1\}$, let 
\be
\label{integraltransform}
F_-(\l;N+a)\,F_+(\l;-N-b)=\sum_{k\geq 0}P_{a,b}^k(N)\l^{-\frac{3k}{2}}.
\ee
Then the polynomials   $P_{a,b}^k(N)$ in the indeterminate $N$ coincide with those  in \eqref{polynomialsK}.
\end{lemma}

\noindent {\bf Proof.}
The expression $\l^{-\frac{a-b+1}{2}}F_-(\l;N+a)\,F_+(\l;-N-b)$ is the formal expansion of the product of a solution to $\partial^3-\l\pa+N+a-1$ and of a solution to $\pa^3-\l\pa-N-b+1$. As such it is annihilated by the following ninth-order differential operator;
\begin{equation}
\label{ninthorder}
\begin{gathered}
\left( a-b-2 \right)  \left( a-b+1 \right)  \left( a-b+4 \right)
 + \left( 3(a-b)^2-3(a-b)-46\right) \lambda\partial_\l -30\,{\lambda}^{2}\pa_\l^2 +
 \\
 + \left(
102-3\,(a-b)+3\,{a}^{2}+21\,ab+3\,{b}^{2}+27\,N\,\left( a+b+N \right) N-4\,{\lambda}^{3} \right) \pa_\l^3
+3\, \left( 23-a+b \right)\lambda\pa_\l^4
\\
+9\,{\lambda}^{2}\pa_\l^5 +3\, \left( -11+a-b \right)\pa_\l^6 -6\,\lambda\pa_\l^7 +\pa_\l^9.
\end{gathered}
\end{equation}
Introduce the power series (a formal Laplace transform)
\be
G(x):=\sum_{k\geq 0}P_{a,b}^k(N)\frac{x^{\frac{3k+a-b-1}{2}}}{\G\le(\frac{3k+a-b+1}{2}\ri)}.
\ee
Then $G(x)$ is annihilated by the third-order differential operator
\begin{equation}
\label{thirdorder}
\begin{gathered}
\pa_x^3-\frac{3\left( 3x^3-2 \right)}{4x}\pa_x^2
+\frac{3(2x^6-\left(7+a-b \right) x^3-((a-b)^2-a+b+2))}{4x^2}\pa_x+
 \\
 +\frac{-x^9+3(a-b+3)x^6-(3(a+b)^2+15ab+9(a-b)+6+27N(a+b+N))x^3+(a-b+2)(a-b-2)}{4x^3}
\end{gathered}
\end{equation}
obtained from \eqref{ninthorder} by replacing $\l$ with $\pa_x$ and $\pa_\l$ with $-x$. We are therefore interested in power-series solutions around the Fuchsian singularity $x=0$ of \eqref{thirdorder}. It is easily checked that for $a,b\in\{0,\pm 1\}$ the equation \eqref{thirdorder} is resonant and the Frobenius solutions at $x=0$ span a two-dimensional space\footnote{The expansion of the third solution involves logarithms.} generated by the two series below;
\be
\begin{gathered}
\renewcommand*{\arraystretch}{1.25}
G_1(x):=x^{\frac{a-b-1}{2}}\e^{\frac{x^3}{3}}\,_2F_2\le(\le.\begin{matrix}
\frac{1-a-b-2N}{2} & \frac{1+a+b+2N}{2} \\
\frac{1}{2} & \frac{1+a-b}{2}\end{matrix}\ri|-\frac{x^3}{4}\ri)
\\
G_2(x):=x^{\frac{a-b+2}{2}}\e^{\frac{x^3}{3}}\,_2F_2\le(\le.\begin{matrix}
\frac{2-a-b-2N}{2} &\frac{2+a+b+2N}{2} \\
\frac{3}{2} & \frac{2+a-b}{2}\end{matrix}\ri|-\frac{x^3}{4}\ri).
\end{gathered}
\ee
By matching with $F_-(\l;N+a)F_+(\l;-N-b)=1-\frac{(a-b+2)(2N+a+b)}{4}\l^{-\frac{3}{2}}+\mathcal{O}\le(\l^{-3}\ri)$ we obtain
\be
G(x)=\frac{G_1(x)}{\G\le(\frac{a-b+1}{2}\ri)}-\frac{2N+a+b}{2}\frac{G_2(x)}{\G\le(\frac{a-b+2}{2}\ri)}\ .
\ee
The proof is complete.
\QED
\subsubsection{One--point function: proof of (\ref{1pointresult}) }
The one--point correlation function is simply the generating function of the derivatives \eqref{taujapanese}
\begin{equation}
\label{329}
\begin{gathered}
S_1(\l;\t)=\sum_{d\geq 0}\frac{(-1)^{d+1}(d+1)!!}{2^{\frac{d+1}{3}}\l^{\frac{d}{2}+1}}\left\langle\left\langle\tau_{\frac{d}{2}}\right\rangle\right\rangle=\sum_{d\geq 0}\frac{(-1)^{d+1}(d+1)!!}{2^{\frac{d+1}{3}}\l^{\frac{d}{2}+1}}\frac{\pa\log\tau}{\pa t_{d+1}}=\sum_{d\geq 0}\frac{1}{\l^{\frac{d}{2}+1}}\frac{\pa\log\tau}{\pa T_{d+1}}=
\\
=-\sum_{d\geq 0}\frac{1}{\l^{\frac{d}{2}+1}}\res{\mu=\infty}\Tr\,\le(\G'(\mu;\t) \theta_{d+1} \G^{-1}(\mu;\t)\mu^{\frac{d+1}{2}}\d\mu\ri)=S^+_1(\l;\t)+S_1^-(\l;\t)-N\l^{-\frac{1}{2}}
\end{gathered}
\end{equation}
where
\begin{equation}
\label{Spiumeno}
\begin{gathered}
S^\pm_1(\l;\t)=\Tr\le(\l^{\frac{1}{2}}\G'(\l;\t) \begin{bmatrix}
0&0&0 \\ 0&\pm 1&0 \\ 0&0&1 
\end{bmatrix} \G^{-1}(\l;\t)\ri)
\end{gathered}
\end{equation}
Here and below,  with an abuse of notation, we identify the analytic functions with their regular asymptotic expansion at $\l=\infty$, which are in any case well-defined.
The last equality in \eqref{329} is proved by observing that both $S_1^\pm$ in \eqref{Spiumeno} are integer power series thanks to the symmetry  \eqref{symmetry}; the term $S_1^+$ thus contains the odd--$d$ terms of the sum in \eqref{329}, while the term $S_1^-$ contains the even ones. Adding $S_1^++S_1^-$ now gives the following simplification:
\begin{equation}
\label{1pf}
\begin{gathered}
S_1(\l;\t)=2\,\Tr\le(\l^{\frac{1}{2}}\G'(\l;\t)  \E_{33} \G^{-1}(\l;\t)\ri)-N\l^{-\frac{1}{2}}=
\\
=2\,\Tr\le(\l^{\frac{1}{2}}\Psi'(\l;\t)  \E_{33} \Psi^{-1}(\l;\t)-\l^{\frac{1}{2}}\Theta'(\l;\t)  \E_{33}\ri)-N\l^{-\frac{1}{2}}=
\\
=2\,\Tr\le(\l^{\frac{1}{2}}\Psi'(\l;\t)  \E_{33} \Psi^{-1}(\l;\t)\ri)-2\,\Tr\le(\l^{\frac{1}{2}}\Theta'(\l;\t)  \E_{33}\ri)-N\l^{-\frac{1}{2}}
\end{gathered}
\end{equation}
The correlators are the evaluations of \eqref{taujapanese} at $\T=0$ and then $\Psi_0(\l) := \Psi(\l;\T)\bigg|_{\T=0}$ is explicitly known in terms of contour integrals of Airy type. By applying the differential operator $\frac{\pa}{\pa\l} \l^{-\frac{1}{2}}$ and setting $\T=0$ we obtain
\begin{equation}
\label{1pfb}
\sum_{d\geq 0}\frac{(-1)^{d}(d+3)!!}{2^{\frac{d+4}{3}}\l^{\frac{d+5}{2}}}\left\langle\tau_{\frac{d}{2}}\right\rangle
=2\,\Tr\le(\E_{32} \Psi_0(\l)  \E_{33} \Psi^{-1}_0(\l)\ri)+N\l^{-2}-\l^{-\frac{1}{2}}
\end{equation}
where we have used that $A=\Psi_0' \Psi_0^{-1}=\begin{bmatrix}
0&1&0 \\ 0&0&1 \\ N&\l&0
\end{bmatrix}$ satisfies $A'=\E_{32}$ and $\Tr\le(A\le(\Psi_0\E_{33}\Psi^{-1}_0\ri)'\ri)=0$. We have explicit formul\ae\ for $\Psi_0,\Psi_0^{-1}$, where $F_\pm(\l;\ell)$ have been defined in Prop. \ref{propode0}, see \eqref{defCjN} and \eqref{Fpiu}:
\begin{equation}
\label{psi}
\Psi_0(\l)\sim\frac{1}{\sqrt{2}}\begin{bmatrix}
\star & -\l^{\frac{N}{2}-\frac{3}{4}}F_-(\l;-N+1) & \l^{\frac{N}{2}-\frac{3}{4}}F_+(\l;-N+1) \\
\star & \l^{\frac{N}{2}-\frac{1}{4}}F_-(\l;-N) & \l^{\frac{N}{2}-\frac{1}{4}}F_+(\l;-N)\\
\star & -\l^{\frac{N}{2}+\frac{1}{4}}F_-(\l;-N-1) & \l^{\frac{N}{2}+\frac{1}{4}}F_+(\l;-N-1)
\end{bmatrix}
\cdot 
\begin{bmatrix}
1 &0&0 \\ 0&\e^{-\frac{2}{3}\l^{\frac{3}{2}}} &0\\0&0&\e^{\frac{2}{3}\l^{\frac{3}{2}}}
\end{bmatrix}
\end{equation}
\begin{equation}
\label{psiinverse}
\begin{gathered}
\Psi^{-1}_0(\l)
\sim
\frac{1}{\sqrt{2}}
\begin{bmatrix}
1 & 0&0 \\ 0& \e^{\frac{2}{3}\l^{\frac{3}{2}}}&0 \\ 0&0&\e^{-\frac{2}{3}\l^{\frac{3}{2}}}
\end{bmatrix} 
\\  \begin{bmatrix}
\star & \star & \star \\
-\l^{-\frac{N}{2}-\frac{1}{4}}F_+(\l;N)
&
-\l^{-\frac{N}{2}+\frac{1}{4}}F_+(\l;N-1)
&
-\l^{-\frac{N}{2}+\frac{3}{4}}F_+(\l;N-2)
\\
\l^{-\frac{N}{2}-\frac{1}{4}}F_-(\l;N)
&
-\l^{-\frac{N}{2}+\frac{1}{4}}F_-(\l;N-1)
&
\l^{-\frac{N}{2}+\frac{3}{4}}F_-(\l;N-2)
\end{bmatrix}
\cdot 
\begin{bmatrix}
-\l &0&1 \\ 0&-1&0 \\ 1&0&0
\end{bmatrix}
\end{gathered}
\end{equation}
The formula for $\Psi_0^{-1}$ is obtained as in Proposition \ref{propositionbilinear}. Inserting this into \eqref{1pfb} we get
\begin{equation}
\label{1pfc}
\sum_{d\geq 0}\frac{(-1)^{d}(d+3)!!}{2^{\frac{d+4}{3}}\l^{\frac{d+5}{2}}}\left\langle\tau_{\frac{d}{2}}\right\rangle=\l^{-\frac{1}{2}}F_+(\l;-N)F_-(\l;N)+N\l^{-2}-\l^{-\frac{1}{2}}
\end{equation}
By applying the case $a=0=b$ of Lemma \ref{lemmalaplace} we obtain the identity
\begin{equation}
\partial_\l(\l^{-\frac{1}{2}}S_1(\l))=\sum_{k\geq 2}P_{0,0}^k(N)\l^{-\frac{3k+1}{2}}
\end{equation}
which is equivalent to \eqref{1pointresult}, by inverting the operator $\pa_\l\l^{-\frac{1}{2}}$ and changing summation variable $g=k-1$.
\subsubsection{Two--point function}
\label{twopoints}
Introduce the operator
\begin{equation}
\nabla(\mu):=\sum_{d\geq 0}\frac{(-1)^{d+1}(d+1)!!}{2^{\frac{d+1}{3}}\mu^{\frac{d}{2}+1}}\frac{\pa}{\pa t_{d+1}}=\sum_{d\geq 0}\frac{1}{\mu^{\frac{d}{2}+1}}\frac{\pa}{\pa T_{d+1}}
\end{equation}

We have, using \eqref{1pf}, $\frac{\pa\Psi}{\pa T_k}=\Omega_k\Psi$ and Lemma \ref{lemmaomega}, omitting the dependence on $\T$,
\bea
\label{2pointresulttimes}
S_2(\l_1,&&\hspace{-24pt}\l_2;\T)  =\nabla(\l_2)S_1(\l_1;\T)=2\sum_{d\geq 0}\frac{1}{\l_2^{\frac{d}{2}+1}}\frac{\pa}{\pa T_{d+1}}\Tr\le(\l_1^{\frac{1}{2}}\Psi'(\l_1)\E_{33}\Psi^{-1}(\l_1)-\l_1^{\frac{1}{2}}\Theta'(\l_1)\E_{33}\ri)=
\nonumber
\\
\nonumber
&=&2\sum_{d\geq 0}\frac{1}{\l_2^{\frac{d}{2}+1}}\Tr\le(\cancel{\l_1^{\frac{1}{2}}\Omega_{d+1}(\l_1)\Psi'(\l_1)\E_{33}\Psi^{-1}(\l_1)}+\l_1^{\frac{1}{2}}\Omega_{d+1}'(\l_1)\Psi(\l_1)\E_{33}\Psi^{-1}(\l_1)+\ri.
\nonumber\\
\nonumber
&&\le.-\cancel{\l_1^{\frac{1}{2}}\Psi'(\l_1)\E_{33}\Psi^{-1}(\l_1)\Omega_{d+1}(\l_1)}-\frac{d+1}{2}\l_1^{\frac{d}{2}}\E_{33}\ri)=
\\
\nonumber
&
=&2\sum_{d\geq 0}\frac{1}{\l_2^{\frac{d}{2}+1}}\Tr\le(\l_1^{\frac{1}{2}}\frac{\partial}{\partial\l_1}\le(\l_1^{\frac{d+1}{2}}\Psi(\l_1)\theta_{d+1}\Psi^{-1}(\l_1)\ri)_+\Psi(\l_1)\E_{33}\Psi^{-1}(\l_1)\ri)-\sum_{d\geq 0}(d+1)\frac{\l_1^{\frac{d}{2}}}{\l_2^{\frac{d}{2}+1}}=
\\
\nonumber
&=&2\,\Tr\sum_{d\geq 0}\frac{1}{\l_2^{\frac{d}{2}+1}}\le(\res{\nu=\infty}\le(\frac{\nu^{\frac{d+1}{2}}\Psi(\nu)\theta_{d+1}\Psi^{-1}(\nu)}{(\l_1-\nu)^2}\ri)\l_1^{\frac{1}{2}}\Psi(\l_1)\E_{33}\Psi^{-1}(\l_1)\ri)-\frac{1}{\le(\l_1^{\frac{1}{2}}-\l_2^{\frac{1}{2}}\ri)^{2}}=
\\
&=&\Tr\frac{\mathcal{A}(\l_1;\T)\mathcal{A}(\l_2;\T)}{(\l_1-\l_2)^2}-\frac{1}{\le(\l_1^{\frac{1}{2}}-\l_2^{\frac{1}{2}}\ri)^{2}}
\eea
where we have introduced
\begin{equation}
\mathcal{A}(\l;\T):=2\,\l^{\frac{1}{2}}\,\Psi(\l;\T) \, \E_{33} \, \Psi^{-1}(\l;\T).
\end{equation}
In the last step of \eqref{2pointresulttimes} we have used the same type of reasoning used after \eqref{Spiumeno} to re-sum the odd/even terms of the sum into one series, using again the symmetry \eqref{symmetry}. Evaluating at $\T=0$, using \eqref{psi} and \eqref{psiinverse} and the recursion \eqref{basicidentity2}, and writing $F^\pm_\ell:=F_\pm(\l;\ell)$ for short, we find
\begin{equation}
\begin{gathered}
A(\l):=\mathcal{A}(\l;\T=0)=2\l^{\frac{1}{2}}\Psi_0(\l)  \E_{33} \Psi_0^{-1}(\l)=
\\
=\l^{\frac{1}{2}}\,\begin{bmatrix}
0&0&\l^{\frac{N}{2}-\frac{3}{4}}F^+_{-N+1} \\
0&0&\l^{\frac{N}{2}-\frac{1}{4}}F^+_{-N} \\
0&0&\l^{\frac{N}{2}+\frac{1}{4}}F^+_{-N-1} \\
\end{bmatrix}
\cdot 
\begin{bmatrix}
0&0&0\\ 0&0&0 \\ 
N\l^{-\frac{N}{2}-\frac{3}{4}}F^-_{N+1}&\l^{-\frac{N}{2}+\frac{1}{4}}F^-_{N-1}&\l^{-\frac{N}{2}-\frac{1}{4}}F^-_{N}
\end{bmatrix}=
\\
=\begin{bmatrix}
 \l^{-1}N F^-_{N+1} F^+_{-N+1}& F^-_{N-1} F^+_{-N+1} & \l^{-\frac{1}{2}}F^-_{N} F^+_{-N+1} \\
 \l^{-\frac{1}{2}}N F^-_{N+1} F^+_{-N} & \l^{\frac{1}{2}} F^-_{N-1} F^+_{-N} & F^-_{N} F^+_{-N} \\
 N F^-_{N+1} F^+_{-N-1} & \l F^-_{N-1} F^+_{-N-1} & \l^{\frac{1}{2}}F^-_{N} F^+_{-N-1}
\end{bmatrix}=
\\
=\begin{bmatrix}
N\sum\limits_{k\geq 0}P_{1,-1}^k(N)\l^{-\frac{3k+2}{2}} & \sum\limits_{k\geq 0}P_{-1,-1}^k(N)\l^{-\frac{3k}{2}} & \sum\limits_{k\geq 0}P_{0,-1}^k(N)\l^{-\frac{3k+1}{2}}\\
N\sum\limits_{k\geq 0}P_{1,0}^k(N)\l^{-\frac{3k+1}{2}} & \sum\limits_{k\geq 0}P_{-1,0}^k(N)\l^{-\frac{3k-1}{2}} & \sum\limits_{k\geq 0}P_{0,0}^k(N)\l^{-\frac{3k}{2}}\\
N\sum\limits_{k\geq 0}P_{1,1}^k(N)\l^{-\frac{3k}{2}} & \sum\limits_{k\geq 0}P_{-1,1}^k(N)\l^{-\frac{3k-2}{2}} & \sum\limits_{k\geq 0}P_{0,1}^k(N)\l^{-\frac{3k-1}{2}}
\end{bmatrix}
\end{gathered}
\end{equation}
where, in the last step,  we applied Lemma \ref{lemmalaplace}. The formula  \eqref{npointresult} of Theorem \ref{thm2} is hence established for $n=2$.
\subsubsection{n--point functions}
\begin{lemma}
\label{nablaA}
The following  identity of formal series holds:
\begin{equation}
\nabla(\mu)\mathcal{A}(\l;\T)=\frac{[\mathcal{A}(\mu;\T),\mathcal{A}(\l;\T)]}{\mu-\l}+[\nabla(\mu)c(\T)\E_{32},\mathcal{A}(\l;\T)]
\end{equation}
where $c(\T)$ is as in \eqref{normalizationinftyinfty} and $\E_{32}$ denotes the elementary unit matrix.
\end{lemma}
\noindent {\bf Proof.}  
First notice that $\frac{\pa}{\pa T_k}\mathcal{A}(\l;\T)=[\Omega_k(\l;\T),\mathcal{A}(\l;\T)]$. Using now Lemma \ref{lemmaomega} we write
\begin{equation}
\begin{gathered}
\nabla(\mu)\mathcal{A}(\l;\T)=\sum_{d\geq 0}\frac{1}{\mu^{\frac{d}{2}+1}}[\Omega_{d+1}(\l;\T),\mathcal{A}(\l;\T)]=
\\
=\sum_{d\geq 0}\frac{1}{\mu^{\frac{d}{2}+1}}\res{\nu=\infty}
\le[
\frac{\nu^{\frac{d+1}{2}}\Psi(\nu;\T)\theta_{d+1}\Psi^{-1}(\nu;\T)}{\l-\nu},\mathcal{A}(\l;\T)
\ri]
+\sum_{d\geq 0}\frac{1}{\mu^{\frac{d}{2}+1}}\frac{\pa c(\T)}{\pa T_{d+1}}\le[\E_{32},\mathcal{A}(\l;\T)\ri]=
\\
=\frac{[\mathcal{A}(\mu;\T),\mathcal{A}(\l;\T)]}{\mu-\l}+[\nabla(\mu)c(\T)\E_{32},\mathcal{A}(\l;\T)]
\end{gathered}
\end{equation}
In the last step we have used again the symmetry \eqref{symmetry} and the argument already used after \eqref{Spiumeno} and in the last step of \eqref{2pointresulttimes}.
\QED

\begin{proposition}
For $n\geq 2$, the $n$--point correlation functions are expressed as follows:
\begin{equation}
\label{npointresulttimes}
S_n(\l_1,...,\l_n;\T)=-\frac{1}{n}\,\sum_{i\in S_n}\Tr\frac{\mathcal{A}(\l_{i_1};\T) \cdots\mathcal{A}(\l_{i_n};\T)}{(\l_{i_1}-\l_{i_2}) \cdots (\l_{i_n}-\l_{i_1})}-
\frac {\delta_{n,2}}{\le(\l_1^{\frac{1}{2}}-\l_2^{\frac{1}{2}}\ri)^{2}}.
\end{equation}
\end{proposition}
\noindent {\bf Proof.}  
The proof proceeds by induction on $n$. The initial case $n=2$  is the content of section \ref{twopoints} (see \eqref{2pointresulttimes}). Assume now that \eqref{npointresulttimes} holds true for $n\geq 2$ and compute, using Lemma \ref{nablaA}:
\begin{equation}
\begin{gathered}
S_{n+1}(\l_1,...,\l_{n+1};\T)=\nabla(\l_{n+1})S_n(\l_1,...,\l_n;\T)=-\frac{1}{n}\,\nabla(\l_{n+1})\sum_{i\in S_n}\Tr\frac{\mathcal{A}(\l_{i_1};\T) \cdots\mathcal{A}(\l_{i_n};\T)}{(\l_{i_1}-\l_{i_2}) \cdots(\l_{i_n}-\l_{i_1})}=
\\
=-\frac{1}{n}\,\sum_{j=1}^n\sum_{i\in S_n}\Tr\frac{\mathcal{A}(\l_{i_1};\T) \cdots\nabla(\l_{n+1})\mathcal{A}(\l_{i_j};\T) \cdots\mathcal{A}(\l_{i_n};\T)}{(\l_{i_1}-\l_{i_{2}}) \cdots (\l_{i_n}-\l_{i_1})}=
\\
=-\frac{1}{n}\,\sum_{j=1}^n\sum_{i\in S_n}\Tr\frac{\mathcal{A}(\l_{i_1};\T) \cdots [\mathcal{A}(\l_{n+1};\T),\mathcal{A}(\l_{i_j};\T)] \cdots \mathcal{A}(\l_{i_n};\T)}{(\l_{i_1}-\l_{i_{2}}) \cdots (\l_{i_n}-\l_{i_1})(\l_{n+1}-\l_j)}+
\\
-\frac{1}{n}\,\sum_{j=1}^n\sum_{i\in S_n}\Tr\frac{\mathcal{A}(\l_{i_1};\T) \cdots [\nabla(\l_{n+1})c(\T)\E_{32},\mathcal{A}(\l_{i_j};\T)] \cdots \mathcal{A}(\l_{i_n};\T)}{(\l_{i_1}-\l_{i_{2}}) \cdots (\l_{i_n}-\l_{i_1})}
\end{gathered}
\end{equation}
The second sum in the above expression is zero because
\begin{equation}
\Tr\sum_{j=1}^n\mathcal{A}(\l_{i_1};\T) \cdots[\nabla(\l_{n+1})c(\T)\E_{32},\mathcal{A}(\l_{i_j};\T)] \cdots\mathcal{A}(\l_{i_n};\T)=0
\end{equation}
since this expresses the infinitesimal adjoint action with generator $\nabla(\l_{n+1})c(\T)\E_{32}$ on $\Tr\prod_{j=1}^n\mathcal{A}(\l_{i_j};\T)$, and this latter  is invariant under adjoint actions. Finally we rearrange the sum (omitting the dependence on $\T$):
\begin{equation}
\nonumber
\begin{gathered}
S_{n+1}(\l_1,...,\l_{n+1};\T)=-\frac{1}{n}\,\sum_{i\in S_n}\sum_{j=1}^n\Tr\frac{\mathcal{A}(\l_{i_1}) \cdots(\mathcal{A}(\l_{n+1})\mathcal{A}(\l_{i_j})-\mathcal{A}(\l_{i_j})\mathcal{A}(\l_{n+1})) \cdots\mathcal{A}(\l_{i_n})}{(\l_{i_1}-\l_{i_{2}}) \cdots(\l_{i_n}-\l_{i_1})(\l_{n+1}-\l_{i_j})}=
\\
=-\frac{1}{n}\,\sum_{i\in S_n}\sum_{j=1}^n\Tr\frac{\mathcal{A}(\l_{i_1}) \cdots\mathcal{A}(\l_{i_{j-1}}) \mathcal{A}(\l_{n+1}) \mathcal{A}(\l_{i_j}) \cdots\mathcal{A}(\l_{i_n})}{(\l_{i_1}-\l_{i_{2}}) \cdots(\l_{i_n}-\l_{i_1})}\le(\frac{1}{\l_{n+1}-\l_{i_j}}-\frac{1}{\l_{n+1}-\l_{i_{j-1}}}\ri)=
\end{gathered}
\end{equation}
(here $i_0=i_n$)
\begin{equation}
\begin{gathered}
=-\frac{1}{n}\,\sum_{i\in S_n}\sum_{j=1}^n\Tr\frac{\mathcal{A}(\l_{i_1}) \cdots\mathcal{A}(\l_{i_{j-1}}) \mathcal{A}(\l_{n+1}) \mathcal{A}(\l_{i_j}) \cdots\mathcal{A}(\l_{i_n})}{(\l_{i_1}-\l_{i_{2}}) \cdots(\l_{i_{j-2}}-\l_{i_{j-1}})(\l_{i_{j-1}}-\l_{n+1})(\l_{n+1}-\l_{i_j}) \cdots(\l_{i_n}-\l_{i_1})}=
\\
=-\frac{1}{n+1}\,\sum_{i\in S_{n+1}}\Tr\frac{\mathcal{A}(\l_{i_1}) \cdots\mathcal{A}(\l_{i_{n+1}})}{(\l_{i_1}-\l_{i_{2}}) \cdots(\l_{i_{n+1}}-\l_{i_1})}
\end{gathered}
\end{equation}
where we have used the cyclic property of the trace repeatedly. The proof is complete. 
\QED 
Evaluation of \eqref{npointresulttimes} at $\T=0$ yields \eqref{npointresult}. The proof of Theorem \ref{thm2} is complete.
\subsection{The String and Dilaton equations: proof of Proposition \ref{propositionstringdilaton}}
\label{proofprop18}
The proof of the String and Dilaton equations is a consequence of the  covariance properties of the solution to  Riemann--Hilbert problem \ref{rhp3times3} under translations and dilations of the $\l$-plane.

In order to derive a proof which is not formal, for the purposes of this section we fix an arbitrary positive integer $K$ and truncate the times, so that $\T=(T_1,...,T_{2K+1},0,...)$ and that the consistency condition  \eqref{consist} on $\beta_{0,\pm}$ is fulfilled, so that the Riemann--Hilbert problem \ref{rhp3times3} is well--posed.
 
Let us introduce the following action of the shifts and dilations on the times $\T$,    $\T_S(x;\T)$,  $\T_D(x;\T)$ according to
\begin{equation}
\label{shifteddilatettimes}
\Theta(\l;\T_S(x,\T))=\Theta(\l+x;\T)_+
,\;\;\;\;\;
\Theta(\l;\T_D(x,\T))=\Theta(\e^x\l;\T)_+
\end{equation}
where $+$ denotes the polar part at $\l=\infty$, i.e., we keep only {\it strictly} positive powers of $\l^{\frac{1}{2}}$ in the Puiseux expansion at infinity. At first order in $x$ we have
\begin{equation}
\label{firstorder}
\T_S(x,\T)=\T+x\mathbb{L}_{-1}\T+\mathcal{O}(x^2),\;\;\;\T_D(x,\T)=\T+x\mathbb{L}_0\T+\mathcal{O}(x^2)
\end{equation}
where the vector fields $\mathbb{L}_{-1}$ and $\mathbb{L}_0$ are
\begin{equation}
\label{virasoro}
\mathbb{L}_{-1}:=\sum_{k\geq 3}\frac{k}{2}T_{k}\frac{\pa}{\pa T_{k-2}}+\frac{\pa}{\pa T_1},\;\;\;\mathbb{L}_0:=\sum_{k\geq 1}\frac{k}{2}T_k\frac{\pa}{\pa T_k}+\frac{\pa}{\pa T_3}
\end{equation}
\begin{lemma}
The following identities hold true
\begin{equation}
\label{translationcovariance}
\Psi(\l+x;\T)=\diag(1 , \e^{\eta}, \e^{\eta}) \Psi(\l;\T_S(x;\T)),\;\;\;\eta := \sum_{k\geq 1}x^{k}T_{2k},
\ee
\be
\label{dilationcovariance}
\Psi(\e^x\l;\T)=\e^{x(S+L)}\Psi(\l;\T_D(x;\T)).
\end{equation}
\end{lemma}

\noindent {\bf Proof.}
Consider the sectionally analytic matrix $\wh \Psi(\l;\T):=\Psi(\l+x;\T)$; it has  constant jumps on the sectors  translated by $-x$. In each of these sectors, the restriction admits entire analytic continuation under the assumption that  $\T=(T_1,...,T_{2K+1},0,...)$ and the condition \eqref{consist} on $\beta_{0,\pm}$.
 We denote by the same symbol  $\wh \Psi(\l;\T)$  the piecewise analytic matrix function with the same sectors as $\Psi(\l;\T)$. 
 Now,  the jumps of $\wh\Psi(\l;\T)$ are the same as those of $\Psi(\l;\T)$. Hence the matrix $\wh \Gamma(\l;\T):= \wh \Psi(\l;\T) \e^{-\Theta(\l;\T_S )}$  (with $\T_S=\T_S(x,\T)$ for brevity) necessarily solves a Riemann--Hilbert problem with jumps equal to those of $\Gamma(\l;\T_S)$ but with a different normalization at $\l=\infty$;
\bea
\wh \Gamma(\l;\T) \sim(\l+x)^{S}  \, G \, Y(\l+x;\T) \, (\l+x)^L \, \e^{\Theta(\l+x;\T)-\Theta(\l;\T_S)}.
\eea
The trailing factor has the form: 
\be
 \exp (\Theta(\l+x;\T)-\Theta(\l;\T_S))= \diag(1,\e^\eta,\e^\eta) (\1 + \mathcal O(\l^{-1})),
 \ \ \
 \eta=\sum_{k\geq 1}x^{k}T_{2k}.
 \label{352}
\ee
The prefactor $\diag (1,\e^\eta, \e^\eta)$ in the right side of \eqref{352} commutes with $G$, hence it follows from the uniqueness of the solution to  the Riemann--Hilbert problem \ref{rhp3times3} that $\wh \Gamma(\l,\T) =\diag (1,\e^\eta, \e^\eta) \Gamma(\l, \T_S)$ and \eqref{translationcovariance} is proved. The proof for the dilations follows along the same lines; the sectionally analytic matrix $\wt \G(\l;\T):=\Psi(\e^x\l;\T)\e^{-\Theta(\l;\T_D)}$ (with $\T_D=\T_D(x;\T)$) solves a Riemann--Hilbert problem with jumps equal to those of $\G(\l;\T_D)$ but with a different normalization at $\l=\infty$;
\bea
\wt \Gamma(\l;\T) \sim\e^{xS}\,\l^S  \, G \, Y(\e^x\l;\T) \, \e^{xL} \,\l^L \, \e^{\Theta(\e^x\l;\T)-\Theta(\l;\T_D)}
\eea
and taking $\e^{xL}$ on the left (it commutes with $G$) one obtains $\wt\G(\l;\T)=\e^{x(S+L)}\G(\l;\T_D)$.
\QED 
Now we are in position to derive \eqref{stringeq} and \eqref{dilatoneq}. 
 For the String equation we apply \eqref{translationcovariance} of the Lemma, writing $\T_S=\T_S(x,\T)$ for short,
\begin{equation}
\begin{gathered}
-\frac{\pa}{\pa T_j}\log\tau(\T;N)
=
\res{\l=\infty}\Tr\le(\l^{\frac{j}{2}}\G^{-1}(\l;\T)\G'(\l;\T)\theta_j\ri)
=
\res{\l=\infty}\Tr\le(\l^{\frac{j}{2}}\Psi^{-1}(\l;\T)\Psi'(\l;\T)\theta_j\ri)
=
\\
=
\res{\l=\infty}\Tr\le(\l^{\frac{j}{2}}\Psi^{-1}(\l-x;\T_S)\Psi'(\l-x;\T_S)\theta_j\ri)
=
\res{\l=\infty}\Tr\le((\l+x)^{\frac{j}{2}}\Psi^{-1}(\l;\T_S)\Psi'(\l;\T_S)\theta_j\ri)
\end{gathered}
\end{equation}
The last expression does not depend on $x$ by construction, so its first variation in $x$ vanishes:
\begin{equation}
\res{\l=\infty}\Tr\le(\frac{j}{2}\l^{\frac{j}{2}-1}\Psi^{-1}(\l;\T)\Psi'(\l;\T)\theta_j\ri)
+\mathbb{L}_{-1}\res{\l=\infty}\Tr\le(\l^{\frac{j}{2}}\Psi^{-1}(\l;\T)\Psi'(\l;\T)\theta_j\ri)=0
\end{equation}
In terms of the tau function
\begin{equation}
\frac{j}{2}\frac{\pa}{\pa T_{j-2}}\log\tau(\T;N)+\frac{1}{2}\delta_{j,1}T_1+N\delta_{j,2}+\mathbb{L}_{-1}\frac{\pa}{\pa T_j}\log\tau(\T;N)=0
\end{equation}
which gives
\begin{equation}
\frac{\pa}{\pa T_j}\le(\mathbb{L}_{-1}\log\tau(\T;N)+\frac{T_1^2}{4}+NT_2\ri)=0
\end{equation}
for all $j=1,2,...$. Therefore we conclude that  $\mathbb{L}_{-1}\log\tau(\T;N)+\frac{T_1^2}{4}+NT_2=const$ and the integration constant is easily seen to be $0$ by evaluation at $\T=0$ (we use the identity $\langle\tau_0\rangle=0$ which implies $\le.\frac{\pa}{\pa T_1}\log\tau(\T;N)\ri|_{\T=0}=0$). The String equation \eqref{stringeq} is established.

The Dilaton equation follows by very similar computations. Write $\T_D=\T_D(x;\T)$ and use \eqref{dilationcovariance}:
\begin{equation}
-\frac{\pa}{\pa T_j}\log\tau(\T;N)=\res{\l=\infty}\Tr\le(\e^{\frac{j}{2}x}\l^{\frac{j}{2}}\Psi^{-1}(\l;\T_D)\Psi'(\l;\T_D)\ri)
\end{equation}
The first variation in $x$ of the above vanishes:
\begin{equation}
\res{\l=\infty}\Tr\le(\frac{j}{2}\l^{\frac{j}{2}}\Psi^{-1}(\l;\T)\Psi'(\l;\T)\theta_j\ri)
+\mathbb{L}_0\res{\l=\infty}\Tr\le(\l^{\frac{j}{2}}\Psi^{-1}(\l;\T)\Psi'(\l;\T)\theta_j\ri)=0.
\end{equation}
In terms of the tau function:
\begin{equation}
\le(\frac{j}{2}\frac{\pa}{\pa T_j}+\mathbb{L}_0\frac{\pa}{\pa T_j}\ri)\log\tau(\T;N)=\frac{\pa}{\pa T_j}\mathbb{L}_0\log\tau(\T;N)=0
\end{equation}
Therefore $\mathbb{L}_0\log\tau(\T;N)=const$, and the constant is easily evaluated at $\T=0$ as 
\begin{equation}
\le.\mathbb{L}_0\log\tau(\T;N)\ri|_{\T=0}=\le.\frac{\pa}{\pa T_3}\log\tau(\T;N)\ri|_{\T=0}=-\frac{3}{2}\langle\tau_1\rangle=-\frac{1+12N^2}{16}
\end{equation}
and the Dilaton equation \eqref{dilatoneq} is established as well.

\appendix
\renewcommand{\theequation}{\Alph{section}.\arabic{equation}}
\section{Asymptotics}\label{proofA}
\paragraph{Asymptotics for $f$.}
Let us first consider $\l\in\mathbb{R}_+$, $\l\to+\infty$. According to {\it Laplace's method} the main contributions to $f(\l;N)$ for large $\l$ come from the saddles of the exponent $\i \frac{x^3}{3}+\i x\l$, provided that the contour of integration can be deformed into the curve of steepest descent through some of the saddles in a neighbourhood of the saddle points. In the present case there are two saddles, $\pm\i \sqrt{\l}$. The identity
\begin{equation}
\i \frac{x^3}{3}+\i x\l=\mp\frac{2}{3}\l^{\frac{3}{2}}\mp\sqrt{\l}\le(x\mp\i \sqrt{\l}\ri)^2+\frac{\i }{3}\le(x\mp\i \sqrt{\l}\ri)^3
\end{equation}
shows that the direction of steepest descent at $\i \sqrt{\l}$ is horizontal, while at $-\i \sqrt{\l}$ is vertical. Hence we deform the contour of integration passing through $\i \sqrt{\l}$ along the direction of steepest descent. Using the expansion
\be
x^{-N}\,\exp\left(\frac{\i }{3}(x-\i \sqrt{\l})^3\right)=\sum_{a,b\geq 0}{-N\choose a}\frac{\i ^{b-a-N}}{b!3^b}\l^{-\frac{a+N}{2}}(x-\i \sqrt{\l})^{a+3b}
\ee
we find
\be \label{expansionarg0}
\begin{gathered}
f(\l;N)=\frac{\i ^N}{\sqrt{2\pi}}\exp\le(-\frac{2}{3}\l^{\frac{3}{2}}\ri)\int_{\mathbb{R}+\i \epsilon}x^{-N}\exp\le(-\sqrt{\l}\le(x-\i \sqrt{\l}\ri)^2+\frac{\i }{3}\le(x-\i \sqrt{\l}\ri)^3\ri)\mathrm{d}x\sim  \\
\sim\frac{\exp\le(-\frac{2}{3}\l^{\frac{3}{2}}\ri)}{\sqrt{2\pi}}\sum_{a,b\geq 0}{-N\choose a}\frac{\i ^{b-a}}{b!3^b}\l^{-\frac{a+N}{2}}\int_{-\infty}^{+\infty}\exp\le(-\sqrt{\l}\xi^2\ri)\xi^{a+3b}\mathrm{d}\xi= \\
=\frac{\exp\le(-\frac{2}{3}\l^{\frac{3}{2}}\ri)}{\sqrt{2\pi}\,\l^{\frac{N}{2}+\frac{1}{4}}}\sum_{a,b\ge 0,\;a+b\text{ even}}{-N\choose a}\frac{\i ^{b-a}}{b!3^b}\Gamma\le(\frac{1+a+3b}{2}\ri)\l^{-\frac{3}{4}(a+b)}= \\
=\frac{\exp\le(-\frac{2}{3}\l^{\frac{3}{2}}\ri)}{\sqrt{2}\,\l^{\frac{N}{2}+\frac{1}{4}}}\sum_{j\geq 0}(-1)^jC_j(N)\l^{-\frac{3}{2}j}
\end{gathered}
\ee
where,  in the second line,  $\xi=x-\i \sqrt{\l}$, in the last step $a+b=2j$ and $C_j(N)$ is as in \eqref{defCjN}.

The asymptotic expansion holds in the whole sector $|\arg\l|<\pi$ by standard arguments that are completely parallel to the well--known case of the Airy functions, see e.g. \cite{Wa2002}.
\paragraph{Asymptotics for $g$.}
Use the Cauchy theorem to rotate the contour
\be
g(\l;N)=\frac{(-\i)^{N}}{(N-1)!}\,\int_0^{+\infty}\,x^{N-1}\,\e^{\frac{\i x^3}{3}}\,\e^{\i \l x}\,\mathrm{d}x
\ee
(now the integral is only conditionally convergent). The series expansion $x^{N-1}\exp\frac{\i x^3}{3}=\sum\limits_{a\geq 0}\frac{\i ^a}{3^aa!}x^{3a+N-1}$ together with Watson's lemma (see e.g. \cite{Ol1997}) gives
\be
g(\l;N)\sim\frac{(-\i)^{N}}{(N-1)!}\sum_{a\geq 0}\frac{i^a}{3^aa!}\Gamma(3a+N)(-\i \l)^{-3a-N},\;\;\;-\frac{\pi}{2}<\arg(-\i \l)<\frac{\pi}{2}
\ee
Rotating the contour of integration within the sector $0<\arg x<\frac{\pi}{3}$ we infer that the above asymptotic expansion holds in the bigger sector $-\frac{\pi}{3}<\arg\l<\pi$. This in particular proves Proposition \ref{propode}.
\section{Proof of Proposition \ref{propositionb}}

\label{appendixpropb}
Denote 
\begin{equation}
\label{notationzetak}
\z_k:=\begin{cases}\l_k&1\leq k\leq n_1 \\
\mu_{k-n_1}&n_1+1\leq k\leq n \end{cases}\;\;\;\;\;
A_k:=\begin{cases}\mathbf{e}_{3}^\top\G^{-1}(\z_k)&1\leq k\leq n_1 \\
\mathbf{e}_{2}^\top\G^{-1}(\z_k)&n_1+1\leq k\leq n \end{cases}
\end{equation}
so that we rewrite the characteristic matrix \eqref{characteristicmatrix} as
\begin{equation}
\label{det}
\mathbf{G}_{k,\ell}=\res{\l=\infty}\frac{\l^{\left\lfloor\frac{\ell-1}{2}\right\rfloor}}{\l-\zeta_k}A_kG_\infty(\l)\mathbf{e}_{2+(\ell\!\!\!\!\mod 2)}\;\;\;\;(k,\ell=1,...,n).
\end{equation}

First we compute $A_k$. Consider the pair of mutually adjoint (in the classical sense) differential operators $L, \ \wh L$ given by
\be L:=\pa^3_\l-\l\pa_\l-N,\;\;\;\;\wh{L}:=-\pa^3_\l+\l\pa_\l-N+1.
\ee
According to the general theory (see e.g. \cite{In1956}) there exists a non-degenerate bilinear pairing between the kernels of $L, \wh L$ that uses the {\it bilinear concomitant identity}; to express such identity we introduce the matrix bilinear concomitant
\begin{equation}
\label{bilinearconcomitant}
\mathcal{B}(\l):=\begin{bmatrix} -\l &0&1 \\ 0&-1&0 \\ 1&0&0 \end{bmatrix}
\end{equation}
Given any solution $u$ of $L\,u=0$ and any solution $\wh{u}$ of $\wh{L}\,\wh{u}=0$ we define their {\it bilinear concomitant} as the bilinear expression 
\be
\label{bilco}
\mathscr B[u,\wh u] := \begin{bmatrix}
\wh{u}&{\wh{u}}'&{\wh{u}}''
\end{bmatrix}
\mathcal{B}(\l)
\begin{bmatrix}
u \\ u' \\ u''
\end{bmatrix}=\wh{u}u''+\wh{u}''u-\wh{u}'u'-\l\wh{u}u
\ee
The above expression is, in fact, independent of $\lambda$ and we have:
\begin{proposition}[Lagrange identity]
\label{proplagrange}
The bilinear concomitant \eqref{bilco} is independent of $\lambda$ and gives a non-degenerate pairing between the solution spaces of the operators $L, \wh L$.
\end{proposition}
\noindent {\bf Proof.}  
The independence of $\l$ follows from the identity
\begin{equation}
0=\widehat{u}Lu-u\widehat{L}\widehat{u}=\widehat{u}u'''+\wh{u}''' u-\l\le(\wh{u}u'+\wh{u}'u\ri)-\wh{u}u=(\wh{u}u''+\wh{u}''u-\wh{u}'u'-\l\wh{u}u)'=\le(\mathscr{B}[u,\wh{u}]\ri)'.
\end{equation}
The nondegeneracy of the pairing follows from $\det \mathcal B=1$.
\QED 

\begin{proposition}
\label{propositionbilinear}
Denote
\begin{equation}
\label{phik}
\begin{gathered}
\phi_k:=\begin{cases}
\omega^{N+\frac{1}{2}}f(\omega^{-1}\l_k;N)&1\leq k\leq n_1,\;\l_k\in I \\
f(\l_k;N)&1\leq k\leq n_1,\;\l_k\in II\cup III \\
\omega^{-N-\frac{1}{2}}f(\omega\l_k;N)&1\leq k\leq n_1,\;\l_k\in IV \\
\omega^{\frac{N}{2}+\frac{1}{4}}f(\omega\mu_k;N)&n_1+1\leq k\leq n_1,\;\mu_k\in I\cup II \\
\omega^{-\frac{N}{2}-\frac{1}{4}}f(\omega^{-1}\mu_k;N)&n_1+1\leq k\leq n_1,\;\mu_k\in III\cup IV \\
\end{cases}\;\;\;\;\;\;\;
Q_k:=\begin{cases}\frac{2}{3}\l_k^{\frac{3}{2}}&1\leq k\leq n_1 \\
-\frac{2}{3}\mu_k^{\frac{3}{2}}&n_1+1\leq k\leq n\end{cases}
\end{gathered}
\end{equation}
Then the  row-vectors $A_k$ defined in \eqref{notationzetak} can be expressed as follows;
\begin{equation}
A_k=\e^{Q_k}[\phi_k,\phi_k',\phi_k''] \mathcal{B}(\zeta_k)
\end{equation}
\end{proposition}
\noindent {\bf Proof.}
Let us consider the case $k=1,...,n_1$ with $\l_k\in II\cup III$, the other cases are completely analogous. The Proposition follows from the following identity in which we set $\l=\l_k$:
\begin{equation}
\label{boooh}
[f(\l;N),-f(\l;N-1),f(\l;N-2)] \mathcal{B}(\l) \Psi(\l)=\mathbf{e}_{3}^\top.
\end{equation}
The equation
\eqref{boooh} follows from the fact that the left-hand side is a constant row vector, because of Prop. \ref{proplagrange}, which is asymptotic to $\mathbf{e}^\top_{3}$ when $\l\to+\infty$.
\QED

Therefore we can use  the expansion \eqref{Ginfty}  and  write the characteristic matrix \eqref{characteristicmatrix} as, 
\begin{equation}
\mathbf{G}_{k,\ell}=\res{\l=\infty}\frac{\l^{\left\lfloor\frac{\ell-1}{2}\right\rfloor}}{\l-\zeta_k}\e^{Q_k}[\phi_k,\phi_k',\phi_k'']\,\mathcal{B}(\z_k)\,\l^S\,G\,Y_0\,\wt{D}\,G^{-1}\,\l^{-S}\mathbf{e}_{2+(\ell\!\!\!\!\mod2)}\ \ \ (k,\ell=1,...,n).
\end{equation}

From now on we denote $F^\pm_{r}:=\frac{1}{\sqrt{2}}F_\pm(\l;r)$ for short.

\begin{lemma}
\label{lemmaB}
Let $\phi:=\phi_k$ as in \eqref{phik} and $\z:=\z_k$ as in \eqref{notationzetak}. For any integer $J\geq 0$ the following identities of formal expansions hold true:
\begin{equation}
\label{identity-}
\begin{gathered}
\frac{[\phi,\phi',\phi'']\mathcal{B}(\l)\l^SGY_0(\l)}{\l-\z}\mathbf{e}_2=
-\sum_{r=1}^{J}\l^{-1-\frac{r}{2}}\phi^{(r)}F_{-N+r+1}^-+
\\
-\sum_{m\geq 0}\frac{\z^m}{\l^m}\le(\l^{-\frac{J+3}{2}}\phi^{(J+1)}F^-_{-N+J}+\l^{-\frac{J+4}{2}}\phi^{(J+2)}F^-_{-N+J+1}-\l^{-\frac{J+5}{2}}(N-J-1)\phi^{(J)}F^-_{-N+J+2}\ri)
\end{gathered}
\end{equation}
\begin{equation}
\begin{gathered}
\label{identity+}
\frac{[\phi,\phi',\phi'']\mathcal{B}(\l)\l^SGY_0(\l)}{\l-\z}\mathbf{e}_3=\sum_{r=1}^{J}(-1)^r\l^{-1-\frac{r}{2}}\phi^{(r)}F_{-N+r+1}^++
\\
+\sum_{m\geq 0}\frac{\z^m}{\l^m}\le((-1)^{J+1}\l^{-\frac{J+3}{2}}\phi^{(J+1)}F^+_{-N+J}+(-1)^J\l^{-\frac{J+4}{2}}\phi^{(J+2)}F^+_{-N+J+1}+\ri. 
\\
\le.(-1)^J\l^{-\frac{J+5}{2}}(N-J-1)\Phi^{(J)}F^+_{-N+J+2}\ri).
\end{gathered}
\end{equation}
\end{lemma}
{\bf Proof.}
The proof is inductive with respect to $J$. 
First compute (we are only interested in the second and third columns)
\begin{equation}
\l^SGY_0(\l)\sim\begin{bmatrix}
\star & -\l^{-1}F^-_{-N+1}& \l^{-1}F^+_{-N+1} \\
\star & \l^{-\frac{1}{2}}F^-_{-N} & \l^{-\frac{1}{2}}F^+_{-N}\\
\star & -F^-_{-N-1} & F^+_{-N-1}
\end{bmatrix}
\end{equation}
where $F^\pm_r:=F_\pm(\l;r)=\sum_{j\geq 0}(\pm 1)^jC_j(N)\l^{-\frac{3j}{2}}$ and $C_n(N)$ are  introduced in Prop. \ref{propode0} in formulas \eqref{Fpiu}, \eqref{defCjN}. Now we use the recursions \eqref{basicidentity2} to write
\begin{equation}
\mathcal{B}(\l)\l^SGY_0(\l)\sim\begin{bmatrix}
\star & \l^{-\frac{3}{2}}(N-1)F^-_{-N+2} & \l^{-\frac{3}{2}}(N-1)F^+_{-N+2}\\
\star & -\l^{-\frac{1}{2}}F^-_{-N} & -\l^{-\frac{1}{2}}F^+_{-N}\\
\star & -\l^{-1}F^-_{-N+1} & \l^{-1}F^+_{-N+1}
\end{bmatrix}
\end{equation}
Inserting the last expression into $\frac{[\phi,\phi',\phi'']}{\l-\z}\mathcal{B}(\l)\l^SGY_0(\l)\mathbf{e}_{2,3}$ and expanding $\frac{1}{\l-\z}=\sum\limits_{m\geq 0}\frac{\z^m}{\l^{m+1}}$ gives \eqref{identity-} and \eqref{identity+} with $J=0$.

We now proceed with the inductive step:  we verify \eqref{identity-} only, \eqref{identity+} being completely analogous. Assume that \eqref{identity-} holds true for an integer $J\geq 0$ and substitute
\begin{equation}
(N-J-1)\phi^{(J)}=\z\phi^{(J+1)}-\phi^{(J+3)}
\end{equation}
(obtained by taking $J$ derivatives of $\phi'''-\z\phi'+(N-1)\phi=0$) into \eqref{identity-} to get:
\be
\begin{gathered}
-\sum_{r=1}^{J}\l^{-1-\frac{r}{2}}\phi^{(r)}F_{-N+r+1}^-+
\\
-\sum_{m\geq 0}\frac{\z^m}{\l^m}\le(\l^{-\frac{J+3}{2}}\phi^{(J+1)}F^-_{-N+J}+\l^{-\frac{J+4}{2}}\phi^{(J+2)}F^-_{-N+J+1}+\l^{-\frac{J+5}{2}}(\phi^{(J+3)}-\z\phi^{(J+1)})F^-_{-N+J+2}\ri).
\end{gathered}
\ee
We now re--organize the second summation
\be
\begin{gathered}
-\le(\sum_{r=1}^{J}\l^{-1-\frac{r}{2}}\phi^{(r)}F_{-N+r+1}^-\ri)-\l^{-\frac{J+3}{2}}\phi^{(J+1)}F^-_{-N+J+2}+
\\
-\sum_{m\geq 0}\frac{\z^m}{\l^m}\le(\l^{-\frac{J+3}{2}}\phi^{(J+1)}(F^-_{-N+J}-F^-_{-N+J+2})+\l^{-\frac{J+4}{2}}\phi^{(J+2)}F^-_{-N+J+1}+\l^{-\frac{J+5}{2}}\phi^{(J+3)}F^-_{-N+J+2}\ri).
\end{gathered}
\ee
Finally  we substitute the identity $F^-_{-N+J}-F^-_{-N+J+2}=-\l^{-\frac{3}{2}}(N-J-2)F^-_{-N+J+3}$ obtained from \eqref{basicidentity2} with the replacement $N\to - N+J+2 $. This yields
\be
\begin{gathered}
-\sum_{r=1}^{J+1}\l^{-1-\frac{r}{2}}\phi^{(r)}F_{-N+r+1}^-+
\\
-\sum_{m\geq 0}\frac{\z^m}{\l^m}\le(-\l^{-\frac{J+6}{2}}\phi^{(J+1)}(N-J-2)F^-_{-N+J+3}+\l^{-\frac{J+4}{2}}\phi^{(J+2)}F^-_{-N+J+1}+\l^{-\frac{J+5}{2}}\phi^{(J+3)}F^-_{-N+J+2}\ri).
\end{gathered}
\ee
This is the identity  \eqref{identity-} under the substitution $J\mapsto J+1$. The proof is complete.
\QED

In particular we shall use the following corollary of Lemma \ref{lemmaB}: for any $J\geq 0$ we have
\be
\label{corollary}
\begin{gathered}
\frac{[\phi_k,\phi_k',\phi_k'']\mathcal{B}(\l)\l^SGY_0(\l)}{\l-\z_k}\mathbf{e}_2=
-\sum_{r=1}^{J}\l^{-1-\frac{r}{2}}\phi_k^{(r)}F_{-N+r+1}^-+\mathcal{O}\le(\l^{-\frac{J+3}{2}}\ri)
\\
\frac{[\phi_k,\phi_k',\phi_k'']\mathcal{B}(\l)\l^SGY_0(\l)}{\l-\z_k}\mathbf{e}_3=\sum_{r=1}^{J}(-1)^r\l^{-1-\frac{r}{2}}\phi_k^{(r)}F_{-N+r+1}^++\mathcal{O}\le(\l^{-\frac{J+3}{2}}\ri)
\end{gathered}
\ee

By construction, the columns of the characteristic matrix are obtained as follows: the $(2K-1)$--th and $2K$--th columns of $\mathbf{G}$ correspond to, respectively, the second and first entries of the coefficient in front of $\l^{-K}$ in the 2-dimensional row-vector power-series (at $\l=\infty$) below ($k$ is the row index of $\mathbf{G}$)
\begin{equation}
\label{powerseries}
\frac{1}{\l-\zeta_k}\e^{Q_k}[\phi_k,\phi_k',\phi_k'']\,\mathcal{B}(\z_k)\,\l^S\,G\,Y_0\,\wt{D}\,G^{-1}\,\l^{-S}
\begin{bmatrix}
0&0 \\
1&0 \\
0&1
\end{bmatrix}
\end{equation}
Let us simplify the last expression: first compute
\begin{equation}
\widetilde{D}G^{-1}\l^{-S}\begin{bmatrix}
0&0 \\
1&0 \\ 
0&1 \end{bmatrix}
=\frac{1}{\sqrt{2}}\begin{bmatrix}
0&0 \\
\beta_+\l^{\frac{1}{2}} & \beta_+ \\
-\beta_-\l^{\frac{1}{2}} & \beta_-
\end{bmatrix}
\end{equation}
where $\beta_\pm=\l^{-\frac{n}{2}}\pi_{\pm}$. The power series \eqref{powerseries} can be rewritten using the identity
\be
\frac{\mathcal{B}(\z_k)}{\l-\z_k}=\E_{11}+\frac{\mathcal{B}(\l)}{\l-\z_k}
\ee
(where $\E_{11}$ is the elementary unit matrix).
This gives the equation
\be
\begin{gathered}
\e^{Q_k}[\phi_k,\phi_k',\phi_k'']\,\le(\E_{11}+\frac{\mathcal{B}(\l)}{\l-\zeta_k}\ri)\,\l^S\,G\,Y_0
\frac{1}{\sqrt{2}}\begin{bmatrix}
0&0\\
\beta_+\l^{\frac{1}{2}} & \beta_+ \\
-\beta_-\l^{\frac{1}{2}} & \beta_-
\end{bmatrix}
=
\\
=\frac{\e^{Q_k}}{\sqrt{2}\l}\begin{bmatrix}
-\sum\limits_{r=0}^{n-1}\l^{-\frac{r}{2}}\phi_k^{(r)}F_{-N+r+1}^-+\mathcal{O}\le(\l^{-\frac{n}{2}}\ri), & \sum\limits_{r=0}^{n-1}(-1)^r\l^{-\frac{r}{2}}\phi_k^{(r)}F_{-N+r+1}^++\mathcal{O}\le(\l^{-\frac{n}{2}}\ri)
\end{bmatrix}
\begin{bmatrix}
\beta_+\l^{\frac{1}{2}} & \beta_+ \\
-\beta_-\l^{\frac{1}{2}} & \beta_-
\end{bmatrix}=
\\
=
{\e^{Q_k}}\left[
-\sum\limits_{r\text{ odd}}\l^{-\frac{r+1}{2}}\phi_k^{(r)}\le(1+\mathcal{O}\le(\l^{-1}\ri)\ri)+
\sum\limits_{r\text{ even}}\star\l^{-\frac{r+2}{2}}\phi_k^{(r)}\le(1+\mathcal{O}\le(\l^{-1}\ri)\ri)
+\mathcal{O}\le(\l^{-\frac{n+2}{2}}\ri)
,\ri.
\\ 
\le.
\sum\limits_{r\text{ odd}}\star\l^{-\frac{r+1}{2}}\phi_k^{(r)}\le(1+\mathcal{O}\le(\l^{-1}\ri)\ri)+
\sum\limits_{r\text{ even}}\l^{-\frac{r+2}{2}}\phi_k^{(r)}\le(1+\mathcal{O}\le(\l^{-1}\ri)\ri)
+\mathcal{O}\le(\l^{-\frac{n+2}{2}}\ri)
\right]
\end{gathered}
\label{B23}
\ee
where we have used \eqref{corollary} with $J=n-1$ and then the monodromy properties $\beta_\pm(\l\e^{2\pi\i})=\beta_\mp(\l)$, $F^\pm_\ell(\l\e^{2\pi\i})=F^\mp_\ell(\l)$; the expansions in the last expression contain only integer powers of $\l$. The $\star$ denotes an expression independent of $\l$ and of the index $k$ and irrelevant to the discussion. The $\mathcal O$ expressions are independent of $k$.

From the last expression we obtain that the wedge of the columns in $\mathbf{G}$ is, performing triangular transformations on $\mathbf{G}$ and up to an irrelevant sign,
\begin{equation}
\e^{\sum\limits_{\ell=1}^n Q_\ell}\,
\le[\begin{array}{c}\phi_1\\
\vdots\\
\phi_n\end{array}
\ri]
\wedge 
\le[
\begin{array}{c}\phi'_1\\
\vdots\\
\phi'_n\end{array}
\ri]
\wedge
\cdots
\wedge
\le[
\begin{array}{c}\phi^{(n-1)}_1\\
\vdots\\
\phi^{(n-1)}_n\end{array}
\ri]
\end{equation}
For example, if we look at the $2K$--th column of $G$ we need to extract the coefficient  of  $\l^{-K}$ from the first component of \eqref{B23}: the main  term comes from the term $r=2K-1$ in the first sum and then there are  other terms with $r<2K-1$ coming from both  sums. 
These additional terms correspond to a  linear combination of the previous columns of $G$ and hence do not affect the determinant.
Using $Q=\sum\limits_{\ell=1}^{n}Q_\ell$, the proof of Proposition \ref{propositionb} is complete.\QED
%%%%%%%%%%%%%%%%%%%%%%%%%%%%%%%%%
%%%%%%%%%%%%%%%%%%%%%%%%%%%%%%%%%
%%%%%%%%%%%%%%%%%%%%%%%%%%%%%%%%%
%%%%%%%%%%%%%%%%%%%%%%%%%%%%%%%%%
\newpage
\section{Table of open intersection numbers}\label{appendixtable}
{\bf Nonzero one--point correlators $\left\langle\tau_{\frac{3g-1}{2}}\right\rangle$ for $1\leq g\leq 35$}
\begin{tiny}
\begin{align*}
&\left\langle \tau _1\right\rangle =\frac{N^2}{2}+\frac{1}{24}, \qquad 
\left\langle \tau _{\frac{5}{2}}\right\rangle =\frac{N(N^2+1)}{12},\qquad 
\left\langle \tau _4\right\rangle =\frac{16 N^4+56 N^2+1}{1152},\qquad
\left\langle \tau _{\frac{11}{2}}\right\rangle =\frac{N \left(3 N^4+25 N^2+12\right)}{2880},\qquad
\left\langle \tau _7\right\rangle =\frac{192 N^6+3120 N^4+5508 N^2+25}{2073600}, \\ &
\left\langle \tau _{\frac{17}{2}}\right\rangle =\frac{N \left(3 N^6+84 N^4+357 N^2+116\right)}{725760},
\qquad
\left\langle \tau _{10}\right\rangle =\frac{2304 N^8+102144 N^6+848736 N^4+1030896 N^2+1225}{9754214400},
\\ &
\left\langle \tau _{\frac{23}{2}}\right\rangle =\frac{N \left(N^8+66 N^6+945 N^4+2764 N^2+704\right)}{139345920}, 
\qquad
\left\langle \tau _{13}\right\rangle =\frac{1024 N^{10}+96000 N^8+2174592 N^6+12408800 N^4+11798484 N^2+3675}{3511517184000}, \\ &
\left\langle \tau _{\frac{29}{2}}\right\rangle =\frac{N \left(3 N^{10}+385 N^8+12969 N^6+127215 N^4+289828 N^2+62400\right)}{459841536000}, \\ &
\left\langle \tau _{16}\right\rangle =\frac{4096 N^{12}+698368 N^{10}+33378048 N^8+516817664 N^6+2287011056 N^4+1823347368 N^2+148225}{20394891804672000},\\& 
\left\langle\tau _{\frac{35}{2}}\right\rangle =\frac{N \left(3 N^{12}+663 N^{10}+43329 N^8+995709 N^6+7546968 N^4+14318928 N^2+2720000\right)}{860823355392000}, \\ &
\left\langle \tau _{19}\right\rangle =\frac{49152 N^{14}+13791232 N^{12}+1196203008 N^{10}+38934694656 N^8+464449713728 N^6+1702700165712 N^4+1186262049012 N^2+25050025}{579051768118247424000}, \\ &
\left\langle \tau _{\frac{41}{2}}\right\rangle =\frac{N \left(3 N^{14}+1050 N^{12}+117936 N^{10}+5240950 N^8+92561469 N^6+578869200 N^4+952988592 N^2+164012800\right)}{2530820664852480000}, \\ &
\left\langle \tau _{22}\right\rangle =(196608 N^{16}+84541440 N^{14}+12045336576 N^{12}+709486469120 N^{10}+17711538983424 N^8+173768644707840 N^6+ \\ &+549856542467392 N^4+343891814589600 N^2+1878751875)\Big{/}8338345460902762905600000, \\ &
\left\langle \tau _{\frac{47}{2}}\right\rangle =\frac{N \left(9 N^{16}+4692 N^{14}+833238 N^{12}+63480924 N^{10}+2159911897 N^8+31292109576 N^6+168227776656 N^4+246969815808 N^2+39197491200\right)}{33042394600313978880000}, \\ &
\left\langle \tau _{25}\right\rangle =(786432 N^{18}+491323392 N^{16}+107118526464 N^{14}+10341480873984 N^{12}+465685699454976 N^{10}+9515782193389056 N^8+\\ & +79996540499728128 N^6+224604466722581568 N^4+128475889764619500 N^2+180986430625)\Big{/}173504292350464690539724800000, 
\\ &
\left\langle \tau _{\frac{53}{2}}\right\rangle =N (3 N^{18}+2223 N^{16}+587214 N^{14}+70595526 N^{12}+4106559015 N^{10}+114161660379 N^8+1413489083016 N^6+\\ &+6716420289072 N^4+8963785066752 N^2+1328600268800)\Big{/}67802993719844284661760000, \\ &
\left\langle \tau _{28}\right\rangle =(9437184 N^{20}+8218214400 N^{18}+2599932985344 N^{16}+383417244057600 N^{14}+28231124693950464 N^{12}+1037084389334323200 N^{10}+\\ &+18071584866415010304 N^8+133861937156213875200 N^6+340105859130592186704 N^4+\\ &+180359927192758704600 N^2+65336101455625)\Big{/}15032411889244260788361756672000000,\\ &
\left\langle \tau _{\frac{59}{2}}\right\rangle =N (3 N^{20}+3045 N^{18}+1141938 N^{16}+203887290 N^{14}+18675776343 N^{12}+885062763585 N^{10}+20943687365988 N^8+227906671223280 N^6+\\ &+976427326945728 N^4+1201616243532800 N^2+167890903040000)\Big{/}569545147246691991158784000000, \\ &
\left\langle \tau _{31}\right\rangle =(12582912 N^{22}+14775484416 N^{20}+6508661440512 N^{18}+1390836757364736 N^{16}+156138779873083392 N^{14}+9355246545848426496 N^{12}+\\ &+292079263388984338432 N^{10}+4463715935799807515136 N^8+29725206956587972412352 N^6+\\ &+69348424013248939599216 N^4+34445896534945699724900 N^2+3201468971325625)\Big{/}194459280199263757558247684308992000000, \\ &
\left\langle \tau _{\frac{65}{2}}\right\rangle =N (9 N^{22}+12144 N^{20}+6232149 N^{18}+1577715084 N^{16}+214268934099 N^{14}+15954755869584 N^{12}+641741905745919 N^{10}+13294651110734004 N^8+\\ &+129748137025654224 N^6+508732601242709184 N^4+583565266267673600 N^2+77419952332800000)\Big{/}19020529737450525736738750464000000, \\ &
\left\langle \tau _{34}\right\rangle =(150994944 N^{24}+232683208704 N^{22}+138084895162368 N^{20}+41038414284324864 N^{18}+6665769736718450688 N^{16}+607760659467887837184 N^{14}+\\ &+30870378013054268325888 N^{12}+842480952377247016525824 N^{10}+11523862433261066955187968 N^8+70040459436414792986605824 N^6+\\ &+151735363253374303757178144 N^4+71159276436758220421947600 N^2+1693577085831255625)\Big{/}29626260256918231991514151199843549184000000,\\ &
\left\langle \tau _{\frac{71}{2}}\right\rangle =N (9 N^{24}+15750 N^{22}+10736055 N^{20}+3715305000 N^{18}+714467029815 N^{16}+78742316706750 N^{14}+\\ &+4965413908715385 N^{12}+174350272423288500 N^{10}+3227263905458585520 N^8+28681192085729304000 N^6+\\ &+104107557779136393216 N^4+112264789305722880000 N^2+14223505875927040000)\Big{/}273895628219287570609038006681600000000, 
\end{align*}
\begin{align*}
&\left\langle \tau _{37}\right\rangle =(1811939328 N^{26}+3582355046400 N^{24}+2787744531087360 N^{22}+1115077072807526400 N^{20}+251590011063565025280 N^{18}+\\ &+33138449323254728294400 N^{16}+2556425163971371197726720 N^{14}+113223358101371194076774400 N^{12}+2756808892688324683107824640 N^{10}+\\ &+34272976656685888849615430400 N^8+192354889204987160338838316672 N^6+390398165386846814440372428000 N^4+\\ &+173966964419611918411839517500 N^2+1058485678644534765625)\Big{/}5777120750099055238345259483969492090880000000000, 
\\ &
\left\langle \tau _{\frac{77}{2}}\right\rangle =N (N^{26}+2223 N^{24}+1963845 N^{22}+901893135 N^{20}+236819889735 N^{18}+36906202951605 N^{16}+3438658175455215 N^{14}+\\ &+188880046083711645 N^{12}+5909445554333014980 N^{10}+99255431931246688240 N^8+812826672988328716224 N^6+2756329274408365873152 N^4+\\ &+2813279836964648960000 N^2+341983288874106880000)\Big{/}555460334028715193195129077550284800000000,
\\ &
\left\langle \tau _{40}\right\rangle =(268435456 N^{28}+668001632256 N^{26}+666484888043520 N^{24}+349289336820203520 N^{22}+105959317532820111360 N^{20}+\\ &+19362525622587933327360 N^{18}+2154495766664566978314240 N^{16}+144644590756018910238474240 N^{14}+5701749549787366356514222080 N^{12}+\\ &+125793667873035366533829294080 N^{10}+1438475401975400176548284377344 N^8+7524681881638971477232896908544 N^6+14409670664017625441442987246000 N^4+\\ &+6133179604447412937555105285000 N^2+9526371107800812890625)\Big{/}17470013148299543040756064679523744082821120000000000,\\ &
\left\langle \tau _{\frac{83}{2}}\right\rangle =N (3 N^{28}+8323 N^{26}+9334143 N^{24}+5551270335 N^{22}+1932625559865 N^{20}+410779697815305 N^{18}+54043957766495925 N^{16}+\\ &+4378648185185293845 N^{14}+213848568458638036560 N^{12}+6054930086513129833440 N^{10}+\\ &+93398815010809285798912 N^8+711479288477940634858752 N^6+2270589046750613021294592 N^4+\\ &+2205775537731744849920000 N^2+258262461203262996480000)\Big{/}37886838463430605897453364121549825638400000000, \\ &
\left\langle \tau _{43}\right\rangle=(1073741824 N^{30}+3308466995200 N^{28}+4152361563979776 N^{26}+2788103339114496000 N^{24}+1107214943018821877760 N^{22}+\\ &+271748875036881715200000 N^{20}+41897331291006657103134720 N^{18}+4050945201217520716455936000 N^{16}+241599398869348056375789404160 N^{14}+\\ &+8609344178313376924100539904000 N^{12}+174204426063863194634949737286656 N^{10}+1849889836089218022250893529132800 N^8+\\ &+9087344214453261213296493811015104 N^6+16515956825060389623153892117086000 N^4+6743423432363384359569201005497500 N^2+\\ &+2670559367220161213671875)\Big{/}176307372692638988367310204745753625283830743040000000000,
\\ &
\left\langle \tau _{\frac{89}{2}}\right\rangle =N (3 N^{30}+10230 N^{28}+14310282 N^{26}+10797034950 N^{24}+4863897529920 N^{22}+1369377408631050 N^{20}+245438133160091190 N^{18}+\\ &+28040134114127700450 N^{16}+2016553574286500994045 N^{14}+88945467688125986909400 N^{12}+2307015640350087041087232 N^{10}+\\ &+32998166006340828793019520 N^8+235623064500239301846467328 N^6+711977968054625435950694400 N^4+\\ &+661366619180251727339520000 N^2+74827468199586758656000000)\Big{/}1057042793129713904538948858991240135311360000000000,
\\ &
\left\langle \tau _{46}\right\rangle =(12884901888 N^{32}+48464410968064 N^{30}+75275548892332032 N^{28}+63540993176754978816 N^{26}+32305430292879397355520 N^{24}+\\ &+10370481568264527764520960 N^{22}+2145232738948850922647715840 N^{20}+287033536702127742019405086720 N^{18}+24616567362702008913924113694720 N^{16}+\\ &+1324765778287049017527651851304960 N^{14}+43199836353610510669596725253865472 N^{12}+809509736892681198367241725112303616 N^{10}+\\ &+8045031807244484567934874330207638528 N^8+37344181294978783570591721693139876864 N^6+64730195732729533165125625841012496000 N^4+\\ &+25445077458388890315979323372645960000 N^2+2566407551898574926338671875)\Big{/}650616519005284100432582809960969858167403561195929600000000000,
\\ &
\left\langle \tau _{\frac{95}{2}}\right\rangle =N (3 N^{32}+12408 N^{30}+21323412 N^{28}+20055762232 N^{26}+11454070025970 N^{24}+4169373007077720 N^{22}+988892153638810740 N^{20}+\\ &+153734020659819173640 N^{18}+15568629025743559408995 N^{16}+1009519377988605493054320 N^{14}+40709157127073869063417632 N^{12}+\\ &+976740709911587217252893952 N^{10}+13056845467163879175082824448 N^8+87948086058419995309074505728 N^6+252897236774376421971242188800 N^4+\\ &+225511373602220039295467520000 N^2+24723317733187169288192000000)\Big{/}35719590065439292262180159843031986652441477120000000000,
\\ &
\left\langle \tau _{49}\right\rangle =(51539607552 N^{34}+233719235346432 N^{32}+442934093131284480 N^{30}+462450991762786746368 N^{28}+295390315160740275683328 N^{26}+\\ &+121308641907934024622407680 N^{24}+32792156455715070124322979840 N^{22}+5880715146514830637586668584960 N^{20}+697065636065712916547896085053440 N^{18}+\\ &+53863958762978282913637251929210880 N^{16}+2647951245938155330851503196191981568 N^{14}+79795408414779173454856101590903390208 N^{12}+\\ &+1395760693072968025784367191095072460800 N^{10}+13065678711042577835847524520160601705472 N^8+\\ &+57609003290994717825863703762658802068992 N^6+95626633867412782236501598237356262608000 N^4+36302569641668944177809901723376068567500 N^2+\\ &+931605941339182698260937890625)\Big{/}96358908930758596410467244486459479874025137027361957478400000000000,
\\ &
\left\langle \tau _{\frac{101}{2}}\right\rangle =N (9 N^{34}+44625 N^{32}+92975652 N^{30}+107377251300 N^{28}+76401171605374 N^{26}+35232914199851550 N^{24}+10795521043518018660 N^{22}+\\ &+2218740460554639136500 N^{20}+305406324629174848883265 N^{18}+27849118985207501173025625 N^{16}+1648346494085042490149359536 N^{14}+\\ &+61370767120020648407067132000 N^{12}+1373015572590141786606818715648 N^{10}+17265746172191035097369934035200 N^8+\\ &+110291644376328313331655537401856 N^6++303101327642788954161961161523200 N^4+260319747106223633779631063040000 N^2+\\ &+27720320206225861244878848000000)\Big{/}4335643842143021294783427801747222539873346492825600000000000.
\end{align*}
\end{tiny}
{\bf Nonzero two--point correlators $\le\langle\tau_{\frac{d_1}{2}}\tau_{\frac{d_2}{2}}\ri\rangle$, for $0\leq d_1\leq d_2\leq 30$}
\begin{tiny}
\begin{align*}
&\left\langle \tau _0 \tau _{\frac{1}{2}}\right\rangle =N,\qquad
\left\langle \tau _1^2\right\rangle =\frac{N^2}{2}+\frac{1}{24},\qquad 
\left\langle \tau _{\frac{1}{2}} \tau _{\frac{3}{2}}\right\rangle =\frac{N^2}{2},\qquad
\left\langle \tau _0 \tau _2\right\rangle =\frac{N^2}{2}+\frac{1}{24},\qquad
\left\langle \tau _{\frac{3}{2}} \tau _2\right\rangle =\frac{2 N^3+N}{6},\qquad
\left\langle \tau _1 \tau _{\frac{5}{2}}\right\rangle =\frac{N^3+N}{6} ,\qquad
\left\langle \tau _{\frac{5}{2}}^2\right\rangle =\frac{N^2 \left(N^2+3\right)}{16},
\\ &
\left\langle \tau _{\frac{1}{2}} \tau _3\right\rangle =\frac{N \left(4 N^2+3\right)}{24} ,\qquad 
\left\langle \tau _2 \tau _3\right\rangle =\frac{N^4}{8}+\frac{13 N^2}{48}+\frac{29}{5760} ,\qquad
\left\langle \tau _0 \tau _{\frac{7}{2}}\right\rangle =\frac{N^3+N}{12} ,\qquad
\left\langle \tau _{\frac{3}{2}} \tau _{\frac{7}{2}}\right\rangle =\frac{N^2 \left(3 N^2+7\right)}{48} ,\qquad
\left\langle \tau _3 \tau _{\frac{7}{2}}\right\rangle =\frac{N \left(36 N^4+195 N^2+79\right)}{1440}, \\ &
\left\langle \tau _1 \tau _4\right\rangle =\frac{16 N^4+56 N^2+1}{384} ,\qquad
\left\langle \tau _{\frac{5}{2}} \tau _4\right\rangle =\frac{ N \left(12 N^4+85 N^2+33\right)}{720},\qquad
\left\langle \tau _4^2\right\rangle =\frac{6720 N^6+89040 N^4+130284 N^2+607}{1451520} ,\qquad
\left\langle \tau _{\frac{1}{2}} \tau _{\frac{9}{2}}\right\rangle =\frac{N^2 \left(N^2+3\right)}{48},\\ &
\left\langle \tau _2 \tau _{\frac{9}{2}}\right\rangle =\frac{N \left(6 N^4+35 N^2+13\right)}{360},\qquad 
\left\langle \tau _{\frac{7}{2}} \tau _{\frac{9}{2}}\right\rangle =\frac{N^2 \left(N^4+11 N^2+16\right)}{288} ,\qquad
\left\langle \tau _0 \tau _5\right\rangle =\frac{16 N^4+56 N^2+1}{1152},\qquad
\left\langle \tau _{\frac{3}{2}} \tau _5\right\rangle =\frac{N \left(8 N^4+50 N^2+17\right)}{720},\\ &
\left\langle \tau _3 \tau _5\right\rangle =\frac{6720 N^6+75600 N^4+107436 N^2+503}{1451520} ,\qquad
\left\langle \tau _{\frac{9}{2}} \tau _5\right\rangle =\frac{N \left(40 N^6+770 N^4+2681 N^2+751\right)}{60480} ,\qquad
\left\langle \tau _1 \tau _{\frac{11}{2}}\right\rangle =\frac{N \left(3 N^4+25 N^2+12\right)}{720}, \\ &
\left\langle \tau _{\frac{5}{2}} \tau _{\frac{11}{2}}\right\rangle =\frac{N^2 \left(N^4+14 N^2+21\right)}{576} ,\qquad
\left\langle \tau _4 \tau _{\frac{11}{2}}\right\rangle =\frac{N \left(60 N^6+1365 N^4+4851 N^2+1396\right)}{120960} ,\qquad 
\left\langle \tau _{\frac{11}{2}}^2\right\rangle =\frac{N^2 \left(3 N^6+106 N^4+735 N^2+788\right)}{55296},\\ &
\left\langle \tau _{\frac{1}{2}} \tau _6\right\rangle =\frac{N \left(16 N^4+120 N^2+49\right)}{5760} ,\qquad
\left\langle \tau _2 \tau _6\right\rangle =\frac{960 N^6+11760 N^4+16404 N^2+77}{414720} ,\qquad
\left\langle \tau _{\frac{7}{2}} \tau _6\right\rangle =\frac{N \left(48 N^6+952 N^4+3283 N^2+939\right)}{96768} ,\\ &
\left\langle \tau _5 \tau _6\right\rangle =\frac{1792 N^8+55552 N^6+376992 N^4+398992 N^2+487}{18579456} ,\qquad
\left\langle \tau _0 \tau _{\frac{13}{2}}\right\rangle =\frac{N \left(3 N^4+25 N^2+12\right)}{2880} ,\qquad
\left\langle \tau _{\frac{3}{2}} \tau _{\frac{13}{2}}\right\rangle =\frac{N^2 \left(N^4+13 N^2+18\right)}{1152},\\ &
\left\langle \tau _3 \tau _{\frac{13}{2}}\right\rangle =\frac{N \left(180 N^6+3675 N^4+12509 N^2+3644\right)}{483840} ,\qquad 
\left\langle \tau _{\frac{9}{2}} \tau _{\frac{13}{2}}\right\rangle =\frac{N^2 \left(3 N^6+94 N^4+635 N^2+676\right)}{55296}, \\ &
\left\langle \tau _6 \tau _{\frac{13}{2}}\right\rangle =\frac{N \left(560 N^8+26040 N^6+304059 N^4+775225 N^2+180876\right)}{69672960} ,\qquad
\left\langle \tau _1 \tau _7\right\rangle =\frac{192 N^6+3120 N^4+5508 N^2+25}{414720}, \\ & 
\left\langle \tau _{\frac{5}{2}} \tau _7\right\rangle =\frac{N \left(48 N^6+1176 N^4+4347 N^2+1219\right)}{241920} ,\qquad
\left\langle \tau _4 \tau _7\right\rangle =\frac{5376 N^8+195328 N^6+1367520 N^4+1462448 N^2+1781}{92897280}, \\ & 
\left\langle \tau _{\frac{11}{2}} \tau _7\right\rangle =\frac{N \left(80 N^8+4200 N^6+50229 N^4+129175 N^2+30036\right)}{12441600},\qquad
\left\langle \tau _7^2\right\rangle =\frac{33792 N^{10}+2492160 N^8+47025792 N^6+235320800 N^4+203660116 N^2+64875}{43794432000}, \\ &
\left\langle \tau _{\frac{1}{2}} \tau _{\frac{15}{2}}\right\rangle =\frac{N^2 \left(N^4+15 N^2+24\right)}{5760} ,\qquad
\left\langle \tau _2 \tau _{\frac{15}{2}}\right\rangle =\frac{N \left(18 N^6+399 N^4+1379 N^2+388\right)}{120960} ,\qquad
\left\langle \tau _{\frac{7}{2}} \tau _{\frac{15}{2}}\right\rangle =\frac{N^2 \left(9 N^6+294 N^4+1981 N^2+2116\right)}{276480}, \\ &
\left\langle \tau _5 \tau _{\frac{15}{2}}\right\rangle =\frac{N \left(56 N^8+2646 N^6+30849 N^4+78829 N^2+18300\right)}{8709120} ,\qquad
\left\langle \tau _{\frac{13}{2}} \tau _{\frac{15}{2}}\right\rangle =\frac{N^2 \left(9 N^8+600 N^6+11077 N^4+55050 N^2+47664\right)}{16588800}, \\ & 
\left\langle \tau _0 \tau _8\right\rangle =\frac{192 N^6+3120 N^4+5508 N^2+25}{2073600} ,\qquad
\left\langle \tau _{\frac{3}{2}} \tau _8\right\rangle =\frac{N \left(96 N^6+2240 N^4+7854 N^2+2095\right)}{1209600}, \\ & 
\left\langle \tau _3 \tau _8\right\rangle =\frac{16128 N^8+542976 N^6+3647392 N^4+3883664 N^2+4735}{464486400} ,\qquad
\left\langle \tau _{\frac{9}{2}} \tau _8\right\rangle =\frac{N \left(32 N^8+1536 N^6+17850 N^4+45709 N^2+10563\right)}{6220800}, \\ & 
\left\langle \tau _6 \tau _8\right\rangle =\frac{33792 N^{10}+2266880 N^8+41799296 N^6+207921120 N^4+179741012 N^2+57275}{43794432000}, \\ &
\left\langle \tau _{\frac{15}{2}} \tau _8\right\rangle =\frac{N \left(2016 N^{10}+184800 N^8+5070758 N^6+43268445 N^4+89417801 N^2+17994380\right)}{38320128000}, \\ & 
\left\langle \tau _1 \tau _{\frac{17}{2}}\right\rangle =\frac{N \left(3 N^6+84 N^4+357 N^2+116\right)}{120960} ,\qquad
\left\langle \tau _{\frac{5}{2}} \tau _{\frac{17}{2}}\right\rangle =\frac{N^2 \left(3 N^6+118 N^4+867 N^2+932\right)}{276480}, \\ & 
\left\langle \tau _4 \tau _{\frac{17}{2}}\right\rangle =\frac{N \left(28 N^8+1533 N^6+18648 N^4+47987 N^2+11244\right)}{8709120} ,\qquad
\left\langle \tau _{\frac{11}{2}} \tau _{\frac{17}{2}}\right\rangle =\frac{N^2 \left(N^8+75 N^6+1423 N^4+7125 N^2+6176\right)}{2764800}, 
\\ & 
\left\langle \tau _7 \tau _{\frac{17}{2}}\right\rangle =\frac{N \left(336 N^{10}+33880 N^8+950565 N^6+8161956 N^4+16881139 N^2+3401964\right)}{7664025600}, \\ & 
\left\langle \tau _{\frac{17}{2}}^2\right\rangle =\frac{N^2 \left(N^{10}+133 N^8+5283 N^6+71479 N^4+286416 N^2+212688\right)}{398131200} ,\qquad
\left\langle \tau _{\frac{1}{2}} \tau _9\right\rangle =\frac{N \left(192 N^6+5040 N^4+19908 N^2+5905\right)}{14515200}, \\ & 
\left\langle \tau _2 \tau _9\right\rangle =\frac{2304 N^8+83712 N^6+579936 N^4+609648 N^2+745}{199065600} ,\qquad 
\left\langle \tau _{\frac{7}{2}} \tau _9\right\rangle =\frac{N \left(64 N^8+3216 N^6+37596 N^4+96099 N^2+22375\right)}{24883200}, 
\\& 
\left\langle \tau _5 \tau _9\right\rangle =\frac{33792 N^{10}+2323200 N^8+42869376 N^6+213461600 N^4+184426132 N^2+58775}{65691648000}, 
\end{align*}
\begin{align*}
&\left\langle \tau _{\frac{13}{2}} \tau _9\right\rangle =\frac{N \left(4032 N^{10}+373296 N^8+10244388 N^6+87465081 N^4+180671095 N^2+36402228\right)}{91968307200}, \\ &
\left\langle \tau _8 \tau _9\right\rangle =\frac{5271552 N^{12}+645765120 N^{10}+25139702016 N^8+338269451520 N^6+1354146482832 N^4+1004877745560 N^2+83330375}{1229747650560000},
\\ &
\left\langle \tau _0 \tau _{\frac{19}{2}}\right\rangle =\frac{N \left(3 N^6+84 N^4+357 N^2+116\right)}{725760} ,\qquad
\left\langle \tau _{\frac{3}{2}} \tau _{\frac{19}{2}}\right\rangle =\frac{N^2 \left(3 N^6+114 N^4+807 N^2+836\right)}{829440}, \\ & 
\left\langle \tau _3 \tau _{\frac{19}{2}}\right\rangle =\frac{N \left(140 N^8+7245 N^6+85008 N^4+216155 N^2+50892\right)}{87091200} ,\qquad
\left\langle \tau _{\frac{9}{2}} \tau _{\frac{19}{2}}\right\rangle =\frac{N^2 \left(3 N^8+210 N^6+3873 N^4+19280 N^2+16674\right)}{12441600}, \\ & 
\left\langle \tau _6 \tau _{\frac{19}{2}}\right\rangle =\frac{N \left(1680 N^{10}+157080 N^8+4307457 N^6+36793636 N^4+75965703 N^2+15322684\right)}{45984153600}, \\ & 
\left\langle \tau _{\frac{15}{2}} \tau _{\frac{19}{2}}\right\rangle =\frac{N^2 \left(3 N^{10}+369 N^8+14361 N^6+193279 N^4+773556 N^2+574272\right)}{1194393600}, \\ & 
\left\langle \tau _9 \tau _{\frac{19}{2}}\right\rangle =\frac{N \left(4032 N^{12}+642096 N^{10}+34171956 N^8+681767073 N^6+4665074232 N^4+8210487831 N^2+1480910780\right)}{19564167168000}, \\ & 
\left\langle \tau _1 \tau _{10}\right\rangle =\frac{2304 N^8+102144 N^6+848736 N^4+1030896 N^2+1225}{1393459200} ,\qquad 
\left\langle \tau _{\frac{5}{2}} \tau _{10}\right\rangle =\frac{N \left(64 N^8+3792 N^6+48636 N^4+127623 N^2+29395\right)}{87091200} ,\\ & 
\left\langle \tau _4 \tau _{10}\right\rangle =\frac{3072 N^{10}+241920 N^8+4695936 N^6+23618080 N^4+20498492 N^2+6525}{13934592000} ,\\ & 
\left\langle \tau _{\frac{11}{2}} \tau _{10}\right\rangle =\frac{N \left(576 N^{10}+59664 N^8+1690524 N^6+14541483 N^4+30110245 N^2+6059148\right)}{22992076800}, \\ &
\left\langle \tau _7 \tau _{10}\right\rangle =\frac{5271552 N^{12}+709023744 N^{10}+28265732352 N^8+382716542208 N^6+1534117927056 N^4+1138856194008 N^2+94429175}{1721646710784000}, \\ & 
\left\langle \tau _{\frac{17}{2}} \tau _{10}\right\rangle =\frac{N \left(4032 N^{12}+694512 N^{10}+37710036 N^8+756387021 N^6+5181535632 N^4+9122835267 N^2+1645088300\right)}{22824861696000}, \\ & 
\left\langle \tau _{10}^2\right\rangle =(63258624 N^{14}+13764022272 N^{12}+990207009792 N^{10}+28096602183936 N^8+302427964622784 N^6+\\ &+1026117017098992 N^4+675475244186700 N^2+14517321875)\Big{/}5061641329704960000, \\ & 
\left\langle \tau _{\frac{1}{2}} \tau _{\frac{21}{2}}\right\rangle =\frac{N^2 \left(3 N^6+126 N^4+987 N^2+1124\right)}{5806080} ,\qquad
\left\langle \tau _2 \tau _{\frac{21}{2}}\right\rangle =\frac{N \left(10 N^8+555 N^6+6762 N^4+17245 N^2+3988\right)}{21772800}, \\ & 
\left\langle \tau _{\frac{7}{2}} \tau _{\frac{21}{2}}\right\rangle =\frac{N^2 \left(9 N^8+660 N^6+12327 N^4+61290 N^2+53114\right)}{87091200}, \\ & 
\left\langle \tau _5 \tau _{\frac{21}{2}}\right\rangle =\frac{N \left(120 N^{10}+11550 N^8+318021 N^6+2717198 N^4+5613179 N^2+1130252\right)}{5748019200}, \\ &
\left\langle \tau _{\frac{13}{2}} \tau _{\frac{21}{2}}\right\rangle =\frac{N^2 \left(15 N^{10}+1875 N^8+73077 N^6+984017 N^4+3937928 N^2+2923728\right)}{8360755200}, \\ & 
\left\langle \tau _8 \tau _{\frac{21}{2}}\right\rangle =\frac{N \left(3168 N^{12}+507936 N^{10}+27046734 N^8+539749353 N^6+3693035918 N^4+6500373711 N^2+1172113180\right)}{17933819904000}, \\ & 
\left\langle \tau _{\frac{19}{2}} \tau _{\frac{21}{2}}\right\rangle =\frac{N^2 \left(45 N^{12}+9135 N^{10}+645309 N^8+18221505 N^6+195919962 N^4+664533660 N^2+437476784\right)}{5267275776000}, \\ & 
\left\langle \tau _0 \tau _{11}\right\rangle =\frac{2304 N^8+102144 N^6+848736 N^4+1030896 N^2+1225}{9754214400} ,\qquad
\left\langle \tau _{\frac{3}{2}} \tau _{11}\right\rangle =\frac{N \left(128 N^8+7392 N^6+92232 N^4+235338 N^2+52885\right)}{609638400}, \\ & 
\left\langle \tau _3 \tau _{11}\right\rangle =\frac{27648 N^{10}+2085120 N^8+39252864 N^6+194549280 N^4+168719388 N^2+53725}{292626432000}, \\ &
\left\langle \tau _{\frac{9}{2}} \tau _{11}\right\rangle =\frac{N \left(1152 N^{10}+112992 N^8+3117576 N^6+26611266 N^4+55024937 N^2+11050207\right)}{80472268800}, \\ & 
\left\langle \tau _6 \tau _{11}\right\rangle =\frac{5271552 N^{12}+666851328 N^{10}+25998964992 N^8+350090906880 N^6+1401114473616 N^4+1039931262552 N^2+86233175}{2410305395097600}, \\ & 
\left\langle \tau _{\frac{15}{2}} \tau _{11}\right\rangle =\frac{N \left(3456 N^{12}+557856 N^{10}+29705208 N^8+592895238 N^6+4056257361 N^4+7140410381 N^2+1287173500\right)}{22824861696000}, \\ & 
\left\langle \tau _9 \tau _{11}\right\rangle =(63258624 N^{14}+12878401536 N^{12}+909763126272 N^{10}+25690921521408 N^8+\\ &+276218849210304 N^6+936943113580656 N^4+616743637298700 N^2+13255326875)\Big{/}5061641329704960000, \\ & 
\left\langle \tau _{\frac{21}{2}} \tau _{11}\right\rangle =N (12672 N^{14}+3215520 N^{12}+294114744 N^{10}+11330809710 N^8+180142229421 N^6+\\ &+1040222771070 N^4+1612629270163 N^2+266022526700)\Big{/}24334814085120000,
\end{align*}
\begin{align*}
&\left\langle \tau _1 \tau _{\frac{23}{2}}\right\rangle =\frac{N \left(N^8+66 N^6+945 N^4+2764 N^2+704\right)}{17418240} ,\qquad
\left\langle \tau _{\frac{5}{2}} \tau _{\frac{23}{2}}\right\rangle =\frac{N^2 \left(3 N^8+255 N^6+5229 N^4+27045 N^2+23468\right)}{116121600}, \\ & 
\left\langle \tau _4 \tau _{\frac{23}{2}}\right\rangle =\frac{N \left(180 N^{10}+19635 N^8+571626 N^6+4958283 N^4+10275364 N^2+2077632\right)}{22992076800},
\\ &\left\langle \tau _{\frac{11}{2}} \tau _{\frac{23}{2}}\right\rangle =\frac{N^2 \left(15 N^{10}+2085 N^8+84249 N^6+1144499 N^4+4592176 N^2+3410256\right)}{16721510400}, \\ &
\left\langle \tau _7 \tau _{\frac{23}{2}}\right\rangle =\frac{N \left(1584 N^{12}+277992 N^{10}+15193035 N^8+305210334 N^6+2091903099 N^4+3682991364 N^2+664397632\right)}{14347055923200},\\ &
\left\langle \tau _{\frac{17}{2}} \tau _{\frac{23}{2}}\right\rangle =\frac{N^2 \left(9 N^{12}+1974 N^{10}+142380 N^8+4042482 N^6+43521443 N^4+147664944 N^2+97215568\right)}{1404606873600}, \\ & 
\left\langle \tau _{10} \tau _{\frac{23}{2}}\right\rangle =\frac{N \left(6336 N^{14}+1718640 N^{12}+160058052 N^{10}+6195828045 N^8+98615480238 N^6+569612518965 N^4+883084804124 N^2+145687073600\right)}{13905608048640000}, \\ & 
\left\langle \tau _{\frac{23}{2}}^2\right\rangle =\frac{N^2 \left(45 N^{14}+14940 N^{12}+1762614 N^{10}+90352452 N^8+2030462997 N^6+18369087576 N^4+54598048144 N^2+32592984832\right)}{2696845197312000}, \\ & 
\left\langle \tau _{\frac{1}{2}} \tau _{12}\right\rangle =\frac{N \left(256 N^8+16128 N^6+219744 N^4+610032 N^2+147105\right)}{9754214400}, \\ & 
\left\langle \tau _2 \tau _{12}\right\rangle =\frac{27648 N^{10}+2223360 N^8+43575168 N^6+218808480 N^4+188161884 N^2+60025}{1170505728000}, \\ & 
\left\langle \tau _{\frac{7}{2}} \tau _{12}\right\rangle =\frac{N \left(6912 N^{10}+709632 N^8+19921440 N^6+170296368 N^4+352010043 N^2+70932715\right)}{1287556300800}, \\ & 
\left\langle \tau _5 \tau _{12}\right\rangle =\frac{2027520 N^{12}+264591360 N^{10}+10391724288 N^8+139996716288 N^6+560550803472 N^4+415904716632 N^2+34490575}{1854081073152000} ,\\ & 
\left\langle \tau _{\frac{13}{2}} \tau _{12}\right\rangle =\frac{N \left(6912 N^{12}+1138176 N^{10}+60801312 N^8+1214137392 N^6+8307591435 N^4+14622494497 N^2+2636977516\right)}{73039557427200}, \\ & 
\left\langle \tau _8 \tau _{12}\right\rangle =\frac{63258624 N^{14}+13026004992 N^{12}+921546802176 N^{10}+26032958704896 N^8+279897991429056 N^6+949445980504112 N^4+624960731561644 N^2+13432006175}{6748855106273280000}, \\ & 
\left\langle \tau _{\frac{19}{2}} \tau _{12}\right\rangle =\frac{N \left(2304 N^{14}+587520 N^{12}+53771616 N^{10}+2071903920 N^8+32939951481 N^6+190213875660 N^4+294876634399 N^2+48646940700\right)}{5056584744960000}, \\ & 
\left\langle \tau _{11} \tau _{12}\right\rangle =(1173159936 N^{16}+366808006656 N^{14}+42566381125632 N^{12}+2172367773868032 N^{10}+48766888479149568 N^8+441064598681860608 N^6+\\ &+1310910543316649664 N^4+782533138799601504 N^2+4343472274925)\Big{/}42053956764035973120000, \\ & 
\left\langle \tau _0 \tau _{\frac{25}{2}}\right\rangle =\frac{N \left(N^8+66 N^6+945 N^4+2764 N^2+704\right)}{139345920}, \\ & 
\left\langle \tau _{\frac{3}{2}} \tau _{\frac{25}{2}}\right\rangle =\frac{N^2 \left(9 N^8+750 N^6+15057 N^4+76200 N^2+64784\right)}{1393459200}, \\ & 
\left\langle \tau _3 \tau _{\frac{25}{2}}\right\rangle =\frac{N \left(540 N^{10}+56925 N^8+1614822 N^6+13797861 N^4+28469628 N^2+5769664\right)}{183936614400}, \\ & 
\left\langle \tau _{\frac{9}{2}} \tau _{\frac{25}{2}}\right\rangle =\frac{N^2 \left(15 N^{10}+1995 N^8+78681 N^6+1059281 N^4+4242364 N^2+3147584\right)}{33443020800}, \\ & 
\left\langle \tau _6 \tau _{\frac{25}{2}}\right\rangle =\frac{N \left(1584 N^{12}+264264 N^{10}+14139411 N^8+282305166 N^6+1931924995 N^4+3399687460 N^2+613477440\right)}{22955289477120}, \\ & 
\left\langle \tau _{\frac{15}{2}} \tau _{\frac{25}{2}}\right\rangle =\frac{N^2 \left(27 N^{12}+5607 N^{10}+396921 N^8+11213421 N^6+120563352 N^4+408951872 N^2+269211200\right)}{5618427494400}, \\ & 
\left\langle \tau _9 \tau _{\frac{25}{2}}\right\rangle =\frac{N \left(82368 N^{14}+21106800 N^{12}+1932286356 N^{10}+74460339525 N^8+1183769835534 N^6+6835871986125 N^4+10596984950492 N^2+1748342052800\right)}{206597605294080000}, \\ & 
\left\langle \tau _{\frac{21}{2}} \tau _{\frac{25}{2}}\right\rangle =\frac{N^2 \left(135 N^{14}+42300 N^{12}+4909338 N^{10}+250558164 N^8+5624644839 N^6+50871706872 N^4+151196562288 N^2+90257579264\right)}{8090535591936000},\\ &
\left\langle \tau _{12} \tau _{\frac{25}{2}}\right\rangle =N (14826240 N^{16}+5629029120 N^{14}+814141744416 N^{12}+53714183876496 N^{10}+1642015141238043 N^8+21901314579189318 N^6+\\ &+110374356210766251 N^4+154180715255760516 N^2+23616315157736000)\Big{/}16521197300156989440000, \\ & 
\left\langle \tau _1 \tau _{13}\right\rangle =\frac{1024 N^{10}+96000 N^8+2174592 N^6+12408800 N^4+11798484 N^2+3675}{390168576000} \\ & \left\langle \tau _{\frac{5}{2}} \tau _{13}\right\rangle =\frac{N \left(768 N^{10}+90112 N^8+2772000 N^6+24853488 N^4+52008627 N^2+10410435\right)}{643778150400},
\end{align*}
\begin{align*}
&\left\langle \tau _4 \tau _{13}\right\rangle =\frac{61440 N^{12}+9000960 N^{10}+374817024 N^8+5152775424 N^6+20719746576 N^4+15408973656 N^2+1276975}{168552824832000},\\ &
\left\langle \tau _{\frac{11}{2}} \tau _{13}\right\rangle =\frac{N \left(768 N^{12}+139776 N^{10}+7766304 N^8+156799344 N^6+1075957155 N^4+1895456745 N^2+341634188\right)}{18259889356800},
\\ &\left\langle \tau _7 \tau _{13}\right\rangle =\frac{35143680 N^{14}+7892684800 N^{12}+574444975104 N^{10}+16349468678400 N^8+176138010240192 N^6+597719310879600 N^4+393499090432124 N^2+8456864675}{6748855106273280000},\\ &
\left\langle \tau _{\frac{17}{2}} \tau _{13}\right\rangle =\frac{N \left(768 N^{14}+211200 N^{12}+19761696 N^{10}+765909200 N^8+12194830179 N^6+70442619300 N^4+109213900357 N^2+18016077300\right)}{2528292372480000}, \\ & 
\left\langle \tau _{10} \tau _{13}\right\rangle =(43450368 N^{16}+14512422912 N^{14}+1715496566784 N^{12}+87981873487872 N^{10}+1977472577653248 N^8+17890144965987840 N^6+\\ &+53175435349569600 N^4+31742964860729376 N^2+176189063575)\Big{/}2002569369715998720000, \\ &
\left\langle \tau _{\frac{23}{2}} \tau _{13}\right\rangle =N (126720 N^{16}+50983680 N^{14}+7494977952 N^{12}+496652285712 N^{10}+15198771356351 N^8+202775563645326 N^6+\\ &+1021968028202127 N^4+1427600669098932 N^2+218666563080000)\Big{/}158857666347663360000, \\ & 
\left\langle \tau _{13}^2\right\rangle =(254017536 N^{18}+122118930432 N^{16}+22016271384576 N^{14}+1847393572208640 N^{12}+74762402614622208 N^{10}+1405295219664935424 N^8+\\ &+11057833818706543872 N^6+29472649803214144704 N^4+16208773786311341508 N^2+23165125941475)\Big{/}6638055329215524372480000, \\ & 
\left\langle \tau _{\frac{1}{2}} \tau _{\frac{27}{2}}\right\rangle =\frac{N^2 \left(N^8+90 N^6+1953 N^4+10660 N^2+9696\right)}{1393459200}, \\ & 
\left\langle \tau _2 \tau _{\frac{27}{2}}\right\rangle =\frac{N \left(30 N^{10}+3355 N^8+99132 N^6+863511 N^4+1781428 N^2+357184\right)}{45984153600}, \\ &
\left\langle \tau _{\frac{7}{2}} \tau _{\frac{27}{2}}\right\rangle =\frac{N^2 \left(N^{10}+139 N^8+5595 N^6+75669 N^4+302884 N^2+224992\right)}{6688604160}, \\ & 
\left\langle \tau _5 \tau _{\frac{27}{2}}\right\rangle =\frac{N \left(616 N^{12}+106106 N^{10}+5734729 N^8+114655684 N^6+784829045 N^4+1381309020 N^2+249076160\right)}{20085878292480}, \\ & 
\left\langle \tau _{\frac{13}{2}} \tau _{\frac{27}{2}}\right\rangle =\frac{N^2 \left(15 N^{12}+3185 N^{10}+226611 N^8+6406515 N^6+68895218 N^4+233681000 N^2+153840256\right)}{5618427494400}, \\ & 
\left\langle \tau _8 \tau _{\frac{27}{2}}\right\rangle =\frac{N \left(96096 N^{14}+24984960 N^{12}+2293313022 N^{10}+88413966355 N^8+1405726647468 N^6+8117529664335 N^4+12584149134164 N^2+2075999073600\right)}{361545809264640000}, \\ & 
\left\langle \tau _{\frac{19}{2}} \tau _{\frac{27}{2}}\right\rangle =\frac{N^2 \left(105 N^{14}+33180 N^{12}+3856062 N^{10}+196858996 N^8+4419325041 N^6+39970711368 N^4+118797106192 N^2+70916825856\right)}{8090535591936000}, \\ & 
\left\langle \tau _{11} \tau _{\frac{27}{2}}\right\rangle =N (164736 N^{16}+62778144 N^{14}+9085035960 N^{12}+599477467446 N^{10}+18325965281767 N^8+244434687067116 N^6+\\ &+1231854365268987 N^4+1720772690281444 N^2+263571162702400)\Big{/}206514966251962368000, \\ & 
\left\langle \tau _{\frac{25}{2}} \tau _{\frac{27}{2}}\right\rangle =\frac{N^2 \left(N^{16}+456 N^{14}+80982 N^{12}+6767732 N^{10}+273606337 N^8+5141648028 N^6+40455842896 N^4+107826049984 N^2+59300291584\right)}{38834570841292800}, \\ & 
\left\langle \tau _0 \tau _{14}\right\rangle =\frac{1024 N^{10}+96000 N^8+2174592 N^6+12408800 N^4+11798484 N^2+3675}{3511517184000}, \\ &
\left\langle \tau _{\frac{3}{2}} \tau _{14}\right\rangle =\frac{N \left(512 N^{10}+59136 N^8+1788864 N^6+15763264 N^4+32435634 N^2+6400905\right)}{1931334451200}, \\ & 
\left\langle \tau _3 \tau _{14}\right\rangle =\frac{20480 N^{12}+2918400 N^{10}+118881024 N^8+1610946304 N^6+6436841136 N^4+4786592136 N^2+396725}{168552824832000}, \\ & 
\left\langle \tau _{\frac{9}{2}} \tau _{14}\right\rangle =\frac{N \left(512 N^{12}+89856 N^{10}+4884672 N^8+97669312 N^6+668411250 N^4+1176810297 N^2+211927191\right)}{27389834035200}, \\ & 
\left\langle \tau _6 \tau _{14}\right\rangle =\frac{11714560 N^{14}+2521559040 N^{12}+179902987264 N^{10}+5085877871360 N^8+54696761351232 N^6+185514007624400 N^4+122126285007444 N^2+2624725425}{4049313063763968000}, \\ & 
\left\langle \tau _{\frac{15}{2}} \tau _{14}\right\rangle =\frac{N \left(512 N^{14}+134400 N^{12}+12352704 N^{10}+476259200 N^8+7572499506 N^6+43726558575 N^4+67790074403 N^2+11181581700\right)}{2528292372480000}, \\ & 
\left\langle \tau _9 \tau _{14}\right\rangle =(43450368 N^{16}+13817217024 N^{14}+1606870646784 N^{12}+82040658452480 N^{10}+1841751822862848 N^8+16657689781943808 N^6+\\ &+49508764782222400 N^4+29553916727539488 N^2+164039493975)\Big{/}2574732046777712640000, \\ & 
\left\langle \tau _{\frac{21}{2}} \tau _{14}\right\rangle =N (84480 N^{16}+32313600 N^{14}+4677968064 N^{12}+308697182464 N^{10}+9436800711162 N^8+125870066508627 N^6+\\ &+634332994157594 N^4+886101824741109 N^2+135722199616500)\Big{/}119143249760747520000, \\ & 
\left\langle \tau _{12} \tau _{14}\right\rangle =(254017536 N^{18}+116022509568 N^{16}+20607998164992 N^{14}+1722284855377920 N^{12}+69628530780813312 N^{10}+1308471980620847616 N^8+\\ &+10295380334781670656 N^6+27440124961174140096 N^4+15090933159115966404 N^2+21567564476675)\Big{/}6638055329215524372480000, 
\end{align*}
\begin{align*}
&\left\langle \tau _{\frac{27}{2}} \tau _{14}\right\rangle =N (33280 N^{18}+18021120 N^{16}+3877528512 N^{14}+403365543360 N^{12}+21051926006514 N^{10}+537728602825245 N^8+6223438271413316 N^6+\\ &+28020133661124225 N^4+35854543560011628 N^2+5154450344836800)\Big{/}30215237843067863040000,
\\ &
\left\langle \tau _1 \tau _{\frac{29}{2}}\right\rangle =\frac{N \left(3 N^{10}+385 N^8+12969 N^6+127215 N^4+289828 N^2+62400\right)}{45984153600}, \\ &
\left\langle \tau _{\frac{5}{2}} \tau _{\frac{29}{2}}\right\rangle =\frac{N^2 \left(N^{10}+157 N^8+6903 N^6+98391 N^4+402196 N^2+298752\right)}{33443020800}, \\ &
\left\langle \tau _4 \tau _{\frac{29}{2}}\right\rangle =\frac{N \left(84 N^{12}+16107 N^{10}+924105 N^8+18938049 N^6+130494819 N^4+230059284 N^2+41574592\right)}{9129944678400}, \\ & 
\left\langle \tau _{\frac{11}{2}} \tau _{\frac{29}{2}}\right\rangle =\frac{N^2 \left(15 N^{12}+3500 N^{10}+259602 N^8+7438800 N^6+80277071 N^4+272582100 N^2+179458112\right)}{14046068736000}, \\ & 
\left\langle \tau _7 \tau _{\frac{29}{2}}\right\rangle =\frac{N \left(48048 N^{14}+13573560 N^{12}+1284350067 N^{10}+49948151825 N^8+795995983497 N^6+4599269671815 N^4+7130639025988 N^2+1176545380800\right)}{361545809264640000},\\ &
\left\langle \tau _{\frac{17}{2}} \tau _{\frac{29}{2}}\right\rangle =\frac{N^2 \left(21 N^{14}+7140 N^{12}+849534 N^{10}+43649996 N^8+981569349 N^6+8881326888 N^4+26398908496 N^2+15759173376\right)}{2696845197312000}, \\ & 
\left\langle \tau _{10} \tau _{\frac{29}{2}}\right\rangle =N (82368 N^{16}+33489456 N^{14}+4940715780 N^{12}+327714142041 N^{10}+10031537206691 N^8+133844571059355 N^6+\\ &+674575966648761 N^4+942321411495548 N^2+144339108148800)\Big{/}147510690179973120000, \\ &
\left\langle \tau _{\frac{23}{2}} \tau _{\frac{29}{2}}\right\rangle =\frac{N^2 \left(N^{16}+483 N^{14}+87210 N^{12}+7320854 N^{10}+296301925 N^8+5569675623 N^6+43826415616 N^4+116811229840 N^2+64241941248\right)}{48543213551616000}, \\ & 
\left\langle \tau _{13} \tau _{\frac{29}{2}}\right\rangle =N (49920 N^{18}+28454400 N^{16}+6213800736 N^{14}+648997143600 N^{12}+33906036299187 N^{10}+866276728791825 N^8+\\ &+10026485732300313 N^6+45143425777985775 N^4+57765581384855844 N^2+8304420770878400)\Big{/}50358729738446438400000,\\ & 
\left\langle \tau _{\frac{29}{2}}^2\right\rangle =N^2 (21 N^{18}+14035 N^{16}+3667482 N^{14}+469606950 N^{12}+31035542153 N^{10}+1046985819495 N^8+\\ &+17036454985464 N^6+119397398644720 N^4+290229854906880 N^2+148788466380800)\Big{/}815525987667148800000,\\ &
\left\langle \tau _{\frac{1}{2}} \tau _{15}\right\rangle =\frac{N \left(1024 N^{10}+126720 N^8+4109952 N^6+38778080 N^4+85002324 N^2+17656275\right)}{38626689024000}, \\ & 
\left\langle \tau _2 \tau _{15}\right\rangle =\frac{4096 N^{12}+616448 N^{10}+26140416 N^8+362461952 N^6+1457264368 N^4+1077400680 N^2+89425}{168552824832000}, \\ & 
\left\langle \tau _{\frac{7}{2}} \tau _{15}\right\rangle =\frac{N \left(3072 N^{12}+562432 N^{10}+31264896 N^8+629774496 N^6+4311709532 N^4+7591288497 N^2+1369549125\right)}{547796680704000}, \\ & 
\left\langle \tau _5 \tau _{15}\right\rangle =\frac{180224 N^{14}+40054784 N^{12}+2893011968 N^{10}+81999884032 N^8+882109974592 N^6+2992460957584 N^4+1969582220916 N^2+42332325}{155742810144768000}, \\ & 
\left\langle \tau _{\frac{13}{2}} \tau _{15}\right\rangle =\frac{N \left(1024 N^{14}+275200 N^{12}+25465728 N^{10}+982991200 N^8+15633209972 N^6+90275207175 N^4+139950278401 N^2+23087592300\right)}{10113169489920000} \\ & 
\left\langle \tau _8 \tau _{15}\right\rangle =(43450368 N^{16}+14048952320 N^{14}+1640066727936 N^{12}+83792070369280 N^{10}+1881342382426624 N^8+\\ &+17015818276154880 N^6+50573774162987072 N^4+30189300829055520 N^2+167566788375)\Big{/}4291220077962854400000, \\ & 
\left\langle \tau _{\frac{19}{2}} \tau _{15}\right\rangle =N (3072 N^{16}+1188096 N^{14}+172371840 N^{12}+11379553568 N^{10}+347896904124 N^8+\\ &+4640360127021 N^6+23385601624864 N^4+32667219631515 N^2+5003674500300)\Big{/}6189259727831040000 \\ & 
\left\langle \tau _{11} \tau _{15}\right\rangle =(254017536 N^{18}+116784562176 N^{16}+20768029212672 N^{14}+1736091534016512 N^{12}+70189456526850048 N^{10}+1319023875339330048 N^8+\\ &+10378403715885452544 N^6+27661427644508451264 N^4+15212627545531493700 N^2+21741492539375)\Big{/}8297569161519405465600000, \\ & 
\left\langle \tau _{\frac{25}{2}} \tau _{15}\right\rangle =N (119808 N^{18}+65065728 N^{16}+14007247488 N^{14}+1457288847456 N^{12}+76058239110308 N^{10}+1942761293747079 N^8+\\ &+22484671721467938 N^6+101234075319003487 N^4+129538894825590708 N^2+18622596605256000)\Big{/}120860951372271452160000, \\ & 
\left\langle \tau _{14} \tau _{15}\right\rangle =338690048 N^{20}+215914905600 N^{18}+55648066142208 N^{16}+7098817455718400 N^{14}+468699174801727488 N^{12}+\\ &+15807842961676185600 N^{10}+257209471731512350208 N^8+1802589500524372723200 N^6+\\ &+4381705578810089640048 N^4+2246308095834329272200 N^2+824566063741875)\Big{/}7965666395058629246976000000.
\end{align*}
\end{tiny}

{\bf Nonzero three--point correlators $\le\langle\tau_{\frac{d_1}{2}}\tau_{\frac{d_2}{2}}\tau_{\frac{d_3}{2}}\ri\rangle$, for $0\leq d_1\leq d_2\leq d_3\leq 10$}
\begin{tiny}
\begin{align*}
&\left\langle \tau _0^3\right\rangle =1,\qquad \left\langle \tau _{\frac{1}{2}}^3\right\rangle =N,,\qquad \left\langle \tau _0 \tau _{\frac{1}{2}} \tau _1\right\rangle =N,\qquad \left\langle \tau _1^3\right\rangle =N^2+\frac{1}{12} ,\qquad \left\langle \tau _0^2 \tau _{\frac{3}{2}}\right\rangle =N, \qquad \left\langle \tau _{\frac{1}{2}} \tau _1 \tau _{\frac{3}{2}}\right\rangle =N^2,\qquad \left\langle \tau _0 \tau _{\frac{3}{2}}^2\right\rangle =N^2, \qquad \left\langle \tau _{\frac{3}{2}}^3\right\rangle =N^3+\frac{N}{4},
\\ &
\left\langle \tau _{\frac{1}{2}}^2 \tau _2\right\rangle =N^2, \ \ 
\left\langle \tau _0 \tau _1 \tau _2\right\rangle =N^2+\frac{1}{12}, \ \ 
\left\langle \tau _1 \tau _{\frac{3}{2}} \tau _2\right\rangle =N^3+\frac{N}{2}, \ \
\left\langle \tau _{\frac{1}{2}} \tau _2^2\right\rangle =N^3+\frac{5 N}{12}, \ \
\left\langle \tau _2^3\right\rangle =N^4+\frac{3 N^2}{2}+\frac{7}{240} ,\ \
\left\langle \tau _0 \tau _{\frac{1}{2}} \tau _{\frac{5}{2}}\right\rangle =\frac{N^2}{2},
\\ &
\left\langle \tau _1^2 \tau _{\frac{5}{2}}\right\rangle =\frac{N^3+N}{2},\qquad
\left\langle \tau _{\frac{1}{2}} \tau _{\frac{3}{2}} \tau _{\frac{5}{2}}\right\rangle =\frac{2 N^3+N}{4},\qquad 
\left\langle \tau _0 \tau _2 \tau _{\frac{5}{2}}\right\rangle =\frac{N \left(3 N^2+2\right)}{6} ,\qquad 
\left\langle \tau _{\frac{3}{2}} \tau _2 \tau _{\frac{5}{2}}\right\rangle =\frac{N^2 \left(6 N^2+11\right)}{12} ,\qquad 
\left\langle \tau _1 \tau _{\frac{5}{2}}^2\right\rangle =\frac{1}{4} N^2 \left(N^2+3\right), \\ &
\left\langle \tau _{\frac{5}{2}}^3\right\rangle =\frac{N \left(3 N^4+19 N^2+6\right)}{24} , \qquad 
\left\langle \tau _0^2 \tau _3\right\rangle =\frac{N^2}{2}+\frac{1}{24} ,\qquad
\left\langle \tau _{\frac{1}{2}} \tau _1 \tau _3\right\rangle =\frac{ N \left(4 N^2+3\right)}{8},\qquad 
\left\langle \tau _0 \tau _{\frac{3}{2}} \tau _3\right\rangle =\frac{N \left(12 N^2+7\right)}{24},\qquad 
\left\langle \tau _{\frac{3}{2}}^2 \tau _3\right\rangle =\frac{N^2 \left(12 N^2+19\right)}{24}, 
\\ &
\left\langle \tau _1 \tau _2 \tau _3\right\rangle =\frac{N^4}{2}+\frac{13 N^2}{12}+\frac{29}{1440}, \qquad 
\left\langle \tau _{\frac{1}{2}} \tau _{\frac{5}{2}} \tau _3\right\rangle =\frac{1}{16} N^2 \left(4 N^2+9\right) , \qquad
\left\langle \tau _2 \tau _{\frac{5}{2}} \tau _3\right\rangle =\frac{N \left(36 N^4+171 N^2+56\right)}{144}, \qquad 
\left\langle \tau _0 \tau _3^2\right\rangle =\frac{N^4}{4}+\frac{13 N^2}{24}+\frac{29}{2880},
\\ &
\left\langle \tau _{\frac{3}{2}} \tau _3^2\right\rangle =\frac{N \left(144 N^4+600 N^2+193\right)}{576},\qquad
\left\langle \tau _3^3\right\rangle =\frac{N^6}{8}+\frac{33 N^4}{32}+\frac{481 N^2}{384}+\frac{583}{96768},\qquad
\left\langle \tau _{\frac{1}{2}}^2 \tau _{\frac{7}{2}}\right\rangle =\frac{N \left(3 N^2+2\right)}{12} , \qquad
\left\langle \tau _0 \tau _1 \tau _{\frac{7}{2}}\right\rangle =\frac{N^3+N}{4}, 
\\ &
\left\langle \tau _1 \tau _{\frac{3}{2}} \tau _{\frac{7}{2}}\right\rangle =\frac{N^2 \left(3 N^2+7\right)}{12}, \qquad 
\left\langle \tau _{\frac{1}{2}} \tau _2 \tau _{\frac{7}{2}}\right\rangle =\frac{N^2 \left(N^2+2\right)}{4} ,\qquad
\left\langle \tau _2^2 \tau _{\frac{7}{2}}\right\rangle =\frac{N \left(36 N^4+153 N^2+49\right)}{144}, \qquad
\left\langle \tau _0 \tau _{\frac{5}{2}} \tau _{\frac{7}{2}}\right\rangle =\frac{N^2 \left(3 N^2+8\right)}{24} ,
\\ &
\left\langle \tau _{\frac{3}{2}} \tau _{\frac{5}{2}} \tau _{\frac{7}{2}}\right\rangle =\frac{N (2 N^4+10 N^2+3)}{16} ,\qquad 
\left\langle \tau _1 \tau _3 \tau _{\frac{7}{2}}\right\rangle =\frac{N (36 N^4+195 N^2+79)}{288} , \qquad
\left\langle \tau _{\frac{5}{2}} \tau _3 \tau _{\frac{7}{2}}\right\rangle =\frac{N^2 \left(12 N^4+115 N^2+144\right)}{192},  \\ &
\left\langle \tau _{\frac{1}{2}} \tau _{\frac{7}{2}}^2\right\rangle =\frac{N \left(9 N^4+45 N^2+16\right)}{144}, \qquad  
\left\langle \tau _2 \tau _{\frac{7}{2}}^2\right\rangle =\frac{N^2 \left(9 N^4+78 N^2+95\right)}{144} , \qquad 
\left\langle \tau _{\frac{7}{2}}^3\right\rangle =\frac{N \left(27 N^6+405 N^4+1188 N^2+305\right)}{1728},
\\ &
\left\langle \tau _0 \tau _{\frac{1}{2}} \tau _4\right\rangle =\frac{N \left(4 N^2+3\right)}{24}, \qquad
\left\langle \tau _1^2 \tau _4\right\rangle =\frac{16 N^4+56 N^2+1}{96}, \qquad
\left\langle \tau _{\frac{1}{2}} \tau _{\frac{3}{2}} \tau _4\right\rangle =\frac{N^2 \left(4 N^2+9\right)}{24} ,\qquad
\left\langle \tau _0 \tau _2 \tau _4\right\rangle =\frac{N^4}{6}+\frac{5 N^2}{12}+\frac{11}{1440}, \\ &
\left\langle \tau _{\frac{3}{2}} \tau _2 \tau _4\right\rangle =\frac{N \left(24 N^4+118 N^2+35\right)}{144} , \qquad
\left\langle \tau _1 \tau _{\frac{5}{2}} \tau _4\right\rangle =\frac{N \left(12 N^4+85 N^2+33\right)}{144},\qquad 
\left\langle \tau _{\frac{5}{2}}^2 \tau _4\right\rangle =\frac{N^2 \left(4 N^4+49 N^2+63\right)}{96},
\\ &
\left\langle \tau _{\frac{1}{2}} \tau _3 \tau _4\right\rangle =\frac{N \left(80 N^4+440 N^2+157\right)}{960}, \qquad 
\left\langle \tau _2 \tau _3 \tau _4\right\rangle =\frac{20160 N^6+193200 N^4+232932 N^2+1121}{241920},\qquad
\left\langle \tau _0 \tau _{\frac{7}{2}} \tau _4\right\rangle =\frac{N \left(12 N^4+73 N^2+29\right)}{288}, \\ &
\left\langle \tau _{\frac{3}{2}} \tau _{\frac{7}{2}} \tau _4\right\rangle =\frac{N^2 \left(12 N^4+119 N^2+143\right)}{288},\qquad 
\left\langle \tau _3 \tau _{\frac{7}{2}} \tau _4\right\rangle =\frac{N \left(720 N^6+11880 N^4+34981 N^2+9181\right)}{34560},\qquad 
\left\langle \tau _1 \tau _4^2\right\rangle =\frac{6720 N^6+89040 N^4+130284 N^2+607}{241920}, \\ &
\left\langle \tau _{\frac{5}{2}} \tau _4^2\right\rangle =\frac{N \left(240 N^6+4920 N^4+15447 N^2+3967\right)}{17280},\qquad 
\left\langle \tau _4^3\right\rangle =\frac{3840 N^8+120320 N^6+729312 N^4+702592 N^2+875}{829440},\qquad 
\left\langle \tau _0^2 \tau _{\frac{9}{2}}\right\rangle =\frac{N^3+N}{12}, \\ &
\left\langle \tau _{\frac{1}{2}} \tau _1 \tau _{\frac{9}{2}}\right\rangle =\frac{N^2 \left(N^2+3\right)}{12},\qquad 
\left\langle \tau _0 \tau _{\frac{3}{2}} \tau _{\frac{9}{2}}\right\rangle =\frac{N^2 \left(2 N^2+5\right)}{24},\qquad 
\left\langle \tau _{\frac{3}{2}}^2 \tau _{\frac{9}{2}}\right\rangle =\frac{N \left(4 N^4+19 N^2+5\right)}{48},\qquad 
\left\langle \tau _1 \tau _2 \tau _{\frac{9}{2}}\right\rangle =\frac{N \left(6 N^4+35 N^2+13\right)}{72}, \\ &
\left\langle \tau _{\frac{1}{2}} \tau _{\frac{5}{2}} \tau _{\frac{9}{2}}\right\rangle =\frac{N \left(N^4+6 N^2+2\right)}{24},\qquad 
\left\langle \tau _2 \tau _{\frac{5}{2}} \tau _{\frac{9}{2}}\right\rangle =\frac{ N^2 \left(6 N^4+61 N^2+75\right)}{144},\qquad 
\left\langle \tau _0 \tau _3 \tau _{\frac{9}{2}}\right\rangle =\frac{N \left(60 N^4+335 N^2+131\right)}{1440}, \\ &
\left\langle \tau _{\frac{3}{2}} \tau _3 \tau _{\frac{9}{2}}\right\rangle =\frac{N^2 \left(120 N^4+1090 N^2+1291\right)}{2880} ,\qquad \left\langle \tau _3^2 \tau _{\frac{9}{2}}\right\rangle =\frac{N \left(720 N^6+10920 N^4+31613 N^2+8261\right)}{34560} ,\qquad \left\langle \tau _1 \tau _{\frac{7}{2}} \tau _{\frac{9}{2}}\right\rangle =\frac{1}{48} N^2 \left(N^4+11 N^2+16\right), \\ &
\left\langle \tau _{\frac{5}{2}} \tau _{\frac{7}{2}} \tau _{\frac{9}{2}}\right\rangle =\frac{1}{576} N \left(6 N^6+103 N^4+316 N^2+79\right) ,\qquad \left\langle \tau _{\frac{1}{2}} \tau _4 \tau _{\frac{9}{2}}\right\rangle =\frac{N^2 \left(20 N^4+225 N^2+303\right)}{1440} ,\qquad \left\langle \tau _2 \tau _4 \tau _{\frac{9}{2}}\right\rangle =\frac{N \left(120 N^6+2070 N^4+6167 N^2+1561\right)}{8640}, \\ &
\left\langle \tau _{\frac{7}{2}} \tau _4 \tau _{\frac{9}{2}}\right\rangle =\frac{N^2 \left(60 N^6+1595 N^4+9297 N^2+8908\right)}{17280} ,\qquad \left\langle \tau _0 \tau _{\frac{9}{2}}^2\right\rangle =\frac{1}{144} N^2 \left(N^4+11 N^2+16\right) ,\qquad \left\langle \tau _{\frac{3}{2}} \tau _{\frac{9}{2}}^2\right\rangle =\frac{1}{576} N \left(4 N^6+65 N^4+193 N^2+46\right), \\ &
\left\langle \tau _3 \tau _{\frac{9}{2}}^2\right\rangle =\frac{N^2 \left(60 N^6+1475 N^4+8413 N^2+8024\right)}{17280} ,\qquad \left\langle \tau _{\frac{9}{2}}^3\right\rangle =\frac{N \left(20 N^8+735 N^6+7224 N^4+16535 N^2+3514\right)}{34560} ,\qquad \left\langle \tau _{\frac{1}{2}}^2 \tau _5\right\rangle =\frac{1}{72} N^2 \left(4 N^2+11\right), \\ &
\left\langle \tau _0 \tau _1 \tau _5\right\rangle =\frac{1}{288} \left(16 N^4+56 N^2+1\right) ,\qquad \left\langle \tau _1 \tau _{\frac{3}{2}} \tau _5\right\rangle =\frac{1}{144} N \left(8 N^4+50 N^2+17\right) ,\qquad \left\langle \tau _{\frac{1}{2}} \tau _2 \tau _5\right\rangle =\frac{N \left(80 N^4+440 N^2+141\right)}{1440}, \\ &
\left\langle \tau _2^2 \tau _5\right\rangle =\frac{960 N^6+9040 N^4+10532 N^2+51}{17280} ,\qquad \left\langle \tau _0 \tau _{\frac{5}{2}} \tau _5\right\rangle =\frac{1}{144} N \left(4 N^4+27 N^2+10\right) ,\qquad \left\langle \tau _{\frac{3}{2}} \tau _{\frac{5}{2}} \tau _5\right\rangle =\frac{1}{288} N^2 \left(8 N^4+86 N^2+103\right), \\ &
\left\langle \tau _1 \tau _3 \tau _5\right\rangle =\frac{6720 N^6+75600 N^4+107436 N^2+503}{241920} ,\qquad \left\langle \tau _{\frac{5}{2}} \tau _3 \tau _5\right\rangle =\frac{N \left(240 N^6+4200 N^4+12743 N^2+3248\right)}{17280} ,\\ & \left\langle \tau _{\frac{1}{2}} \tau _{\frac{7}{2}} \tau _5\right\rangle =\frac{1}{864} N^2 \left(12 N^4+125 N^2+166\right), \qquad
\left\langle \tau _2 \tau _{\frac{7}{2}} \tau _5\right\rangle =\frac{N \left(240 N^6+3840 N^4+11261 N^2+2841\right)}{17280} ,\\ & \left\langle \tau _{\frac{7}{2}}^2 \tau _5\right\rangle =\frac{N^2 \left(36 N^6+891 N^4+5098 N^2+4865\right)}{10368}, \qquad
\left\langle \tau _0 \tau _4 \tau _5\right\rangle =\frac{448 N^6+5488 N^4+7924 N^2+37}{48384} ,\qquad
\left\langle \tau _{\frac{3}{2}} \tau _4 \tau _5\right\rangle =\frac{N \left(32 N^6+576 N^4+1722 N^2+421\right)}{3456},
\end{align*}
\begin{align*}
&\left\langle \tau _3 \tau _4 \tau _5\right\rangle =\frac{3840 N^8+103680 N^6+603040 N^4+576080 N^2+719}{829440} ,\qquad \left\langle \tau _1 \tau _{\frac{9}{2}} \tau _5\right\rangle =\frac{N \left(40 N^6+770 N^4+2681 N^2+751\right)}{8640}, \\ &
\left\langle \tau _{\frac{5}{2}} \tau _{\frac{9}{2}} \tau _5\right\rangle =\frac{N^2 \left(40 N^6+1110 N^4+6737 N^2+6431\right)}{17280} ,\qquad \left\langle \tau _4 \tau _{\frac{9}{2}} \tau _5\right\rangle =\frac{N \left(160 N^8+6400 N^6+64690 N^4+148317 N^2+31947\right)}{207360},\\ &
 \left\langle \tau _{\frac{1}{2}} \tau _5^2\right\rangle =\frac{N \left(448 N^6+8176 N^4+26740 N^2+7005\right)}{145152} ,\qquad \left\langle \tau _2 \tau _5^2\right\rangle =\frac{N^8}{324}+\frac{13 N^6}{162}+\frac{2023 N^4}{4320}+\frac{277 N^2}{630}+\frac{533}{967680},\\ &
\left\langle \tau _{\frac{7}{2}} \tau _5^2\right\rangle =\frac{N \left(6720 N^8+251440 N^6+2487436 N^4+5672125 N^2+1220349\right)}{8709120} ,\\ &
\left\langle \tau _5^3\right\rangle =\frac{394240 N^{10}+20993280 N^8+328056960 N^6+1454969120 N^4+1157523444 N^2+376857}{2299207680}.
\end{align*}
\end{tiny}

{\bf Nonzero four--point correlators $\le\langle\tau_{\frac{d_1}{2}}\tau_{\frac{d_2}{2}}\tau_{\frac{d_3}{2}}\tau_{\frac{d_4}{2}}\ri\rangle$, for $0\leq d_1\leq d_2\leq d_3\leq d_4\leq 7$}
\begin{tiny}

% !TEX root =penner-1.4.tex

\begin{align*}
%\displaybreak
&\left\langle {\tau_{{0}}}^{3}\tau_{{1}} \right\rangle =1 
 ,\quad  \left\langle {\tau_{{\frac 1 2 }}}^{3}\tau_{{1}} \right\rangle =2\,N 
 ,\quad  \left\langle \tau_{{0}}\tau_{{\frac 1 2 }}{\tau_{{1}}}^{2} \right\rangle =2\,N 
,\quad 
    \left\langle {\tau_{{1}}}^{4} \right\rangle =\frac 1 4 +3\,{N}^{2} 
 ,\quad  \left\langle \tau_{{0}}{\tau_{{\frac 1 2 }}}^{2}\tau_{{\frac 3 2 }} \right\rangle =N 
 ,\quad  \left\langle {\tau_{{0}}}^{2}\tau_{{1}}\tau_{{\frac 3 2 }} \right\rangle =2\,N ,
    \left\langle \tau_{{\frac 1 2 }}{\tau_{{1}}}^{2}\tau_{{\frac 3 2 }} \right\rangle =3\,{N}^{2}
,\\ &\left\langle {\tau_{{\frac 1 2 }}}^{2}{\tau_{{\frac 3 2 }}}^{2} \right\rangle =2\,{N}^{2} 
 ,\quad  \left\langle \tau_{{0}}\tau_{{1}}{\tau_{{\frac 3 2 }}}^{2} \right\rangle =3\,{N}^{2} 
,\quad 
    \left\langle \tau_{{1}}{\tau_{{\frac 3 2 }}}^{3} \right\rangle =N \left( 4\,{N}^{2}+1 \right)  
 ,\quad  \left\langle {\tau_{{0}}}^{2}\tau_{{\frac 1 2 }}\tau_{{2}} \right\rangle =N 
 ,\quad  \left\langle {\tau_{{\frac 1 2 }}}^{2}\tau_{{1}}\tau_{{2}} \right\rangle =3\,{N}^{2},
    \quad \left\langle \tau_{{0}}{\tau_{{1}}}^{2}\tau_{{2}} \right\rangle =\frac 1 4 +3\,{N}^{2} 
 ,\\ &  \left\langle \tau_{{0}}\tau_{{\frac 1 2 }}\tau_{{\frac 3 2 }}\tau_{{2}} \right\rangle =2\,{N}^{2} 
 ,\quad  \left\langle {\tau_{{1}}}^{2}\tau_{{\frac 3 2 }}\tau_{{2}} \right\rangle =2\,N \left( 2\,{N}^{2}+1 \right)  
,\quad 
    \left\langle \tau_{{\frac 1 2 }}{\tau_{{\frac 3 2 }}}^{2}\tau_{{2}} \right\rangle =N \left( 3\,{N}^{2}+1 \right)  
 ,\quad  \left\langle {\tau_{{0}}}^{2}{\tau_{{2}}}^{2} \right\rangle =\frac 1 6 +2\,{N}^{2} 
,\quad
\left\langle \tau_{{\frac 1 2 }}\tau_{{1}}{\tau_{{2}}}^{2} \right\rangle =\frac 1 3 \,N \left( 12\,{N}^{2}+5 \right)  
,\\ &    \left\langle \tau_{{0}}\tau_{{\frac 3 2 }}{\tau_{{2}}}^{2} \right\rangle =\frac 1 {12}\,N \left( 36\,{N}^{2}+17 \right)  
 ,\quad  \left\langle {\tau_{{\frac 3 2 }}}^{2}{\tau_{{2}}}^{2} \right\rangle =\frac 1 {12}\,{N}^{2} \left( 48\,{N}^{2}+65 \right)  
 ,\quad  \left\langle \tau_{{1}}{\tau_{{2}}}^{3} \right\rangle =5\,{N}^{4}+\frac {15}2\,{N}^{2}+{\frac {7}{48}} 
,\quad   \left\langle {\tau_{{0}}}^{3}\tau_{{\frac 5 2 }} \right\rangle =N 
 ,\quad  \left\langle {\tau_{{\frac 1 2 }}}^{3}\tau_{{\frac 5 2 }} \right\rangle =\frac 3 2 \,{N}^{2} 
 ,\\ &  \left\langle \tau_{{0}}\tau_{{\frac 1 2 }}\tau_{{1}}\tau_{{\frac 5 2 }} \right\rangle =\frac 3 2 \,{N}^{2} 
,\quad  
    \left\langle {\tau_{{1}}}^{3}\tau_{{\frac 5 2 }} \right\rangle =2\,N \left( {N}^{2}+1 \right)  
 ,\quad  \left\langle {\tau_{{0}}}^{2}\tau_{{\frac 3 2 }}\tau_{{\frac 5 2 }} \right\rangle =\frac 3 2 \,{N}^{2} 
,\quad
\left\langle \tau_{{\frac 1 2 }}\tau_{{1}}\tau_{{\frac 3 2 }}\tau_{{\frac 5 2 }} \right\rangle =N \left( 2\,{N}^{2}+1 \right)  
,\quad 
    \left\langle \tau_{{0}}{\tau_{{\frac 3 2 }}}^{2}\tau_{{\frac 5 2 }} \right\rangle =\frac 1 4 \,N \left( 8\,{N}^{2}+3 \right)  
 ,\\ &  \left\langle {\tau_{{\frac 3 2 }}}^{3}\tau_{{\frac 5 2 }} \right\rangle =5/8\,{N}^{2} \left( 4\,{N}^{2}+5 \right)  
 ,\quad  \left\langle {\tau_{{\frac 1 2 }}}^{2}\tau_{{2}}\tau_{{\frac 5 2 }} \right\rangle =\frac 1 6 \,N \left( 12\,{N}^{2}+5 \right)  
,\quad
    \left\langle \tau_{{0}}\tau_{{1}}\tau_{{2}}\tau_{{\frac 5 2 }} \right\rangle =\frac 2 3 \,N \left( 3\,{N}^{2}+2 \right)  
 ,\quad  \left\langle \tau_{{1}}\tau_{{\frac 3 2 }}\tau_{{2}}\tau_{{\frac 5 2 }} \right\rangle ={\frac {5}{12}}\,{N}^{2} \left( 6\,{N}^{2}+11 \right)  
 ,\\ &  \left\langle \tau_{{\frac 1 2 }}{\tau_{{2}}}^{2}\tau_{{\frac 5 2 }} \right\rangle =\frac 1 8 \,{N}^{2} \left( 20\,{N}^{2}+31 \right)  
 ,\quad  
    \left\langle {\tau_{{2}}}^{3}\tau_{{\frac 5 2 }} \right\rangle =\frac 1 {24} \,N \left( 72\,{N}^{4}+258\,{N}^{2}+73 \right)  
 ,\quad
 \left\langle \tau_{{0}}\tau_{{\frac 1 2 }}{\tau_{{\frac 5 2 }}}^{2} \right\rangle =\frac 1 2 \,N \left( 2\,{N}^{2}+1 \right)  
 ,\quad  \left\langle {\tau_{{1}}}^{2}{\tau_{{\frac 5 2 }}}^{2} \right\rangle =\frac 5 4 \,{N}^{2} \left( {N}^{2}+3 \right)  
,\\ &
    \left\langle \tau_{{\frac 1 2 }}\tau_{{\frac 3 2 }}{\tau_{{\frac 5 2 }}}^{2} \right\rangle =\frac 1 4 \,{N}^{2} \left( 5\,{N}^{2}+9 \right)  
 ,\quad  \left\langle \tau_{{0}}\tau_{{2}}{\tau_{{\frac 5 2 }}}^{2} \right\rangle =\frac 1 {12}\,{N}^{2} \left( 15\,{N}^{2}+31 \right)  
,\quad 
 \left\langle \tau_{{\frac 3 2 }}\tau_{{2}}{\tau_{{\frac 5 2 }}}^{2} \right\rangle =\frac 1 6 \,N \left( 9\,{N}^{4}+38\,{N}^{2}+10 \right)  
,\quad 
    \left\langle \tau_{{1}}{\tau_{{\frac 5 2 }}}^{3} \right\rangle =\frac 1 4 \,N \left( {N}^{2}+6 \right)  \left( 3\,{N}^{2}+1 \right)  
 ,\\ &  \left\langle {\tau_{{\frac 5 2 }}}^{4} \right\rangle =\frac {{N}^{2} \left( 14\,{N}^{4}+157\,{N}^{2}+177 \right)}{32} 
 ,\quad  \left\langle \tau_{{0}}{\tau_{{\frac 1 2 }}}^{2}\tau_{{3}} \right\rangle ={N}^{2} 
,\quad
    \left\langle {\tau_{{0}}}^{2}\tau_{{1}}\tau_{{3}} \right\rangle =\frac 1 8 +\frac 3 2 \,{N}^{2} 
 ,\quad  \left\langle \tau_{{\frac 1 2 }}{\tau_{{1}}}^{2}\tau_{{3}} \right\rangle =\frac{N \left( 4\,{N}^{2}+3 \right)}{2}  
 ,\quad  \left\langle {\tau_{{\frac 1 2 }}}^{2}\tau_{{\frac 3 2 }}\tau_{{3}} \right\rangle =\frac{N \left( 36\,{N}^{2}+19 \right) }{24} 
,\\ &
    \left\langle \tau_{{0}}\tau_{{1}}\tau_{{\frac 3 2 }}\tau_{{3}} \right\rangle =\frac 1 6 \,N \left( 12\,{N}^{2}+7 \right)  
 ,\quad  \left\langle \tau_{{1}}{\tau_{{\frac 3 2 }}}^{2}\tau_{{3}} \right\rangle ={\frac {5}{24}}\,{N}^{2} \left( 12\,{N}^{2}+19 \right)  
 ,\quad  \left\langle \tau_{{0}}\tau_{{\frac 1 2 }}\tau_{{2}}\tau_{{3}} \right\rangle =\frac 1 {24} \,N \left( 36\,{N}^{2}+19 \right)  
,\quad
    \left\langle {\tau_{{1}}}^{2}\tau_{{2}}\tau_{{3}} \right\rangle ={\frac {29}{288}}+\frac 5 2 \,{N}^{4}+{\frac {65}{12}}\,{N}^{2} 
 ,\\ & \left\langle \tau_{{\frac 1 2 }}\tau_{{\frac 3 2 }}\tau_{{2}}\tau_{{3}} \right\rangle =\frac 2 3 \,{N}^{2} \left( 3\,{N}^{2}+5 \right)  
 ,\quad  \left\langle \tau_{{0}}{\tau_{{2}}}^{2}\tau_{{3}} \right\rangle ={\frac {5}{72}}+2\,{N}^{4}+\frac {11} 3\,{N}^{2} 
,\quad \left\langle \tau_{{\frac 3 2 }}{\tau_{{2}}}^{2}\tau_{{3}} \right\rangle ={\frac {1}{288}}\,N \left( 720\,{N}^{4}+2688\,{N}^{2}+731 \right)  
 ,\quad  \left\langle {\tau_{{0}}}^{2}\tau_{{\frac 5 2 }}\tau_{{3}} \right\rangle =\frac 1 8 \,N \left( 8\,{N}^{2}+5 \right)  
 ,\\ & \left\langle \tau_{{\frac 1 2 }}\tau_{{1}}\tau_{{\frac 5 2 }}\tau_{{3}} \right\rangle ={\frac {5}{16}}\,{N}^{2} \left( 4\,{N}^{2}+9 \right)  
,\quad  \left\langle \tau_{{0}}\tau_{{\frac 3 2 }}\tau_{{\frac 5 2 }}\tau_{{3}} \right\rangle =\frac 1{48}\,{N}^{2} \left( 60\,{N}^{2}+109 \right)  
 ,\quad  \left\langle {\tau_{{\frac 3 2 }}}^{2}\tau_{{\frac 5 2 }}\tau_{{3}} \right\rangle ={\frac {1}{96}}\,N \left( 144\,{N}^{4}+532\,{N}^{2}+137 \right)  
 ,\\ &  \left\langle \tau_{{1}}\tau_{{2}}\tau_{{\frac 5 2 }}\tau_{{3}} \right\rangle =\frac 1 {24} \,N \left( 36\,{N}^{4}+171\,{N}^{2}+56 \right)  
,\quad
    \left\langle \tau_{{\frac 1 2 }}{\tau_{{\frac 5 2 }}}^{2}\tau_{{3}} \right\rangle =\frac 1{16} \,N \left( 12\,{N}^{4}+58\,{N}^{2}+17 \right)  
 ,\quad  \left\langle \tau_{{2}}{\tau_{{\frac 5 2 }}}^{2}\tau_{{3}} \right\rangle ={\frac {1}{96}}\,{N}^{2} \left( 84\,{N}^{4}+725\,{N}^{2}+789 \right)  
 ,\\ &  \left\langle {\tau_{{\frac 1 2 }}}^{2}{\tau_{{3}}}^{2} \right\rangle =\frac 1 {12}\,{N}^{2} \left( 12\,{N}^{2}+25 \right)  
,\quad
    \left\langle \tau_{{0}}\tau_{{1}}{\tau_{{3}}}^{2} \right\rangle ={\frac {29}{576}}+\frac 5 4 \,{N}^{4}+{\frac {65}{24}}\,{N}^{2} 
 ,\quad  \left\langle \tau_{{1}}\tau_{{\frac 3 2 }}{\tau_{{3}}}^{2} \right\rangle ={\frac {1}{96}}\,N \left( 144\,{N}^{4}+600\,{N}^{2}+193 \right)  
 ,\\ & \left\langle \tau_{{\frac 1 2 }}\tau_{{2}}{\tau_{{3}}}^{2} \right\rangle ={\frac {1}{576}}\,N \left( 720\,{N}^{4}+3096\,{N}^{2}+925 \right)  
,\quad
    \left\langle {\tau_{{2}}}^{2}{\tau_{{3}}}^{2} \right\rangle =\frac 3 2 \,{N}^{6}+{\frac {275}{24}}\,{N}^{4}+{\frac {1151}{96}}\,{N}^{2}+{\frac {205}{3456}} 
 ,\quad  \left\langle \tau_{{0}}\tau_{{\frac 5 2 }}{\tau_{{3}}}^{2} \right\rangle ={\frac {1}{576}}\,N \left( 432\,{N}^{4}+1968\,{N}^{2}+641 \right)  
 ,\\ & \left\langle \tau_{{\frac 3 2 }}\tau_{{\frac 5 2 }}{\tau_{{3}}}^{2} \right\rangle ={\frac {{N}^{2} \left( 1008\,{N}^{4}+7704\,{N}^{2}+8147 \right)  }{1152}}
,\quad \left\langle \tau_{{1}}{\tau_{{3}}}^{3} \right\rangle ={\frac {7}{8}}\,{N}^{6}+{\frac {231}{32}}\,{N}^{4}+{\frac {3367}{384}}\,{N}^{2}+{\frac {583}{13824}} 
 ,\quad  \left\langle \tau_{{\frac 5 2 }}{\tau_{{3}}}^{3} \right\rangle ={\frac {N \left( 6912\,{N}^{6}+92016\,{N}^{4}+239544\,{N}^{2}+56671 \right)}{13824}}  
 ,\\ & \left\langle {\tau_{{0}}}^{2}\tau_{{\frac 1 2 }}\tau_{{\frac 7 2 }} \right\rangle =\frac{N^2}{2} 
,\quad  \left\langle {\tau_{{\frac 1 2 }}}^{2}\tau_{{1}}\tau_{{\frac 7 2 }} \right\rangle =\frac{N \left( 3\,{N}^{2}+2 \right)} {3}   
 ,\quad  \left\langle \tau_{{0}}{\tau_{{1}}}^{2}\tau_{{\frac 7 2 }} \right\rangle =N \left( {N}^{2}+1 \right)  
 ,\quad  \left\langle \tau_{{0}}\tau_{{\frac 1 2 }}\tau_{{\frac 3 2 }}\tau_{{\frac 7 2 }} \right\rangle =\frac{N \left( 9\,{N}^{2}+5 \right)}{12}  
,
\end{align*}
\begin{align*}
&
    \left\langle {\tau_{{1}}}^{2}\tau_{{\frac 3 2 }}\tau_{{\frac 7 2 }} \right\rangle ={\frac {5}{12}}\,{N}^{2} \left( 3\,{N}^{2}+7 \right)  
 ,\quad  \left\langle \tau_{{\frac 1 2 }}{\tau_{{\frac 3 2 }}}^{2}\tau_{{\frac 7 2 }} \right\rangle =\frac 1 {24} \,{N}^{2} \left( 24\,{N}^{2}+41 \right)  
 ,\quad  \left\langle {\tau_{{0}}}^{2}\tau_{{2}}\tau_{{\frac 7 2 }} \right\rangle =\frac 1 {12}\,N \left( 9\,{N}^{2}+7 \right)  
,\quad
    \left\langle \tau_{{\frac 1 2 }}\tau_{{1}}\tau_{{2}}\tau_{{\frac 7 2 }} \right\rangle =\frac 5 4 \,{N}^{2} \left( {N}^{2}+2 \right)  
 ,\\&  \left\langle \tau_{{0}}\tau_{{\frac 3 2 }}\tau_{{2}}\tau_{{\frac 7 2 }} \right\rangle ={N}^{2} \left( {N}^{2}+2 \right)  
 ,\quad  \left\langle {\tau_{{\frac 3 2 }}}^{2}\tau_{{2}}\tau_{{\frac 7 2 }} \right\rangle =\frac 1 {24} \,N \left( 30\,{N}^{4}+118\,{N}^{2}+29 \right)  
,\quad
    \left\langle \tau_{{1}}{\tau_{{2}}}^{2}\tau_{{\frac 7 2 }} \right\rangle =\frac 1 {24} \,N \left( 36\,{N}^{4}+153\,{N}^{2}+49 \right)  
 ,\\&  \left\langle {\tau_{{\frac 1 2 }}}^{2}\tau_{{\frac 5 2 }}\tau_{{\frac 7 2 }} \right\rangle =\frac{5}{8}\,{N}^{2} \left( {N}^{2}+2 \right)  
 ,\quad  \left\langle \tau_{{0}}\tau_{{1}}\tau_{{\frac 5 2 }}\tau_{{\frac 7 2 }} \right\rangle ={\frac {5}{24}}\,{N}^{2} \left( 3\,{N}^{2}+8 \right)  
,\quad
    \left\langle \tau_{{1}}\tau_{{\frac 3 2 }}\tau_{{\frac 5 2 }}\tau_{{\frac 7 2 }} \right\rangle =\frac{3}{8}\,N \left( 2\,{N}^{4}+10\,{N}^{2}+3 \right)  
 ,\\ &  \left\langle \tau_{{\frac 1 2 }}\tau_{{2}}\tau_{{\frac 5 2 }}\tau_{{\frac 7 2 }} \right\rangle ={\frac {1}{72}}\,N \left( 54\,{N}^{4}+234\,{N}^{2}+67 \right)  
 ,\quad  \left\langle {\tau_{{2}}}^{2}\tau_{{\frac 5 2 }}\tau_{{\frac 7 2 }} \right\rangle ={\frac {1}{288}}\,{N}^{2} \left( 252\,{N}^{4}+1959\,{N}^{2}+2080 \right)  
,\quad
    \left\langle \tau_{{0}}{\tau_{{\frac 5 2 }}}^{2}\tau_{{\frac 7 2 }} \right\rangle =\frac 1 {24} \,N \left( 9\,{N}^{4}+49\,{N}^{2}+15 \right)  
 ,\\&  \left\langle \tau_{{\frac 3 2 }}{\tau_{{\frac 5 2 }}}^{2}\tau_{{\frac 7 2 }} \right\rangle ={\frac {1}{96}}\,{N}^{2} \left( 42\,{N}^{4}+377\,{N}^{2}+401 \right)  
 ,\quad  \left\langle \tau_{{0}}\tau_{{\frac 1 2 }}\tau_{{3}}\tau_{{\frac 7 2 }} \right\rangle =\frac 1{16} \,{N}^{2} \left( 8\,{N}^{2}+17 \right)  
,\quad
    \left\langle {\tau_{{1}}}^{2}\tau_{{3}}\tau_{{\frac 7 2 }} \right\rangle =\frac 1{48}\,N \left( 36\,{N}^{4}+195\,{N}^{2}+79 \right)  
 ,\\&  \left\langle \tau_{{\frac 1 2 }}\tau_{{\frac 3 2 }}\tau_{{3}}\tau_{{\frac 7 2 }} \right\rangle ={\frac {1}{288}}\,N \left( 180\,{N}^{4}+783\,{N}^{2}+239 \right)  
 ,\quad  \left\langle \tau_{{0}}\tau_{{2}}\tau_{{3}}\tau_{{\frac 7 2 }} \right\rangle ={\frac {1}{288}}\,N \left( 180\,{N}^{4}+843\,{N}^{2}+289 \right)  
,\quad
    \left\langle \tau_{{\frac 3 2 }}\tau_{{2}}\tau_{{3}}\tau_{{\frac 7 2 }} \right\rangle =\frac 1{48}\,{N}^{2} \left( 36\,{N}^{4}+283\,{N}^{2}+297 \right)  
 ,\\&  \left\langle \tau_{{1}}\tau_{{\frac 5 2 }}\tau_{{3}}\tau_{{\frac 7 2 }} \right\rangle ={\frac {7{N}^{2} \left( 12\,{N}^{4}+115\,{N}^{2}+144 \right)  }{192}}
 ,\quad  \left\langle {\tau_{{\frac 5 2 }}}^{2}\tau_{{3}}\tau_{{\frac 7 2 }} \right\rangle ={\frac {N \left( 144\,{N}^{6}+2199\,{N}^{4}+5963\,{N}^{2}+1373 \right)  }{576}}
,\quad
    \left\langle \tau_{{\frac 1 2 }}{\tau_{{3}}}^{2}\tau_{{\frac 7 2 }} \right\rangle ={\frac {{N}^{2} \left( 432\,{N}^{4}+3696\,{N}^{2}+4313 \right)}{1152}}  
 ,\\&  \left\langle \tau_{{2}}{\tau_{{3}}}^{2}\tau_{{\frac 7 2 }} \right\rangle ={\frac {1}{6912}}\,N \left( 3024\,{N}^{6}+40968\,{N}^{4}+104997\,{N}^{2}+24955 \right)  
 ,\quad  \left\langle {\tau_{{0}}}^{2}{\tau_{{\frac 7 2 }}}^{2} \right\rangle =\frac 1 {12}\,{N}^{2} \left( 3\,{N}^{2}+8 \right)  
,\quad \left\langle \tau_{{\frac 1 2 }}\tau_{{1}}{\tau_{{\frac 7 2 }}}^{2} \right\rangle =\frac 1 {24} \,N \left( 9\,{N}^{4}+45\,{N}^{2}+16 \right)  
 ,\\&  \left\langle \tau_{{0}}\tau_{{\frac 3 2 }}{\tau_{{\frac 7 2 }}}^{2} \right\rangle ={\frac {5}{144}}\,N \left( 3\,{N}^{2}+14 \right)  \left( 3\,{N}^{2}+1 \right)  
 ,\quad  \left\langle {\tau_{{\frac 3 2 }}}^{2}{\tau_{{\frac 7 2 }}}^{2} \right\rangle ={\frac {1}{72}}\,{N}^{2} \left( 27\,{N}^{4}+222\,{N}^{2}+226 \right)  
,\quad
    \left\langle \tau_{{1}}\tau_{{2}}{\tau_{{\frac 7 2 }}}^{2} \right\rangle ={\frac {7}{144}}\,{N}^{2} \left( 9\,{N}^{4}+78\,{N}^{2}+95 \right)  
 ,\\&  \left\langle \tau_{{\frac 1 2 }}\tau_{{\frac 5 2 }}{\tau_{{\frac 7 2 }}}^{2} \right\rangle ={\frac {1}{96}}\,{N}^{2} \left( 21\,{N}^{4}+185\,{N}^{2}+214 \right)  
 ,\quad  \left\langle \tau_{{2}}\tau_{{\frac 5 2 }}{\tau_{{\frac 7 2 }}}^{2} \right\rangle ={\frac {1}{864}}\,N \left( 216\,{N}^{6}+2997\,{N}^{4}+7893\,{N}^{2}+1802 \right)  
,\quad
    \left\langle \tau_{{0}}\tau_{{3}}{\tau_{{\frac 7 2 }}}^{2} \right\rangle ={\frac {1}{288}}\,{N}^{2} \left( 54\,{N}^{4}+501\,{N}^{2}+622 \right)  
 ,\\&  \left\langle \tau_{{\frac 3 2 }}\tau_{{3}}{\tau_{{\frac 7 2 }}}^{2} \right\rangle ={\frac {1}{3456}}\,N \left( 756\,{N}^{6}+10539\,{N}^{4}+27099\,{N}^{2}+6226 \right)  
,\quad
\left\langle {\tau_{{3}}}^{2}{\tau_{{\frac 7 2 }}}^{2} \right\rangle ={\frac {1}{2304}}\,{N}^{2} \left( 288\,{N}^{6}+6228\,{N}^{4}+31129\,{N}^{2}+27236 \right)  
,\\ &
    \left\langle \tau_{{1}}{\tau_{{\frac 7 2 }}}^{3} \right\rangle ={\frac {1}{216}}\,N \left( 27\,{N}^{6}+405\,{N}^{4}+1188\,{N}^{2}+305 \right)
,\quad
 \left\langle \tau_{{\frac 5 2 }}{\tau_{{\frac 7 2 }}}^{3} \right\rangle ={\frac {1}{1152}}\,{N}^{2} \left( 81\,{N}^{6}+1791\,{N}^{4}+9294\,{N}^{2}+8084 \right).
 \end{align*}

\end{tiny}
{\bf Nonzero five--point correlators $\le\langle\tau_{\frac{d_1}{2}}\dots\tau_{\frac{d_5}{2}}\ri\rangle$, for $0\leq d_1 \leq \dots\leq d_5 \leq 6$}
\begin{tiny}
% !TEX root =penner-1.4.tex

\begin{align*}
&
\left\langle {\tau_{{0}}}^{3}{\tau_{{1}}}^{2} \right\rangle =2 
,\quad  
    \left\langle {\tau_{{\frac 1 2 }}}^{3}{\tau_{{1}}}^{2} \right\rangle =6\,N 
 ,\quad  \left\langle \tau_{{0}}\tau_{{\frac 1 2 }}{\tau_{{1}}}^{3} \right\rangle =6\,N 
 ,\quad  \left\langle {\tau_{{1}}}^{5} \right\rangle =12\,{N}^{2}+1 
,\quad 
    \left\langle {\tau_{{\frac 1 2 }}}^{4}\tau_{{\frac 3 2 }} \right\rangle =3\,N 
 ,\quad  \left\langle \tau_{{0}}{\tau_{{\frac 1 2 }}}^{2}\tau_{{1}}\tau_{{\frac 3 2 }} \right\rangle =3\,N 
 ,\quad  \left\langle {\tau_{{0}}}^{2}{\tau_{{1}}}^{2}\tau_{{\frac 3 2 }} \right\rangle =6\,N 
, \\ & 
    \left\langle \tau_{{\frac 1 2 }}{\tau_{{1}}}^{3}\tau_{{\frac 3 2 }} \right\rangle =12\,{N}^{2} 
 ,\quad  \left\langle {\tau_{{0}}}^{2}\tau_{{\frac 1 2 }}{\tau_{{\frac 3 2 }}}^{2} \right\rangle =2\,N 
 ,\quad  \left\langle {\tau_{{\frac 1 2 }}}^{2}\tau_{{1}}{\tau_{{\frac 3 2 }}}^{2} \right\rangle =8\,{N}^{2} 
,\quad  
    \left\langle \tau_{{0}}{\tau_{{1}}}^{2}{\tau_{{\frac 3 2 }}}^{2} \right\rangle =12\,{N}^{2} 
 ,\quad  \left\langle \tau_{{0}}\tau_{{\frac 1 2 }}{\tau_{{\frac 3 2 }}}^{3} \right\rangle =6\,{N}^{2} 
 ,\quad  \left\langle {\tau_{{1}}}^{2}{\tau_{{\frac 3 2 }}}^{3} \right\rangle =5\,N \left( 4\,{N}^{2}+1 \right)  
, \\ & 
    \left\langle \tau_{{\frac 1 2 }}{\tau_{{\frac 3 2 }}}^{4} \right\rangle =3\,N \left( 4\,{N}^{2}+1 \right)  
 ,\quad  \left\langle {\tau_{{0}}}^{4}\tau_{{2}} \right\rangle =1 
 ,\quad  \left\langle \tau_{{0}}{\tau_{{\frac 1 2 }}}^{3}\tau_{{2}} \right\rangle =2\,N 
,\quad 
    \left\langle {\tau_{{0}}}^{2}\tau_{{\frac 1 2 }}\tau_{{1}}\tau_{{2}} \right\rangle =3\,N 
 ,\quad  \left\langle {\tau_{{\frac 1 2 }}}^{2}{\tau_{{1}}}^{2}\tau_{{2}} \right\rangle =12\,{N}^{2} 
 ,\quad  \left\langle \tau_{{0}}{\tau_{{1}}}^{3}\tau_{{2}} \right\rangle =12\,{N}^{2}+1 
, \\ & 
    \left\langle {\tau_{{0}}}^{3}\tau_{{\frac 3 2 }}\tau_{{2}} \right\rangle =3\,N 
 ,\quad  \left\langle {\tau_{{\frac 1 2 }}}^{3}\tau_{{\frac 3 2 }}\tau_{{2}} \right\rangle =7\,{N}^{2} 
 ,\quad  \left\langle \tau_{{0}}\tau_{{\frac 1 2 }}\tau_{{1}}\tau_{{\frac 3 2 }}\tau_{{2}} \right\rangle =8\,{N}^{2} 
,\quad 
    \left\langle {\tau_{{1}}}^{3}\tau_{{\frac 3 2 }}\tau_{{2}} \right\rangle =10\,N \left( 2\,{N}^{2}+1 \right)  
 ,\quad  \left\langle {\tau_{{0}}}^{2}{\tau_{{\frac 3 2 }}}^{2}\tau_{{2}} \right\rangle =7\,{N}^{2} 
, \\ &
 \left\langle \tau_{{\frac 1 2 }}\tau_{{1}}{\tau_{{\frac 3 2 }}}^{2}\tau_{{2}} \right\rangle =5\,N \left( 3\,{N}^{2}+1 \right)  
,\quad  
    \left\langle \tau_{{0}}{\tau_{{\frac 3 2 }}}^{3}\tau_{{2}} \right\rangle =N \left( 13\,{N}^{2}+4 \right)  
 ,\quad  \left\langle {\tau_{{\frac 3 2 }}}^{4}\tau_{{2}} \right\rangle ={N}^{2} \left( 21\,{N}^{2}+22 \right)  
 ,\quad  \left\langle \tau_{{0}}{\tau_{{\frac 1 2 }}}^{2}{\tau_{{2}}}^{2} \right\rangle =6\,{N}^{2} 
, \quad 
    \left\langle {\tau_{{0}}}^{2}\tau_{{1}}{\tau_{{2}}}^{2} \right\rangle =\frac 2 3 +8\,{N}^{2} 
 ,\\&  \left\langle \tau_{{\frac 1 2 }}{\tau_{{1}}}^{2}{\tau_{{2}}}^{2} \right\rangle =\frac 5 3 \,N \left( 12\,{N}^{2}+5 \right)  
 ,\quad  \left\langle {\tau_{{\frac 1 2 }}}^{2}\tau_{{\frac 3 2 }}{\tau_{{2}}}^{2} \right\rangle =\frac 1 {12}\,N \left( 156\,{N}^{2}+53 \right)  
, \quad
    \left\langle \tau_{{0}}\tau_{{1}}\tau_{{\frac 3 2 }}{\tau_{{2}}}^{2} \right\rangle ={\frac {5}{12}}\,N \left( 36\,{N}^{2}+17 \right)  
 ,\quad  \left\langle \tau_{{1}}{\tau_{{\frac 3 2 }}}^{2}{\tau_{{2}}}^{2} \right\rangle =\frac 1 2 \,{N}^{2} \left( 48\,{N}^{2}+65 \right)  
 ,\\ & \left\langle \tau_{{0}}\tau_{{\frac 1 2 }}{\tau_{{2}}}^{3} \right\rangle =N \left( 12\,{N}^{2}+5 \right)  
, \quad
    \left\langle {\tau_{{1}}}^{2}{\tau_{{2}}}^{3} \right\rangle =30\,{N}^{4}+45\,{N}^{2}+{\frac {7}{8}} 
 ,\quad  \left\langle \tau_{{\frac 1 2 }}\tau_{{\frac 3 2 }}{\tau_{{2}}}^{3} \right\rangle =\frac 1 4 \,{N}^{2} \left( 84\,{N}^{2}+109 \right)  
 ,\quad  \left\langle \tau_{{0}}{\tau_{{2}}}^{4} \right\rangle =20\,{N}^{4}+30\,{N}^{2}+{\frac {7}{12}} 
, \\ & 
    \left\langle \tau_{{\frac 3 2 }}{\tau_{{2}}}^{4} \right\rangle =\frac 1 {48}\,N \left( 1488\,{N}^{4}+4584\,{N}^{2}+1129 \right)  
 ,\quad  \left\langle {\tau_{{0}}}^{2}{\tau_{{\frac 1 2 }}}^{2}\tau_{{\frac 5 2 }} \right\rangle =N 
 ,\quad  \left\langle {\tau_{{0}}}^{3}\tau_{{1}}\tau_{{\frac 5 2 }} \right\rangle =3\,N 
, \quad
    \left\langle {\tau_{{\frac 1 2 }}}^{3}\tau_{{1}}\tau_{{\frac 5 2 }} \right\rangle =6\,{N}^{2} 
 ,\quad  \left\langle \tau_{{0}}\tau_{{\frac 1 2 }}{\tau_{{1}}}^{2}\tau_{{\frac 5 2 }} \right\rangle =6\,{N}^{2} 
 ,\\&  \left\langle {\tau_{{1}}}^{4}\tau_{{\frac 5 2 }} \right\rangle =10\,N \left( {N}^{2}+1 \right)  
, \quad
    \left\langle \tau_{{0}}{\tau_{{\frac 1 2 }}}^{2}\tau_{{\frac 3 2 }}\tau_{{\frac 5 2 }} \right\rangle =\frac 7 2 \,{N}^{2} 
 ,\quad  \left\langle {\tau_{{0}}}^{2}\tau_{{1}}\tau_{{\frac 3 2 }}\tau_{{\frac 5 2 }} \right\rangle =6\,{N}^{2} 
 ,\quad  \left\langle \tau_{{\frac 1 2 }}{\tau_{{1}}}^{2}\tau_{{\frac 3 2 }}\tau_{{\frac 5 2 }} \right\rangle =5\,N \left( 2\,{N}^{2}+1 \right)  
, \\ & 
    \left\langle {\tau_{{\frac 1 2 }}}^{2}{\tau_{{\frac 3 2 }}}^{2}\tau_{{\frac 5 2 }} \right\rangle =\frac 1 4 \,N \left( 28\,{N}^{2}+11 \right)  
 ,\quad  \left\langle \tau_{{0}}\tau_{{1}}{\tau_{{\frac 3 2 }}}^{2}\tau_{{\frac 5 2 }} \right\rangle =\frac 5 4 \,N \left( 8\,{N}^{2}+3 \right)  
 ,\quad  \left\langle \tau_{{1}}{\tau_{{\frac 3 2 }}}^{3}\tau_{{\frac 5 2 }} \right\rangle ={\frac {15}{4}}\,{N}^{2} \left( 4\,{N}^{2}+5 \right)  
, \quad
    \left\langle {\tau_{{0}}}^{2}\tau_{{\frac 1 2 }}\tau_{{2}}\tau_{{\frac 5 2 }} \right\rangle =\frac 7 2 \,{N}^{2} 
 ,\\& \left\langle {\tau_{{\frac 1 2 }}}^{2}\tau_{{1}}\tau_{{2}}\tau_{{\frac 5 2 }} \right\rangle =\frac 5 6 \,N \left( 12\,{N}^{2}+5 \right)  
 ,\quad  \left\langle \tau_{{0}}{\tau_{{1}}}^{2}\tau_{{2}}\tau_{{\frac 5 2 }} \right\rangle =\frac {10} 3\,N \left( 3\,{N}^{2}+2 \right)  
, \quad     \left\langle \tau_{{0}}\tau_{{\frac 1 2 }}\tau_{{\frac 3 2 }}\tau_{{2}}\tau_{{\frac 5 2 }} \right\rangle =\frac 1 6 \,N \left( 42\,{N}^{2}+17 \right)  
 ,\quad  \left\langle {\tau_{{1}}}^{2}\tau_{{\frac 3 2 }}\tau_{{2}}\tau_{{\frac 5 2 }} \right\rangle =\frac 5 2 \,{N}^{2} \left( 6\,{N}^{2}+11 \right)  
 ,\\&  \left\langle \tau_{{\frac 1 2 }}{\tau_{{\frac 3 2 }}}^{2}\tau_{{2}}\tau_{{\frac 5 2 }} \right\rangle =\frac 1 {12}\,{N}^{2} \left( 138\,{N}^{2}+191 \right)  
, \quad 
    \left\langle {\tau_{{0}}}^{2}{\tau_{{2}}}^{2}\tau_{{\frac 5 2 }} \right\rangle ={\frac {7}{12}}\,N \left( 12\,{N}^{2}+7 \right)  
 ,\quad  \left\langle \tau_{{\frac 1 2 }}\tau_{{1}}{\tau_{{2}}}^{2}\tau_{{\frac 5 2 }} \right\rangle =\frac 3 4 \,{N}^{2} \left( 20\,{N}^{2}+31 \right)  
 ,\\& \left\langle \tau_{{0}}\tau_{{\frac 3 2 }}{\tau_{{2}}}^{2}\tau_{{\frac 5 2 }} \right\rangle =\frac 1 {24}\,{N}^{2} \left( 276\,{N}^{2}+443 \right)  
\quad
    \left\langle {\tau_{{\frac 3 2 }}}^{2}{\tau_{{2}}}^{2}\tau_{{\frac 5 2 }} \right\rangle =\frac 1 {48}\,N \left( 816\,{N}^{4}+2716\,{N}^{2}+623 \right)  
 ,\quad  \left\langle \tau_{{1}}{\tau_{{2}}}^{3}\tau_{{\frac 5 2 }} \right\rangle ={\frac {7}{24}}\,N \left( 72\,{N}^{4}+258\,{N}^{2}+73 \right)  
 ,\quad  \left\langle {\tau_{{0}}}^{3}{\tau_{{\frac 5 2 }}}^{2} \right\rangle =3\,{N}^{2} 
, 
\\ & 
    \left\langle {\tau_{{\frac 1 2 }}}^{3}{\tau_{{\frac 5 2 }}}^{2} \right\rangle =N \left( 5\,{N}^{2}+2 \right)  
 ,\quad  \left\langle \tau_{{0}}\tau_{{\frac 1 2 }}\tau_{{1}}{\tau_{{\frac 5 2 }}}^{2} \right\rangle =\frac 5 2 \,N \left( 2\,{N}^{2}+1 \right)  
 ,\quad  \left\langle {\tau_{{1}}}^{3}{\tau_{{\frac 5 2 }}}^{2} \right\rangle =\frac {15} 2\,{N}^{2} \left( {N}^{2}+3 \right)  
,\quad\left\langle {\tau_{{0}}}^{2}\tau_{{\frac 3 2 }}{\tau_{{\frac 5 2 }}}^{2} \right\rangle =N \left( 5\,{N}^{2}+2 \right),
\end{align*}
%%%%%%%%%%%%%%%%%%%%%%%%%%%%%%%%%%%%%%%%%%%%%%%%%%%%%%%%%%%%
\begin{align*} 
    & \left\langle \tau_{{\frac 1 2 }}\tau_{{1}}\tau_{{\frac 3 2 }}{\tau_{{\frac 5 2 }}}^{2} \right\rangle =\frac 3 2 \,{N}^{2} \left( 5\,{N}^{2}+9 \right)  
 ,\quad  \left\langle \tau_{{0}}{\tau_{{\frac 3 2 }}}^{2}{\tau_{{\frac 5 2 }}}^{2} \right\rangle =\frac 1 4 \,{N}^{2} \left( 30\,{N}^{2}+43 \right)  
, \quad
    \left\langle {\tau_{{\frac 3 2 }}}^{3}{\tau_{{\frac 5 2 }}}^{2} \right\rangle =\frac 1 4 \,N \left( 42\,{N}^{4}+133\,{N}^{2}+30 \right)  
 ,\quad  \left\langle {\tau_{{\frac 1 2 }}}^{2}\tau_{{2}}{\tau_{{\frac 5 2 }}}^{2} \right\rangle =\frac 1 2 \,{N}^{2} \left( 15\,{N}^{2}+23 \right)  
 ,\\&  \left\langle \tau_{{0}}\tau_{{1}}\tau_{{2}}{\tau_{{\frac 5 2 }}}^{2} \right\rangle =\frac 1 2 \,{N}^{2} \left( 15\,{N}^{2}+31 \right)  
, \quad
    \left\langle \tau_{{1}}\tau_{{\frac 3 2 }}\tau_{{2}}{\tau_{{\frac 5 2 }}}^{2} \right\rangle =\frac{7N \left( 9\,{N}^{4}+38\,{N}^{2}+10 \right) }{6} 
 ,\quad  \left\langle \tau_{{\frac 1 2 }}{\tau_{{2}}}^{2}{\tau_{{\frac 5 2 }}}^{2} \right\rangle ={\frac {1}{72}}\,N \left( 756\,{N}^{4}+2742\,{N}^{2}+691 \right)  
 ,\\&  \left\langle {\tau_{{2}}}^{3}{\tau_{{\frac 5 2 }}}^{2} \right\rangle =\frac {{N}^{2} \left( 336\,{N}^{4}+2292\,{N}^{2}+2177 \right)  }{24}
, \quad
    \left\langle \tau_{{0}}\tau_{{\frac 1 2 }}{\tau_{{\frac 5 2 }}}^{3} \right\rangle =\frac{3{N}^{2} \left( 5\,{N}^{2}+9 \right)  }{4}
 ,\quad  \left\langle {\tau_{{1}}}^{2}{\tau_{{\frac 5 2 }}}^{3} \right\rangle =\frac{7N \left( {N}^{2}+6 \right)  \left( 3\,{N}^{2}+1 \right) }{4} 
 ,\\&  \left\langle \tau_{{\frac 1 2 }}\tau_{{\frac 3 2 }}{\tau_{{\frac 5 2 }}}^{3} \right\rangle =\frac{3N \left( 14\,{N}^{4}+58\,{N}^{2}+15 \right)  }{8}
, \quad 
    \left\langle \tau_{{0}}\tau_{{2}}{\tau_{{\frac 5 2 }}}^{3} \right\rangle =\frac 1 4 \,N \left( 21\,{N}^{4}+95\,{N}^{2}+26 \right)  
 ,\quad  \left\langle \tau_{{\frac 3 2 }}\tau_{{2}}{\tau_{{\frac 5 2 }}}^{3} \right\rangle ={\frac {7}{8}}\,{N}^{2} \left( 8\,{N}^{4}+63\,{N}^{2}+60 \right)  
 ,\\&  \left\langle \tau_{{1}}{\tau_{{\frac 5 2 }}}^{4} \right\rangle =\frac 1 4 \,{N}^{2} \left( 14\,{N}^{4}+157\,{N}^{2}+177 \right)  
, \quad
    \left\langle {\tau_{{\frac 5 2 }}}^{5} \right\rangle =1/8\,N \left( 18\,{N}^{6}+321\,{N}^{4}+828\,{N}^{2}+175 \right)  
 ,\quad  \left\langle {\tau_{{0}}}^{3}\tau_{{\frac 1 2 }}\tau_{{3}} \right\rangle =N 
 ,\quad  \left\langle {\tau_{{\frac 1 2 }}}^{4}\tau_{{3}} \right\rangle =4\,{N}^{2} 
, \\ & 
    \left\langle \tau_{{0}}{\tau_{{\frac 1 2 }}}^{2}\tau_{{1}}\tau_{{3}} \right\rangle =4\,{N}^{2} 
 ,\quad  \left\langle {\tau_{{0}}}^{2}{\tau_{{1}}}^{2}\tau_{{3}} \right\rangle =\frac 1 2 +6\,{N}^{2} 
 ,\quad  \left\langle \tau_{{\frac 1 2 }}{\tau_{{1}}}^{3}\tau_{{3}} \right\rangle =\frac 5 2 \,N \left( 4\,{N}^{2}+3 \right)  
,\quad
    \left\langle {\tau_{{0}}}^{2}\tau_{{\frac 1 2 }}\tau_{{\frac 3 2 }}\tau_{{3}} \right\rangle =3\,{N}^{2} 
 ,\quad  \left\langle {\tau_{{\frac 1 2 }}}^{2}\tau_{{1}}\tau_{{\frac 3 2 }}\tau_{{3}} \right\rangle ={\frac {5}{24}}\,N \left( 36\,{N}^{2}+19 \right)  
 ,\\&  \left\langle \tau_{{0}}{\tau_{{1}}}^{2}\tau_{{\frac 3 2 }}\tau_{{3}} \right\rangle =\frac 5 6 \,N \left( 12\,{N}^{2}+7 \right)  
, \quad
    \left\langle \tau_{{0}}\tau_{{\frac 1 2 }}{\tau_{{\frac 3 2 }}}^{2}\tau_{{3}} \right\rangle =\frac 1 {12}\,N \left( 72\,{N}^{2}+31 \right)  
 ,\quad  \left\langle {\tau_{{1}}}^{2}{\tau_{{\frac 3 2 }}}^{2}\tau_{{3}} \right\rangle =\frac 5 4 \,{N}^{2} \left( 12\,{N}^{2}+19 \right)  
 ,\quad  \left\langle \tau_{{\frac 1 2 }}{\tau_{{\frac 3 2 }}}^{3}\tau_{{3}} \right\rangle =\frac 5 4 \,{N}^{2} \left( 8\,{N}^{2}+11 \right)  
, \\ & 
    \left\langle {\tau_{{0}}}^{3}\tau_{{2}}\tau_{{3}} \right\rangle ={\frac {7}{24}}+\frac 7 2 \,{N}^{2} 
 ,\quad  \left\langle {\tau_{{\frac 1 2 }}}^{3}\tau_{{2}}\tau_{{3}} \right\rangle =\frac 1 {12}\,N \left( 84\,{N}^{2}+37 \right)  
 ,\quad  \left\langle \tau_{{0}}\tau_{{\frac 1 2 }}\tau_{{1}}\tau_{{2}}\tau_{{3}} \right\rangle ={\frac {5}{24}}\,N \left( 36\,{N}^{2}+19 \right)  
, \quad 
    \left\langle {\tau_{{1}}}^{3}\tau_{{2}}\tau_{{3}} \right\rangle =15\,{N}^{4}+{\frac {65}{2}}\,{N}^{2}+{\frac {29}{48}} 
 ,\\&  \left\langle {\tau_{{0}}}^{2}\tau_{{\frac 3 2 }}\tau_{{2}}\tau_{{3}} \right\rangle =\frac{1}{8}\,N \left( 52\,{N}^{2}+27 \right)  
 ,\quad  \left\langle \tau_{{\frac 1 2 }}\tau_{{1}}\tau_{{\frac 3 2 }}\tau_{{2}}\tau_{{3}} \right\rangle =4\,{N}^{2} \left( 3\,{N}^{2}+5 \right)  
, \quad 
    \left\langle \tau_{{0}}{\tau_{{\frac 3 2 }}}^{2}\tau_{{2}}\tau_{{3}} \right\rangle ={\frac {7}{24}}\,{N}^{2} \left( 36\,{N}^{2}+55 \right)  
 ,\quad  \left\langle {\tau_{{\frac 3 2 }}}^{3}\tau_{{2}}\tau_{{3}} \right\rangle =\frac 1 {24}\,N \left( 372\,{N}^{4}+1183\,{N}^{2}+262 \right)  
 ,\\&  \left\langle {\tau_{{\frac 1 2 }}}^{2}{\tau_{{2}}}^{2}\tau_{{3}} \right\rangle =\frac 1 {12}\,{N}^{2} \left( 132\,{N}^{2}+203 \right) , 
\quad
    \left\langle \tau_{{0}}\tau_{{1}}{\tau_{{2}}}^{2}\tau_{{3}} \right\rangle =12\,{N}^{4}+{\frac {5}{12}}+22\,{N}^{2} 
 ,\quad  \left\langle \tau_{{1}}\tau_{{\frac 3 2 }}{\tau_{{2}}}^{2}\tau_{{3}} \right\rangle ={\frac {7}{288}}\,N \left( 720\,{N}^{4}+2688\,{N}^{2}+731 \right)  
 ,\\& \left\langle \tau_{{\frac 1 2 }}{\tau_{{2}}}^{3}\tau_{{3}} \right\rangle =\frac 1 3 \,N \left( 48\,{N}^{4}+168\,{N}^{2}+43 \right)  
, \quad
    \left\langle {\tau_{{2}}}^{4}\tau_{{3}} \right\rangle =22\,{N}^{6}+{\frac {857}{6}}\,{N}^{4}+{\frac {3163}{24}}\,{N}^{2}+{\frac {193}{288}} 
 ,\quad  \left\langle \tau_{{0}}{\tau_{{\frac 1 2 }}}^{2}\tau_{{\frac 5 2 }}\tau_{{3}} \right\rangle =\frac{1}{8}\,N \left( 28\,{N}^{2}+13 \right)  
 ,\\&  \left\langle {\tau_{{0}}}^{2}\tau_{{1}}\tau_{{\frac 5 2 }}\tau_{{3}} \right\rangle =\frac 5 8 \,N \left( 8\,{N}^{2}+5 \right)  
, \quad 
    \left\langle \tau_{{\frac 1 2 }}{\tau_{{1}}}^{2}\tau_{{\frac 5 2 }}\tau_{{3}} \right\rangle ={\frac {15}{8}}\,{N}^{2} \left( 4\,{N}^{2}+9 \right)  
 ,\quad  \left\langle {\tau_{{\frac 1 2 }}}^{2}\tau_{{\frac 3 2 }}\tau_{{\frac 5 2 }}\tau_{{3}} \right\rangle =\frac 1 {16} \,{N}^{2} \left( 92\,{N}^{2}+159 \right)  
 ,\quad  \left\langle \tau_{{0}}\tau_{{1}}\tau_{{\frac 3 2 }}\tau_{{\frac 5 2 }}\tau_{{3}} \right\rangle =\frac{1}{8}\,{N}^{2} \left( 60\,{N}^{2}+109 \right)  
\\ & 
   \left\langle \tau_{{1}}{\tau_{{\frac 3 2 }}}^{2}\tau_{{\frac 5 2 }}\tau_{{3}} \right\rangle ={\frac {7}{96}}\,N \left( 144\,{N}^{4}+532\,{N}^{2}+137 \right)  
 ,\quad  \left\langle \tau_{{0}}\tau_{{\frac 1 2 }}\tau_{{2}}\tau_{{\frac 5 2 }}\tau_{{3}} \right\rangle =\frac 1 {48}\,{N}^{2} \left( 276\,{N}^{2}+481 \right)  
, \quad
 \left\langle {\tau_{{1}}}^{2}\tau_{{2}}\tau_{{\frac 5 2 }}\tau_{{3}} \right\rangle ={\frac {7}{24}}\,N \left( 36\,{N}^{4}+171\,{N}^{2}+56 \right)  
,\\& 
    \left\langle \tau_{{\frac 1 2 }}\tau_{{\frac 3 2 }}\tau_{{2}}\tau_{{\frac 5 2 }}\tau_{{3}} \right\rangle ={\frac {1}{144}}\,N \left( 1224\,{N}^{4}+4650\,{N}^{2}+1199 \right)  
, \quad
 \left\langle \tau_{{0}}{\tau_{{2}}}^{2}\tau_{{\frac 5 2 }}\tau_{{3}} \right\rangle ={\frac {1}{288}}\,N \left( 2448\,{N}^{4}+9888\,{N}^{2}+2951 \right)  
 ,\\&  \left\langle \tau_{{\frac 3 2 }}{\tau_{{2}}}^{2}\tau_{{\frac 5 2 }}\tau_{{3}} \right\rangle ={\frac {1}{576}}\,{N}^{2} \left( 6768\,{N}^{4}+47472\,{N}^{2}+44689 \right)  
, \quad 
    \left\langle {\tau_{{0}}}^{2}{\tau_{{\frac 5 2 }}}^{2}\tau_{{3}} \right\rangle =\frac{3}{8}\,{N}^{2} \left( 10\,{N}^{2}+19 \right)  
 ,\quad  \left\langle \tau_{{\frac 1 2 }}\tau_{{1}}{\tau_{{\frac 5 2 }}}^{2}\tau_{{3}} \right\rangle ={\frac {7}{16}}\,N \left( 12\,{N}^{4}+58\,{N}^{2}+17 \right)  
, \\&
 \left\langle \tau_{{0}}\tau_{{\frac 3 2 }}{\tau_{{\frac 5 2 }}}^{2}\tau_{{3}} \right\rangle =\frac 1 {24}\,N \left( 18\,{N}^{2}+67 \right)  \left( 7\,{N}^{2}+2 \right)  
,\quad
    \left\langle {\tau_{{\frac 3 2 }}}^{2}{\tau_{{\frac 5 2 }}}^{2}\tau_{{3}} \right\rangle ={\frac {1}{96}}\,{N}^{2} \left( 672\,{N}^{4}+4678\,{N}^{2}+4329 \right)  
, \quad
 \left\langle \tau_{{1}}\tau_{{2}}{\tau_{{\frac 5 2 }}}^{2}\tau_{{3}} \right\rangle =\frac 1 {12}\,{N}^{2} \left( 84\,{N}^{4}+725\,{N}^{2}+789 \right)  
 ,\\&  \left\langle \tau_{{\frac 1 2 }}{\tau_{{\frac 5 2 }}}^{3}\tau_{{3}} \right\rangle =\frac 1 {32} \,{N}^{2} \left( 112\,{N}^{4}+975\,{N}^{2}+1011 \right)  
\quad  \left\langle \tau_{{2}}{\tau_{{\frac 5 2 }}}^{3}\tau_{{3}} \right\rangle ={\frac {1}{96}}\,N \left( 432\,{N}^{6}+6045\,{N}^{4}+14707\,{N}^{2}+3114 \right)  
 ,\quad  \left\langle {\tau_{{0}}}^{2}\tau_{{\frac 1 2 }}{\tau_{{3}}}^{2} \right\rangle =\frac 1 {12}\,N \left( 36\,{N}^{2}+19 \right)  
 ,\\& \left\langle {\tau_{{\frac 1 2 }}}^{2}\tau_{{1}}{\tau_{{3}}}^{2} \right\rangle =\frac 1 2 \,{N}^{2} \left( 12\,{N}^{2}+25 \right)  
, \quad
    \left\langle \tau_{{0}}{\tau_{{1}}}^{2}{\tau_{{3}}}^{2} \right\rangle =\frac {15} 2\,{N}^{4}+{\frac {65}{4}}\,{N}^{2}+{\frac {29}{96}} 
 ,\quad  \left\langle \tau_{{0}}\tau_{{\frac 1 2 }}\tau_{{\frac 3 2 }}{\tau_{{3}}}^{2} \right\rangle =\frac 5 4 \,{N}^{2} \left( 4\,{N}^{2}+7 \right)  
 ,\quad  \left\langle {\tau_{{1}}}^{2}\tau_{{\frac 3 2 }}{\tau_{{3}}}^{2} \right\rangle ={\frac {7}{96}}\,N \left( 144\,{N}^{4}+600\,{N}^{2}+193 \right)  
, \\ & 
    \left\langle \tau_{{\frac 1 2 }}{\tau_{{\frac 3 2 }}}^{2}{\tau_{{3}}}^{2} \right\rangle ={\frac {1}{288}}\,N \left( 2160\,{N}^{4}+7920\,{N}^{2}+2113 \right)  
 ,\quad  \left\langle {\tau_{{0}}}^{2}\tau_{{2}}{\tau_{{3}}}^{2} \right\rangle ={\frac {21}{4}}\,{N}^{4}+{\frac {241}{24}}\,{N}^{2}+{\frac {109}{576}} 
, \quad
\left\langle \tau_{{\frac 1 2 }}\tau_{{1}}\tau_{{2}}{\tau_{{3}}}^{2} \right\rangle ={\frac {7}{576}}\,N \left( 720\,{N}^{4}+3096\,{N}^{2}+925 \right)  
,\\& 
    \left\langle \tau_{{0}}\tau_{{\frac 3 2 }}\tau_{{2}}{\tau_{{3}}}^{2} \right\rangle ={\frac {1}{192}}\,N \left( 1488\,{N}^{4}+5816\,{N}^{2}+1669 \right)  
, \quad
\left\langle {\tau_{{\frac 3 2 }}}^{2}\tau_{{2}}{\tau_{{3}}}^{2} \right\rangle ={\frac {1}{576}}\,{N}^{2} \left( 6192\,{N}^{4}+41544\,{N}^{2}+38299 \right)  
 ,\\&  \left\langle \tau_{{1}}{\tau_{{2}}}^{2}{\tau_{{3}}}^{2} \right\rangle =12\,{N}^{6}+{\frac {275}{3}}\,{N}^{4}+{\frac {1151}{12}}\,{N}^{2}+{\frac {205}{432}} 
, \quad 
\left\langle {\tau_{{\frac 1 2 }}}^{2}\tau_{{\frac 5 2 }}{\tau_{{3}}}^{2} \right\rangle ={\frac {1}{576}}\,N \left( 2448\,{N}^{4}+10824\,{N}^{2}+3041 \right)  
 ,\quad  \left\langle \tau_{{0}}\tau_{{1}}\tau_{{\frac 5 2 }}{\tau_{{3}}}^{2} \right\rangle ={\frac {7}{576}}\,N \left( 432\,{N}^{4}+1968\,{N}^{2}+641 \right)  
, \\ & 
 \left\langle \tau_{{1}}\tau_{{\frac 3 2 }}\tau_{{\frac 5 2 }}{\tau_{{3}}}^{2} \right\rangle ={\frac {1}{144}}\,{N}^{2} \left( 1008\,{N}^{4}+7704\,{N}^{2}+8147 \right)  
,\quad  
    \left\langle \tau_{{\frac 1 2 }}\tau_{{2}}\tau_{{\frac 5 2 }}{\tau_{{3}}}^{2} \right\rangle ={\frac {1}{1152}}\,{N}^{2} \left( 6768\,{N}^{4}+52728\,{N}^{2}+53659 \right)  
, \\ &
 \left\langle {\tau_{{2}}}^{2}\tau_{{\frac 5 2 }}{\tau_{{3}}}^{2} \right\rangle ={\frac {1}{2304}}\,N \left( 17856\,{N}^{6}+222672\,{N}^{4}+518084\,{N}^{2}+112379 \right)  
 ,\quad  \left\langle \tau_{{0}}{\tau_{{\frac 5 2 }}}^{2}{\tau_{{3}}}^{2} \right\rangle ={\frac {1}{576}}\,{N}^{2} \left( 2016\,{N}^{4}+16404\,{N}^{2}+17615 \right)  
, \\ & 
    \left\langle \tau_{{\frac 3 2 }}{\tau_{{\frac 5 2 }}}^{2}{\tau_{{3}}}^{2} \right\rangle ={\frac {1}{576}}\,N \left( 2592\,{N}^{6}+32364\,{N}^{4}+76145\,{N}^{2}+16018 \right)  
 ,\quad  \left\langle \tau_{{0}}\tau_{{\frac 1 2 }}{\tau_{{3}}}^{3} \right\rangle ={\frac {1}{192}}\,N \left( 720\,{N}^{4}+3096\,{N}^{2}+925 \right)  
, \\ & 
 \left\langle {\tau_{{1}}}^{2}{\tau_{{3}}}^{3} \right\rangle =7\,{N}^{6}+{\frac {231}{4}}\,{N}^{4}+{\frac {3367}{48}}\,{N}^{2}+{\frac {583}{1728}} 
,\quad  
    \left\langle \tau_{{\frac 1 2 }}\tau_{{\frac 3 2 }}{\tau_{{3}}}^{3} \right\rangle ={\frac {1}{64}}\,{N}^{2} \left( 336\,{N}^{4}+2488\,{N}^{2}+2573 \right)  
, \\ &
 \left\langle \tau_{{0}}\tau_{{2}}{\tau_{{3}}}^{3} \right\rangle ={\frac {43}{8}}\,{N}^{6}+{\frac {1331}{32}}\,{N}^{4}+{\frac {17179}{384}}\,{N}^{2}+{\frac {3043}{13824}} 
 ,\quad  \left\langle \tau_{{\frac 3 2 }}\tau_{{2}}{\tau_{{3}}}^{3} \right\rangle ={\frac {1}{4608}}\,N \left( 32832\,{N}^{6}+390096\,{N}^{4}+893388\,{N}^{2}+191695 \right)  
, \\ & 
    \left\langle \tau_{{1}}\tau_{{\frac 5 2 }}{\tau_{{3}}}^{3} \right\rangle ={\frac {1}{1536}}\,N \left( 6912\,{N}^{6}+92016\,{N}^{4}+239544\,{N}^{2}+56671 \right)  
,\quad 
 \left\langle {\tau_{{\frac 5 2 }}}^{2}{\tau_{{3}}}^{3} \right\rangle ={\frac {1}{1536}}\,{N}^{2} \left( 4320\,{N}^{6}+86112\,{N}^{4}+403102\,{N}^{2}+324543 \right)  
, \\ & 
 \left\langle \tau_{{\frac 1 2 }}{\tau_{{3}}}^{4} \right\rangle ={\frac {1}{3456}}\,N \left( 12096\,{N}^{6}+155088\,{N}^{4}+388476\,{N}^{2}+87751 \right)  
,\quad 
    \left\langle \tau_{{2}}{\tau_{{3}}}^{4} \right\rangle ={\frac {73}{16}}\,{N}^{8}+{\frac {4141}{48}}\,{N}^{6}+{\frac {148945}{
384}}\,{N}^{4}+{\frac {2144437}{6912}}\,{N}^{2}+{\frac {134233}{331776}}.
\end{align*}
\end{tiny}
{\bf Nonzero six--point correlators $\le\langle\tau_{\frac{d_1}{2}}\dots\tau_{\frac{d_6}{2}}\ri\rangle$, for $0\leq d_1 \leq \dots\leq d_6 \leq 4$}
\begin{tiny}
% !TEX root =penner-1.4.tex

\begin{align*}
&\left\langle{\tau_{{0}}}^{3}{\tau_{{1}}}^{3} \right\rangle =6 
,\quad 
    \left\langle{\tau_{{\frac 1 2 }}}^{3}{\tau_{{1}}}^{3} \right\rangle =24\,N 
 ,\quad  \left\langle\tau_{{0}}\tau_{{\frac 1 2 }}{\tau_{{1}}}^{4} \right\rangle =24\,N 
 ,\quad  \left\langle{\tau_{{1}}}^{6} \right\rangle =60\,{N}^{2}+5 
,\quad 
    \left\langle{\tau_{{\frac 1 2 }}}^{4}\tau_{{1}}\tau_{{\frac 3 2 }} \right\rangle =12\,N 
 ,\quad  \left\langle\tau_{{0}}{\tau_{{\frac 1 2 }}}^{2}{\tau_{{1}}}^{2}\tau_{{\frac 3 2 }} \right\rangle =12\,N 
,\quad
 \left\langle{\tau_{{0}}}^{2}{\tau_{{1}}}^{3}\tau_{{\frac 3 2 }} \right\rangle =24\,N 
,\\&
    \left\langle\tau_{{\frac 1 2 }}{\tau_{{1}}}^{4}\tau_{{\frac 3 2 }} \right\rangle =60\,{N}^{2} 
 ,\quad  \left\langle\tau_{{0}}{\tau_{{\frac 1 2 }}}^{3}{\tau_{{\frac 3 2 }}}^{2} \right\rangle =6\,N 
 ,\quad  \left\langle{\tau_{{0}}}^{2}\tau_{{\frac 1 2 }}\tau_{{1}}{\tau_{{\frac 3 2 }}}^{2} \right\rangle =8\,N 
,\quad  
    \left\langle{\tau_{{\frac 1 2 }}}^{2}{\tau_{{1}}}^{2}{\tau_{{\frac 3 2 }}}^{2} \right\rangle =40\,{N}^{2} 
,\quad
 \left\langle\tau_{{0}}{\tau_{{1}}}^{3}{\tau_{{\frac 3 2 }}}^{2} \right\rangle =60\,{N}^{2} 
 ,\quad \left\langle{\tau_{{0}}}^{3}{\tau_{{\frac 3 2 }}}^{3} \right\rangle =6\,N 
,\\& \left\langle{\tau_{{\frac 1 2 }}}^{3}{\tau_{{\frac 3 2 }}}^{3} \right\rangle =24\,{N}^{2} 
,\quad 
 \left\langle\tau_{{0}}\tau_{{\frac 1 2 }}\tau_{{1}}{\tau_{{\frac 3 2 }}}^{3} \right\rangle =30\,{N}^{2} 
 ,\quad  \left\langle{\tau_{{1}}}^{3}{\tau_{{\frac 3 2 }}}^{3} \right\rangle =30\,N \left( 4\,{N}^{2}+1 \right)  
,\quad
    \left\langle{\tau_{{0}}}^{2}{\tau_{{\frac 3 2 }}}^{4} \right\rangle =24\,{N}^{2} 
 ,\quad \left\langle\tau_{{\frac 1 2 }}\tau_{{1}}{\tau_{{\frac 3 2 }}}^{4} \right\rangle =18\,N \left( 4\,{N}^{2}+1 \right)  
 ,\\& \left\langle\tau_{{0}}{\tau_{{\frac 3 2 }}}^{5} \right\rangle =15\,N \left( 4\,{N}^{2}+1 \right)  
,\quad  
    \left\langle{\tau_{{\frac 3 2 }}}^{6} \right\rangle =15\,{N}^{2} \left( 8\,{N}^{2}+7 \right)  
 ,\quad  \left\langle{\tau_{{\frac 1 2 }}}^{5}\tau_{{2}} \right\rangle =8\,N 
,\quad
 \left\langle{\tau_{{0}}}^{4}\tau_{{1}}\tau_{{2}} \right\rangle =3 
,\quad 
    \left\langle\tau_{{0}}{\tau_{{\frac 1 2 }}}^{3}\tau_{{1}}\tau_{{2}} \right\rangle =8\,N 
 ,\quad  \left\langle{\tau_{{0}}}^{2}\tau_{{\frac 1 2 }}{\tau_{{1}}}^{2}\tau_{{2}} \right\rangle =12\,N 
 ,\\&  \left\langle{\tau_{{\frac 1 2 }}}^{2}{\tau_{{1}}}^{3}\tau_{{2}} \right\rangle =60\,{N}^{2} 
,\quad  
    \left\langle\tau_{{0}}{\tau_{{1}}}^{4}\tau_{{2}} \right\rangle =60\,{N}^{2}+5 
,\quad
 \left\langle{\tau_{{0}}}^{2}{\tau_{{\frac 1 2 }}}^{2}\tau_{{\frac 3 2 }}\tau_{{2}} \right\rangle =5\,N 
 ,\quad  \left\langle{\tau_{{0}}}^{3}\tau_{{1}}\tau_{{\frac 3 2 }}\tau_{{2}} \right\rangle =12\,N 
,\quad 
    \left\langle{\tau_{{\frac 1 2 }}}^{3}\tau_{{1}}\tau_{{\frac 3 2 }}\tau_{{2}} \right\rangle =35\,{N}^{2} 
 ,\\&  \left\langle\tau_{{0}}\tau_{{\frac 1 2 }}{\tau_{{1}}}^{2}\tau_{{\frac 3 2 }}\tau_{{2}} \right\rangle =40\,{N}^{2} 
 ,\quad  \left\langle{\tau_{{1}}}^{4}\tau_{{\frac 3 2 }}\tau_{{2}} \right\rangle =60\,N \left( 2\,{N}^{2}+1 \right)  
,\quad
    \left\langle\tau_{{0}}{\tau_{{\frac 1 2 }}}^{2}{\tau_{{\frac 3 2 }}}^{2}\tau_{{2}} \right\rangle =22\,{N}^{2} 
 ,\quad  \left\langle{\tau_{{0}}}^{2}\tau_{{1}}{\tau_{{\frac 3 2 }}}^{2}\tau_{{2}} \right\rangle =35\,{N}^{2} 
 ,\quad  \left\langle\tau_{{\frac 1 2 }}{\tau_{{1}}}^{2}{\tau_{{\frac 3 2 }}}^{2}\tau_{{2}} \right\rangle =30\,N \left( 3\,{N}^{2}+1 \right)  
,\\&
    \left\langle{\tau_{{\frac 1 2 }}}^{2}{\tau_{{\frac 3 2 }}}^{3}\tau_{{2}} \right\rangle =N \left( 57\,{N}^{2}+17 \right)  
 ,\quad  \left\langle\tau_{{0}}\tau_{{1}}{\tau_{{\frac 3 2 }}}^{3}\tau_{{2}} \right\rangle =6\,N \left( 13\,{N}^{2}+4 \right)  
 ,\quad \left\langle\tau_{{1}}{\tau_{{\frac 3 2 }}}^{4}\tau_{{2}} \right\rangle =7\,{N}^{2} \left( 21\,{N}^{2}+22 \right)  
,\quad
    \left\langle{\tau_{{0}}}^{3}\tau_{{\frac 1 2 }}{\tau_{{2}}}^{2} \right\rangle =6\,N 
 ,\quad  \left\langle{\tau_{{\frac 1 2 }}}^{4}{\tau_{{2}}}^{2} \right\rangle =28\,{N}^{2} 
 ,\\& \left\langle\tau_{{0}}{\tau_{{\frac 1 2 }}}^{2}\tau_{{1}}{\tau_{{2}}}^{2} \right\rangle =30\,{N}^{2} 
,\quad
    \left\langle{\tau_{{0}}}^{2}{\tau_{{1}}}^{2}{\tau_{{2}}}^{2} \right\rangle =40\,{N}^{2}+\frac {10} 3 
 ,\quad \left\langle\tau_{{\frac 1 2 }}{\tau_{{1}}}^{3}{\tau_{{2}}}^{2} \right\rangle =10\,N \left( 12\,{N}^{2}+5 \right)  
 ,\quad  \left\langle{\tau_{{0}}}^{2}\tau_{{\frac 1 2 }}\tau_{{\frac 3 2 }}{\tau_{{2}}}^{2} \right\rangle =22\,{N}^{2} 
,\\&
    \left\langle{\tau_{{\frac 1 2 }}}^{2}\tau_{{1}}\tau_{{\frac 3 2 }}{\tau_{{2}}}^{2} \right\rangle =\frac 1 2 \,N \left( 156\,{N}^{2}+53 \right)  
 ,\quad  \left\langle\tau_{{0}}{\tau_{{1}}}^{2}\tau_{{\frac 3 2 }}{\tau_{{2}}}^{2} \right\rangle =\frac 5 2 \,N \left( 36\,{N}^{2}+17 \right)  
 ,\quad  \left\langle\tau_{{0}}\tau_{{\frac 1 2 }}{\tau_{{\frac 3 2 }}}^{2}{\tau_{{2}}}^{2} \right\rangle =\frac{1}{6}\,N \left( 336\,{N}^{2}+113 \right)  
,\quad
    \left\langle{\tau_{{1}}}^{2}{\tau_{{\frac 3 2 }}}^{2}{\tau_{{2}}}^{2} \right\rangle =\frac 7 2 \,{N}^{2} \left( 48\,{N}^{2}+65 \right)  
 ,\\&  \left\langle\tau_{{\frac 1 2 }}{\tau_{{\frac 3 2 }}}^{3}{\tau_{{2}}}^{2} \right\rangle =\frac 3 2 \,{N}^{2} \left( 76\,{N}^{2}+87 \right)  
 ,\quad  \left\langle{\tau_{{0}}}^{3}{\tau_{{2}}}^{3} \right\rangle =24\,{N}^{2}+2 
,\quad
    \left\langle{\tau_{{\frac 1 2 }}}^{3}{\tau_{{2}}}^{3} \right\rangle =22\,N \left( 3\,{N}^{2}+1 \right)  
 ,\quad  \left\langle\tau_{{0}}\tau_{{\frac 1 2 }}\tau_{{1}}{\tau_{{2}}}^{3} \right\rangle =6\,N \left( 12\,{N}^{2}+5 \right)  
 ,\\& \left\langle{\tau_{{1}}}^{3}{\tau_{{2}}}^{3} \right\rangle =210\,{N}^{4}+315\,{N}^{2}+{\frac {49}{8}}
,\quad
    \left\langle{\tau_{{0}}}^{2}\tau_{{\frac 3 2 }}{\tau_{{2}}}^{3} \right\rangle =\frac 3 4 \,N \left( 76\,{N}^{2}+35 \right)  
 ,\quad \left\langle\tau_{{\frac 1 2 }}\tau_{{1}}\tau_{{\frac 3 2 }}{\tau_{{2}}}^{3} \right\rangle =7/4\,{N}^{2} \left( 84\,{N}^{2}+109 \right)  
 ,\quad  \left\langle\tau_{{0}}{\tau_{{\frac 3 2 }}}^{2}{\tau_{{2}}}^{3} \right\rangle =38\,{N}^{2} \left( 3\,{N}^{2}+4 \right)  
,\\&
    \left\langle{\tau_{{\frac 3 2 }}}^{3}{\tau_{{2}}}^{3} \right\rangle =  \frac 1 4 \,N \left( 804\,{N}^{4}+2261\,{N}^{2}+467 \right)
 ,\quad  \left\langle{\tau_{{\frac 1 2 }}}^{2}{\tau_{{2}}}^{4} \right\rangle =32\,{N}^{2} \left( 4\,{N}^{2}+5 \right)  
 ,\quad \left\langle\tau_{{0}}\tau_{{1}}{\tau_{{2}}}^{4} \right\rangle =140\,{N}^{4}+210\,{N}^{2}+{\frac {49}{12}}
,\\&
    \left\langle\tau_{{1}}\tau_{{\frac 3 2 }}{\tau_{{2}}}^{4} \right\rangle =\frac 1 6\,N \left( 1488\,{N}^{4}+4584\,{N}^{2}+1129 \right)
 ,\quad  \left\langle\tau_{{\frac 1 2 }}{\tau_{{2}}}^{5} \right\rangle = {\frac {5}{12}}\,N \left( 528\,{N}^{4}+1560\,{N}^{2}+361 \right)  
 ,\quad  \left\langle{\tau_{{2}}}^{6} \right\rangle = 348\,{N}^{6}+1955\,{N}^{4}+{\frac {6533}{4}}\,{N}^{2}+{\frac {1225}{
144}}.
\end{align*}
\end{tiny}
%\bibliographystyle{alpha}
%\bibliography{bib}
%
\newcommand{\etalchar}[1]{$^{#1}$}

\end{document}